\documentclass[12pt,preprint]{aastex}

\newcommand{\SIRTF}{{\em SIRTF}}

\shorttitle{SEDs of Debris Disks}
\shortauthors{Wolf et al.}

\begin{document}


\title{
  Model Spectral Energy Distributions\\ of Circumstellar Debris Disks\\
  \vspace*{3mm}
  {\normalsize I. Analytic Disk Density Distributions}
  \vspace*{3mm}
}


\author{Sebastian Wolf, Lynne A. Hillenbrand}
\affil{California Institute of Technology, 1201 E California Blvd,\\ Mail code 105-24, Pasadena, CA 91125, USA}

%

%




\begin{abstract}
We present results of a study aimed at deriving fundamental properties of circumstellar debris disks from observed
infrared to submillimeter spectral energy distributions. This investigation is motivated by 
increasing telescope/detector sensitivity, in particular the expected availability of the
{\em Space Infrared Telescope Facility} (\SIRTF) 
followed by the {\em Stratospheric Observatory for Infrared Astronomy} ({\em SOFIA}), 
which will enable detailed studies with large source samples of late stage circumstellar disk
and planetary system evolution.
We base our study on an analytic model of the disk density distribution and geometry, taking into account 
existing constraints from observations and results of theoretical investigations of debris disks.
We also outline the effects of the most profound characteristics of circumstellar dust 
including the grain size distribution and dust chemical composition.  
In particular we find that an increasing iron content in silicates mainly causes an increase of the dust absorption
effiency and thus increases the dust reemission continuum. Furthermore, the influence of the sp$^2$/sp$^3$
hybridization ratio in carbon grains on the spectral energy distribution is examined.
We investigate the influence 
of various parameters on the resulting dust scattering and absorption/reemission spectral energy distribution
and discuss the possibility for distinguishing between different disks from their infrared
to submillimeter spectra.  
The strength and shape of amorphous silicate may be particularly diagnostic of debris disk evolutionary stages.
Since the appearance of these features at 10$\mu$m and 20$\mu$m
depends on the relative abundance of small grains - and therefore
the minimum grain size and slope of the grain size distribution - they can be used to trace recent
collisional processes in debris disks. Thus, debris disk surveys containing statistically large numbers
of objects should reveal the likelihood of collisions and therefore the evolution of
dust/planetesimals in debris disks. 
The results of our study underline the importance of knowledge of the stellar photospheric flux, especially
in the near to mid-infrared wavelength range, for a proper analysis of debris disk spectral energy
distributions:
While the quality of subtraction of the direct stellar light at far-infrared wavelengths determines
the accuracy of the mass estimate in the disk, our simulations show that the remaining stellar contribution
due to scattering at near- to mid-infrared wavelengths constrains the dust grain size
and chemical composition, e.g. the iron abundance in silicate grains.
\end{abstract}




\section{Introduction}

Debris disks are solar system sized dust disks with micron-sized grains produced as by-products of collisions
between asteroid-like bodies left over from the planet formation process.
In the case of our solar system, the debris
of Jupiter-family short-period comets and colliding asteroids represents the dominant source of zodiacal dust
located between Mars and Jupiter.  A second belt of dust
is located beyond the orbit of Neptune (see, e.g., Dermott et al.~1992, Liou, Dermott, \& Xu~1995). 
Besides the solar system, optical to mid-infrared images of $\beta$~Pic
(see, e.g., Kalas \& Jewitt~1995; Weinberger, Becklin, \& Zuckerman~2003)
and submillimeter images of Vega, Fomalhaut, and $\epsilon$~Eri 
(Holland et al.~1998; Greaves et al.~1998) have revealed spatially resolved debris disks which were first
inferred from observations of infrared flux excesses above photospheric values with IRAS.
Based on studies with {\em ISO},
the disk fraction is thought to decrease dramatically with age, amounting to much less than 10\% for stars 
with ages $\ge 1$Gyr (e.g., Spangler et al.~2001; see also Habing et al.~2001).
Planetary debris disks are assumed to represent the almost final stage of the circumstellar disk evolution process, 
i.e., they are the evolutionary products of ongoing planet formation.

In contrast to optically thick young circumstellar disks around Herbig Ae/Be 
and T\,Tauri stars with spatial structures dominated
by gas dynamics, the much lower optical depth and lower gas-to-dust mass ratio in debris disks 
(Zuckermann, Forveille, \& Kastner~1995; Dent et al.~1995; 
Artymowicz~1997; Liseau \& Artymowicz~1998; Greaves, Coulson, \& Holland~2000; 
Thi et al.~2001) let the stellar radiation  -- in addition to gravitation -- be responsible for the disk structure.
The Poynting-Robertson
effect, stellar wind drag, and - if existing - gravitational stirring by embedded planets
are all important in determining the dust population and disk structure
(Liou \& Zook~1999, Grady et al.~2000, Moro-Mart\'{\i}n \& Malhotra~2002).
Similar to T\,Tauri-like disks, however, embedded planets may alter the debris disk structure remarkably
(see, e.g., Gor'kavyi et al.~1997, Kenyon \& Bromley~2001).
While planets may open gaps in gas-dominated young circumstellar disks 
(see, e.g., Bryden et al.~1999, 2000; Kley~2000),
they decrease  the particle density within their orbits in the case of a debris disk,
provided that the dust sources are located outside the planet's orbit.

Since the mass of small grains in debris disks and therefore the thermal dust reemission from these disks 
is much smaller than in case of T\,Tauri disks,
only a very limited sample of observations exists so far. However, because of the high sensitivity of the mid-infrared
detectors aboard the {\em Space Infrared Telescope Facility} (\SIRTF) a substantial
increase in the total number and in the specific information about debris disks is expected 
(c.f. Meyer~2001). In order to allow a quick classification of the basic characteristics of observed objects we perform
a preparatory study with the goal of revealing the influence of major disk and dust parameters,
such as the inner/outer radius, density distribution, grain size distribution, composition, etc.,
on the emergent spectral energy distributions (SED) of debris disk systems.

Our first investigation, described in this publication, is based on a simply-structured, optically thin
analytic disk model with a limited number of basic parameters. This approach allows us to distinguish clearly between
the influences of different model parameters. Furthermore, it provides an easy to reproduce, flux-scalable
(flux is directly proportional to dust mass in the very optically thin limit)
database of debris disk SEDs\footnote{A database with all dust SEDs presented in this publication
is available at\\{\tt http://mc.caltech.edu/$\sim$swolf/downloads/debris/} (or contact the authors)}.
In order to consider a reasonable model setup and parameter ranges we use boundary constraints given by
previous observations and numerical simulations.

In \S~\ref{sec_sedcalc}
we describe the basic model and the method by which the SEDs have been calculated.
In \S~\ref{efficiencies}, \S~\ref{sec_parstudy}, and \S~\ref{sec-seddis}
we discuss specific properties of different species expected in debris disks and provide a detailed
analysis of the influence of the different model parameters on the SED.
Finally, we derive the SIRTF-appropriate colors $m(8\mu$m$)-m(24\mu$m$)$
and $m(24\mu$m$)-m(70\mu$m$)$ from the simulated debris disk SEDs (\S~\ref{colors}).

\section{Model description}\label{sec_sedcalc}

According to the expected structure of debris disks, we base our simulations on the assumption of an optically
thin disk, whereby the optical depth is measured in the disk midplane as seen from the star.
In this case, multiple scattering of radiation and dust heating due to dust-reemission can be neglected.
The only radiative processes to be considered are scattering, absorption, and reemission of stellar radiation
by dust grains. The two-dimensional radiative transfer problem can then be reduced to a one-dimensional problem.

In all simulations we chose the shortest wavelength of stellar emission to measure the optical depth.
This wavelength amounts to 0.2\,$\mu$m in practice due to the focus on solar-type stars and the availability
of optical constants.
In order to satisfy the optically thin assumption in our calculations
we apply a limit for the optical depth of $\tau_{\rm max}=0.05$,
which translates into a maximum attenuation of stellar light by 4.8\,\% at 0.2\,$\mu$m.
Depending on the grain size and chemical composition, the absorption efficiency decreases towards longer wavelengths,
but may reach a local maximum in the mid-infrared wavelength region for particular dust chemistries
(see, e.g., Fig.~\ref{scaabs-1} 
and Fig.~\ref{scaabs-5} for the wavelength-dependent absorption efficiency of MgSiO$_3$ and crystalline Olivine,
respectively). However, the flux of the considered stellar radiation source is small in the mid-infrared compared to 
that in the ultraviolet/optical wavelength region 
($S_{\nu}(10\mu$m$)/S_{\nu}(0.55\mu$m$) \approx 1.3\times10^{-4}$ for the solar SED: see below). 
Due to the small amount of dust in the disk/shell for $\tau_{\rm max}=0.05$, 
the flux reemitted in the mid-infrared wavelength range 
by hot dust is also negligible for further disk heating.

Each dust grain is heated by direct stellar radiation only. Thus the dust grain temperature is a function of
the optical parameters of the grains, the incident stellar radiation, and the distance $d$ from the star.
In this case the radiative transfer equation has a simple solution which allows one to derive the distance from
the star at which the dust has a certain temperature. Let
\begin{equation}\label{eq_dist1} 
  L_{\lambda, \rm s}         = 4\pi R_{\rm s}^2                         \pi B_{\lambda}(T_{\rm s})
\end{equation}
be the monochromatic luminosity of the star (radius $R_{\rm s}$, effective temperature $T_{\rm s}$) 
at wavelength $\lambda$ and
\begin{equation}\label{eq_dist2}
  L_{\lambda, \rm g}^{\rm abs} = L_{\rm s}        Q_{\lambda}^{\rm abs} \frac{\pi a^2}{4\pi d^2}
  \hspace*{1cm}{\rm and}
\end{equation}
\begin{equation}\label{eq_dist3}
  L_{\lambda, \rm g}^{\rm emi} = 4\pi a^2 Q_{\lambda}^{\rm abs} \pi B_{\lambda}(T_{\rm g})
\end{equation}
be the absorbed and reemitted luminosity of a dust grain with radius $a$ and resulting temperature $T_{\rm g}$
at the (unknown) distance $d$ from the star. Using the constraint of energy conservation
\begin{equation}\label{eq_dist4}
  \int_{0}^{\infty} L_{\lambda, \rm g}^{\rm emi} d\lambda = \int_{0}^{\infty} L_{\lambda, \rm g}^{\rm abs} d\lambda
\end{equation}
one derives the distance of the grain from the star as
\begin{equation}\label{eq_dist5}
  d(T_{\rm g}) = \frac{R_{\rm s}}{2} 
  \left[
    \frac{\int_0^{\infty} d\lambda Q_{\lambda}^{\rm abs}(a) B_{\lambda}(T_{\rm s})}
	 {\int_0^{\infty} d\lambda Q_{\lambda}^{\rm abs}(a) B_{\lambda}(T_{\rm g})}
	 \right]^{1/2}.
\end{equation}
If the dust sublimation temperature is known, Eq.~\ref{eq_dist4} also allows one to estimate the sublimation
radius for each dust component in the shell (characterized by the grain radius and chemical composition).
The flux of the light scattered by a single dust grain amounts to
\begin{equation}\label{eq_scatt}
  L_{\lambda, \rm g}^{\rm sca} = L_{\lambda, {\rm s}}A Q_{\lambda}^{\rm sca} 
  \left(
  \frac{a}{2d}
  \right)^2,
\end{equation}
where $A$ is the dust grain's albedo.

We calculate the temperature
distribution of the dust based on 500 logarithmically equidistantly distributed wavelengths
in the interval [0.2$\mu$m, 500$\mu$m]. 
According to Eq.~\ref{eq_dist5}, distances $d(T)$ corresponding to 500 temperatures equidistantly
distributed between 2.73~K and the dust grain sublimation temperature of the grains are determined.
The dust reemission and scattering is calculated at 200 logarithmically
equidistantly distributed wavelengths in the interval [2$\mu$m, 200$\mu$m]. The latter interval was chosen
according to the wavelength range covered by broad band filters aboard \SIRTF: 
3.6$\mu$m - 160$\mu$m.
The net spectral energy distribution results from a simple summation of the reemitted and scattered light
contributions from all grains.  

Since the SED depends on the radial density distribution only, we consider a spherical shell instead
of a circumstellar disk. This simplification has no effect on the resulting temperature distribution 
and/or SED, but reduces the number of free parameters.
If a disk with an opening angle from the midplane $\theta$ - but the same radial density distribution - is considered instead, 
the SEDs shown in the subsequent sections
have to be multiplied by the factor
\begin{equation}
  f_{\theta} = \frac{1}{2}
  \left[ 
    \cos\left(\frac{\pi-\theta}{2}\right) - \cos\left(\frac{\pi+\theta}{2}\right)
    \right],
\end{equation}
but the relative shape of the dust contribution to the net SED by scattering, absorption and reemission
remains unchanged.  For $\theta = 10^\circ$, $f_{\theta} \approx 0.087$ while for 
$\theta = 45^\circ$, $f_{\theta} \approx 0.38$. 

Our parameters are as follows, unless specified otherwise in subsequent sections.
The applied radial density profile in the disk/shell follows a power-law
\begin{equation}\label{eq_radprof}
  n(r) \propto r^{-q}.
\end{equation}
Except for a parameter study in \S~\ref{diskdendis} we use $q=1$, representing a debris-disk with no perturbations
by embedded planets and no particular assumptions about dust production processes (see \S~\ref{diskdendis}
for a more detailed description). 
If not otherwise noted, the mass of the disk amounts to $10^{-10}\,$M$_{\sun}$,
the inner disk radius is equivalent to the dust sublimation radius, and the outer radius amounts to 100\,AU. 
The central star has a solar-type SED (we use the solar SED published by Labs \& Neckel~1968,
extended by a blackbody SED beyond 151\,$\mu$m).
The distance to the system is assumed to be 50\,pc.

\section{Dust grain parameters}\label{efficiencies}

The dust found in young, massive circumstellar disks around T\,Tauri and Herbig Ae/Be stars consists mainly
of two distinct species: a silicate component and a carbonaceous component (Savage \& Mathis~1979, Draine \& Lee~1984;
Malfait et al.~1999).
Based on the investigation of the interstellar extinction Weingartner \& Draine (2001; see also
Draine \& Lee 1984) derived optical properties and relative abundances of both.
Despite a different size distribution of the grains in the circumstellar vs.\ the interstellar
environment (see, e.g., Mathis, Rumpl, \& Nordsieck~1977; Beckwith \& Sargent~1991; Miyake \& Nakagawa~1993), 
the chemical composition turned out to be very similar, allowing detailed studies of the circumstellar environment 
and disks of young stellar objects.
However, the very complex, (magneto-) hydrodynamically dominated optically thick disks permits
deriviation of only very basic further information about the dust, such as the grain size distribution
exponent and minimum and approximate maximum upper grain size 
(e.g., Wolf, Padgett, \& Stapelfeldt and references therein).
In contrast to the situation for young optically thick disks, older optically thin
debris disks are expected to allow a more detailed investigation of dust grain parameters,
in particular the chemical components and the grain sizes.  An important caveat to this claim of 
simpler modelling for debris disks is that because the dust dynamics are determined by a combination
of radiation pressure, the Poynting-Robertson effect, and the 
(possible) production of dust due to collision processes and gravitational interaction with
possibly embedded planets, the disk density distribution may be dictated by
several additional parameters, the influence of which 
on the dust reemission/scattered light SED will be discussed in detail in Sect.~\ref{sec_parstudy}
and explored more thoroughly in a forthcoming paper.

Fortunately, existing laboratory measurements provide a rich database of dust properties
expected in different environments (defined by the dust formation conditions, temperature,
abundances of particular elements, etc.). 
A compilation of a large number of measured optical dust parameters has been published by Henning et al.~(1999).
Before applying selected chemical compositions in
our investigation of the influence of disk model parameters on the observable SED, we 
outline in brief the most characteristic differences between the dust compositions expected to dominate.


\begin{figure}[t]
  \begin{center}
    \resizebox{0.47\hsize}{!}{\includegraphics{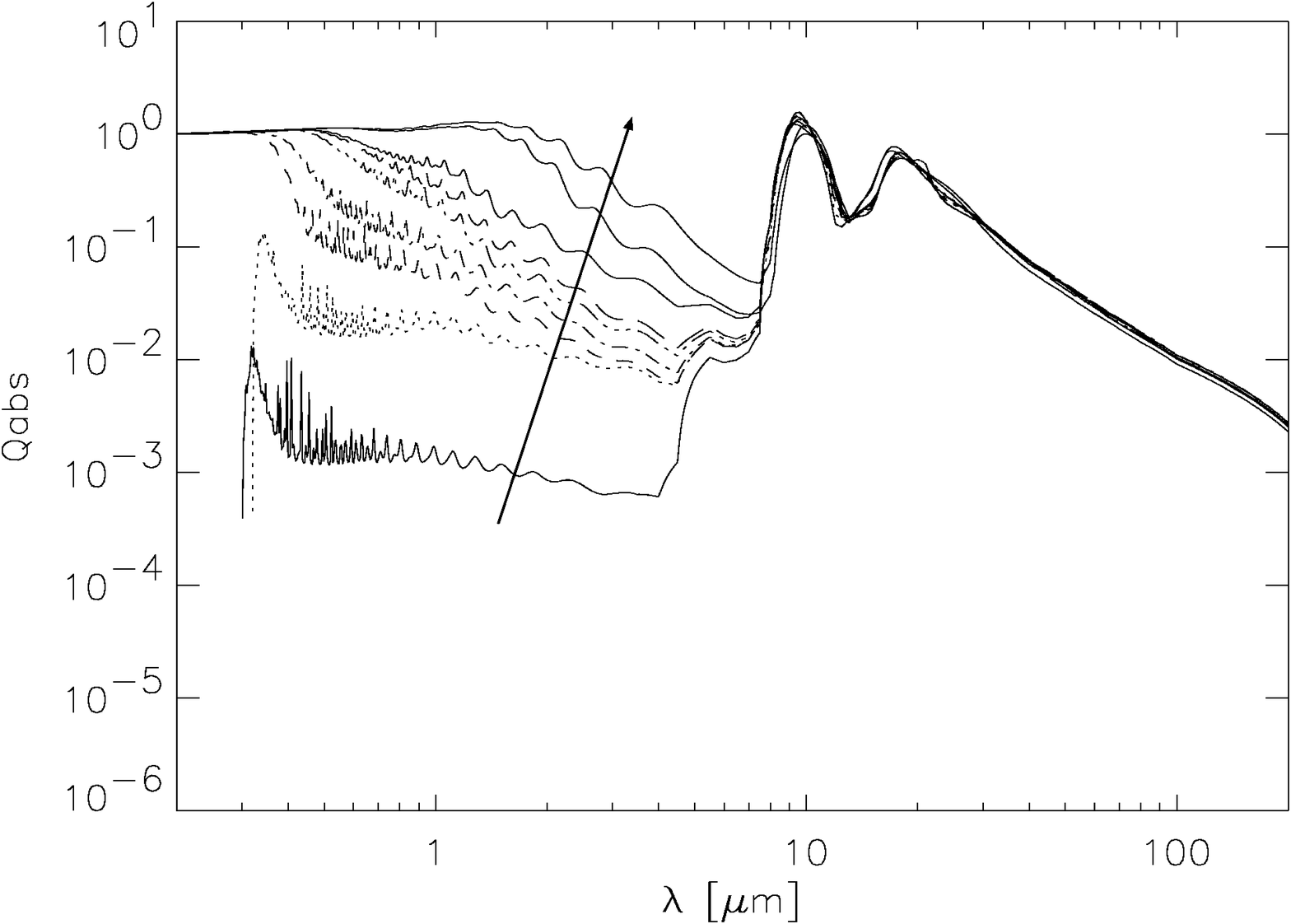}}
    \resizebox{0.47\hsize}{!}{\includegraphics{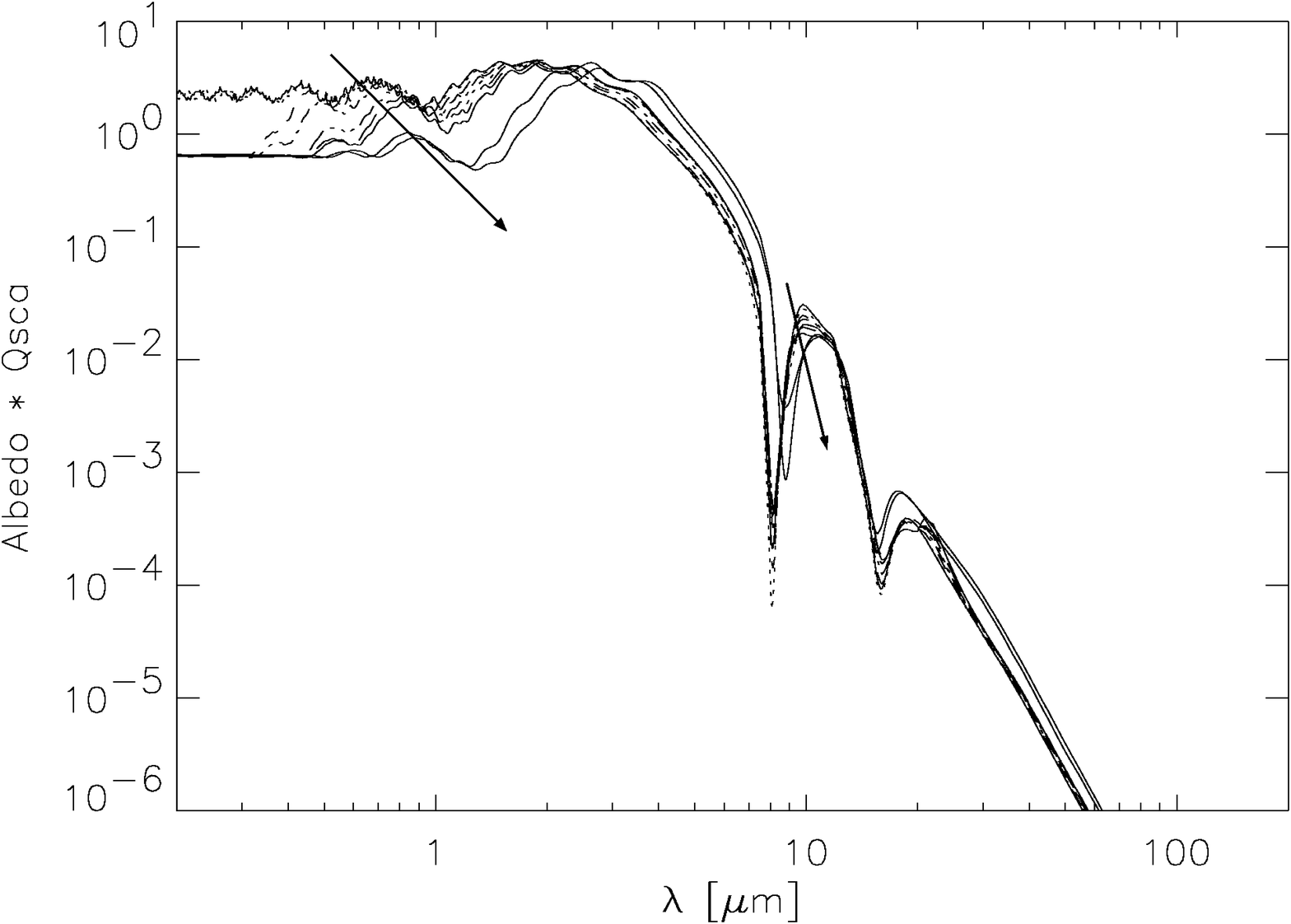}}
  \end{center}
  \caption{Absorption and scattering coefficients over the 
    transition from Fe-deficient to Fe-rich Silicates.  Arrow indicates increasing iron content from 
    Mg SiO$_3$
    $\rightarrow$
    Mg$_{0.95}$Fe$_{0.05}$SiO$_3$ 
    $\rightarrow$
    Mg$_{0.8}$Fe$_{0.2}$SiO$_3$ 
    $\rightarrow$
    Mg$_{0.7}$Fe$_{0.3}$SiO$_3$ 
    $\rightarrow$
    Mg$_{0.6}$Fe$_{0.4}$SiO$_3$ 
    $\rightarrow$
    Mg$_{0.5}$Fe$_{0.5}$SiO$_3$
    $\rightarrow$
    Mg$_{0.4}$Fe$_{0.6}$SiO$_3$
    $\rightarrow$
    MgFeSiO$_4$ 
    $\rightarrow$
    Mg$_{0.8}$Fe$_{1.2}$ SiO$_4$.
    The optical parameters have been calculated for grains with a radius of 1\,$\mu$m.}
  \label{si-reihe-q}
  \bigskip
\end{figure}

\begin{figure}[t]
  \begin{center}
    \resizebox{0.5\hsize}{!}{\includegraphics{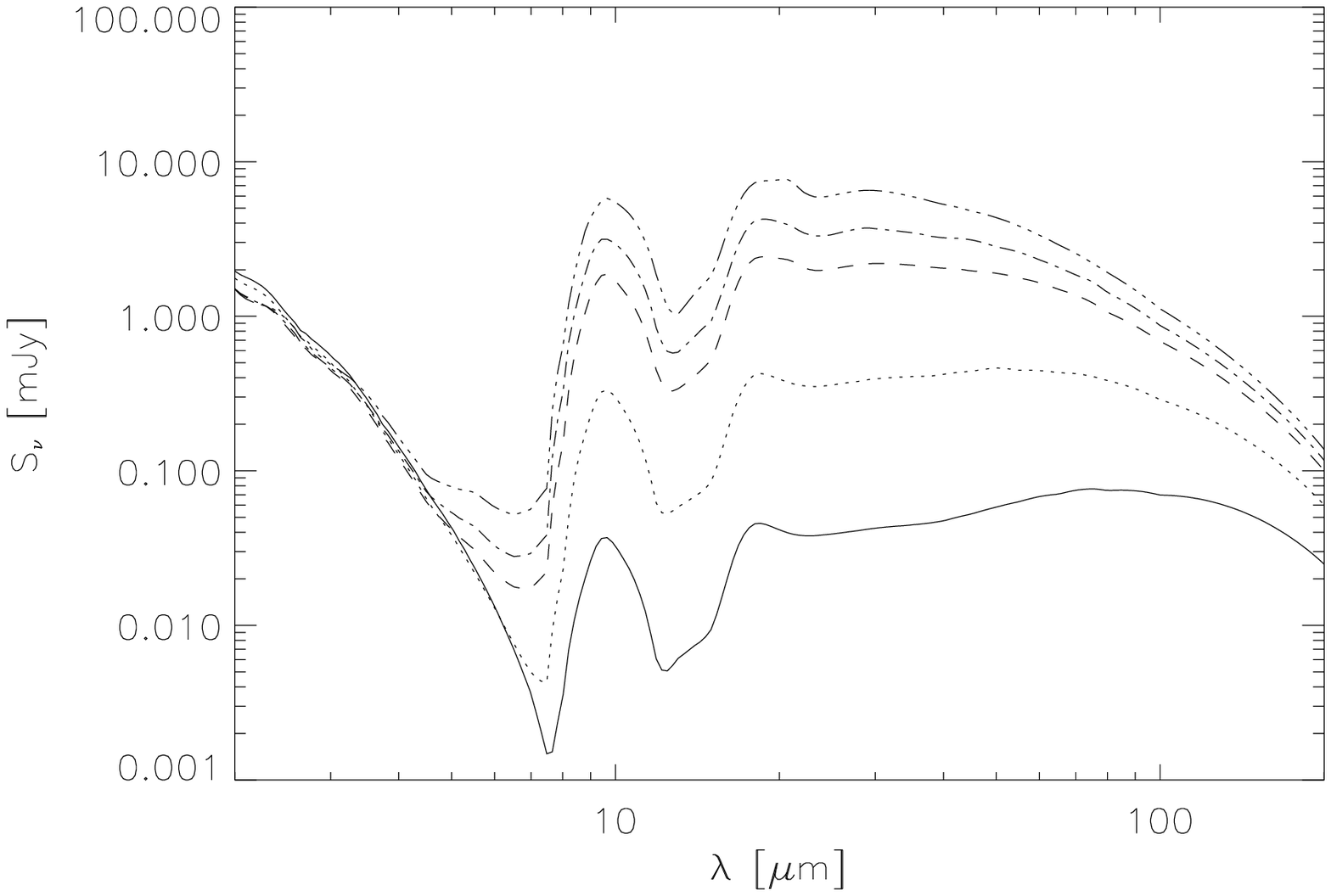}}
    \resizebox{0.5\hsize}{!}{\includegraphics{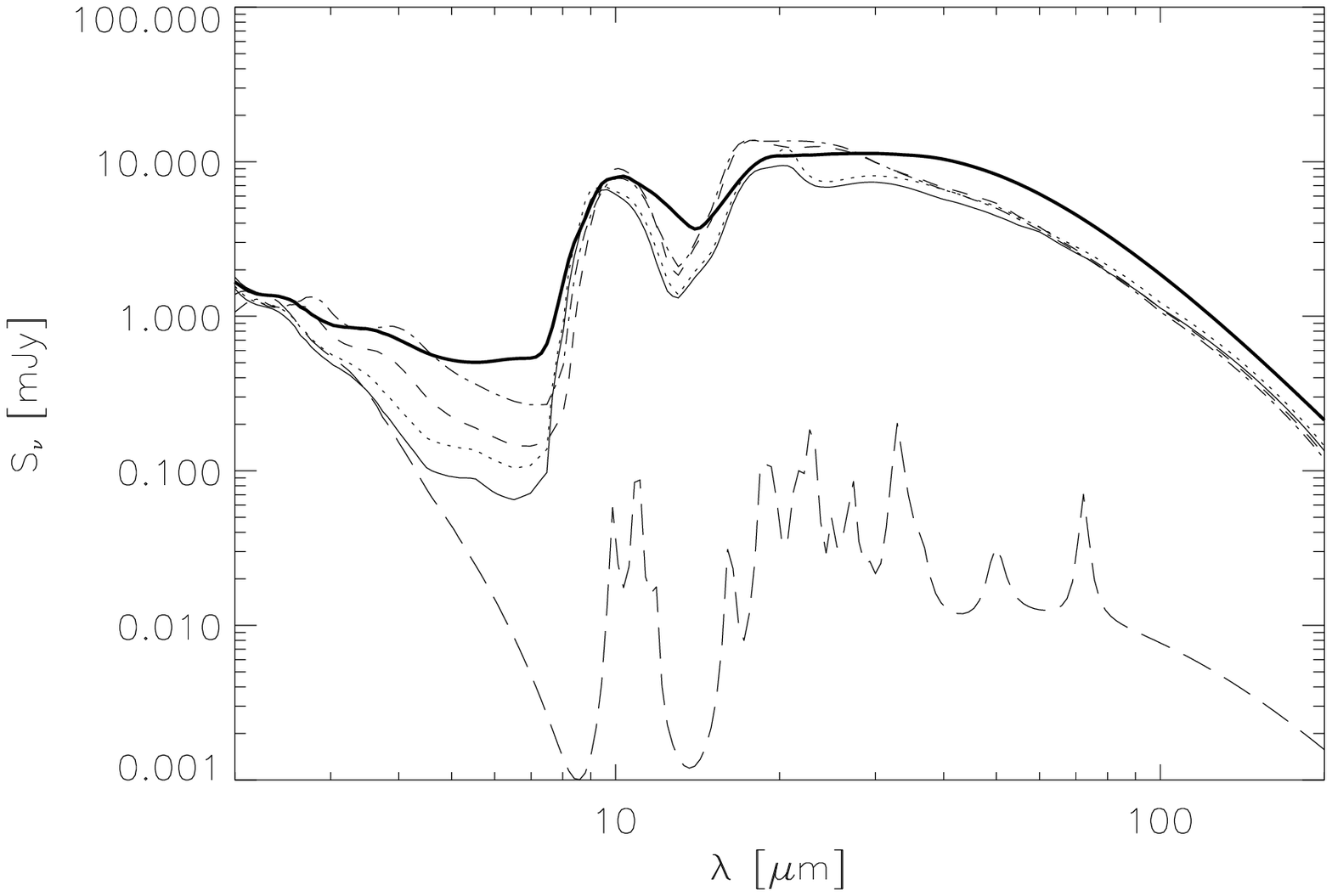}}
    \resizebox{0.5\hsize}{!}{\includegraphics{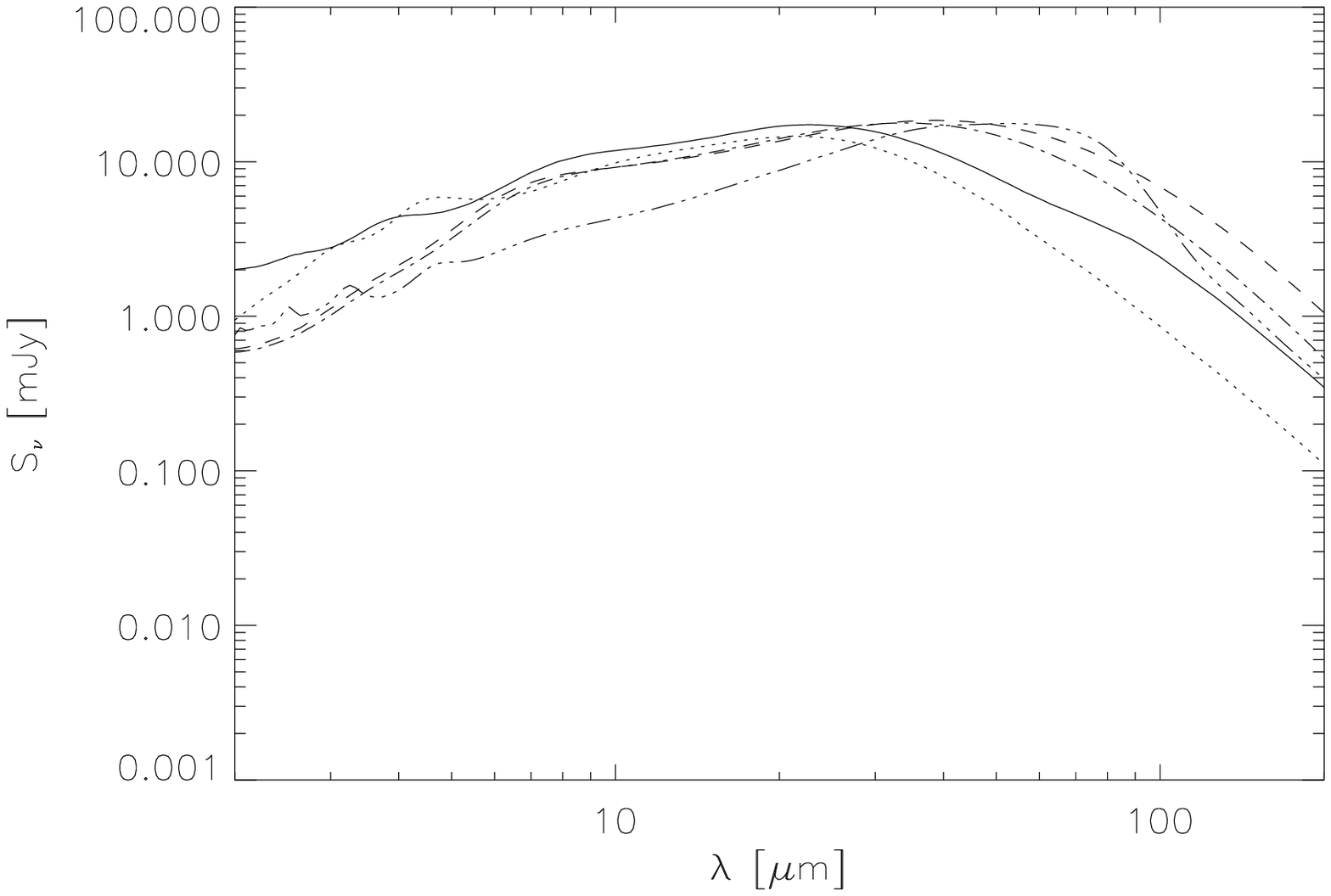}}
  \end{center}
  \caption{
    Contributions from single dust compositions. 
    {\bf Top:}
    Mg SiO$_3$                     (solid line),
    Mg$_{0.95}$Fe$_{0.05}$SiO$_3$  (dotted line),
    Mg$_{0.8}$Fe$_{0.2}$SiO$_3$    (dashed line),
    Mg$_{0.7}$Fe$_{0.3}$SiO$_3$    (dash-dotted line),
    Mg$_{0.6}$Fe$_{0.4}$SiO$_3$    (dash-dot-dot-dotted line).
    {\bf Middle:}
    Mg$_{0.5}$Fe$_{0.5}$SiO$_3$    (solid line),
    Mg$_{0.4}$Fe$_{0.6}$SiO$_3$    (dotted line),
    MgFeSiO$_4$                    (dashed line),
    Mg$_{0.8}$Fe$_{1.2}$SiO$_4$    (dash-dotted line),
    ``Astronomical Silicate''      (fat solid line).
    Crystalline Olivine            (long-dashed line).
    {\bf Bottom:}
    C (400\,K  - solid line),
    C (600\,K  - dotted line),
    C (800\,K  - dashed line),
    C (1000\,K - dash-dotted line),
    Graphite (fat solid line).
    In all cases a single grain radius of $a$ = 1$\mu$m
    and a disk mass of $10^{-10}$M$_{\sun}$ are assumed.
  }
  \label{si-reihe-sed}
  \bigskip
\end{figure}

\begin{figure}[t]
  \begin{center}
    \resizebox{0.44\hsize}{!}{\includegraphics{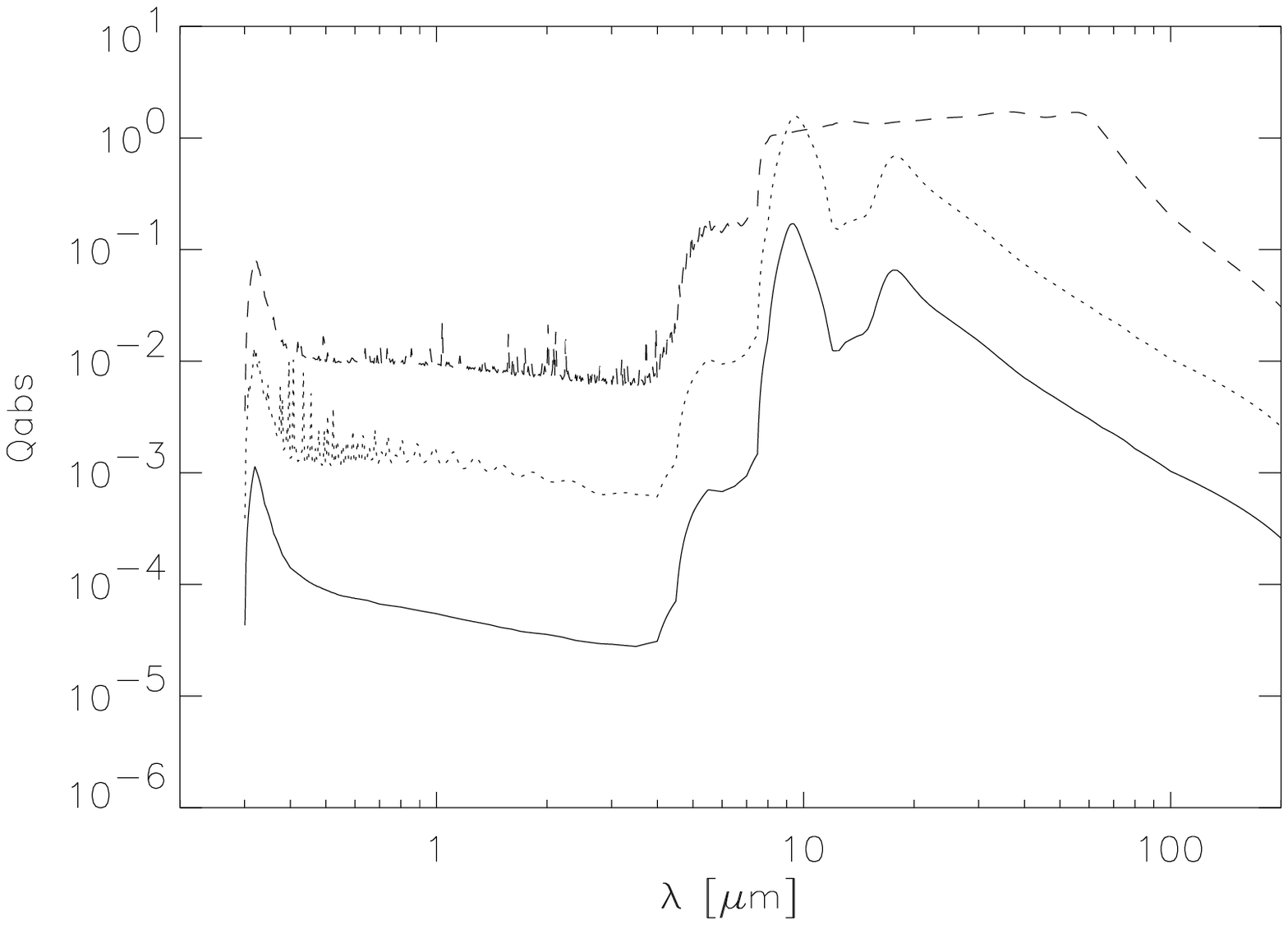}}\hspace*{10mm}
    \resizebox{0.44\hsize}{!}{\includegraphics{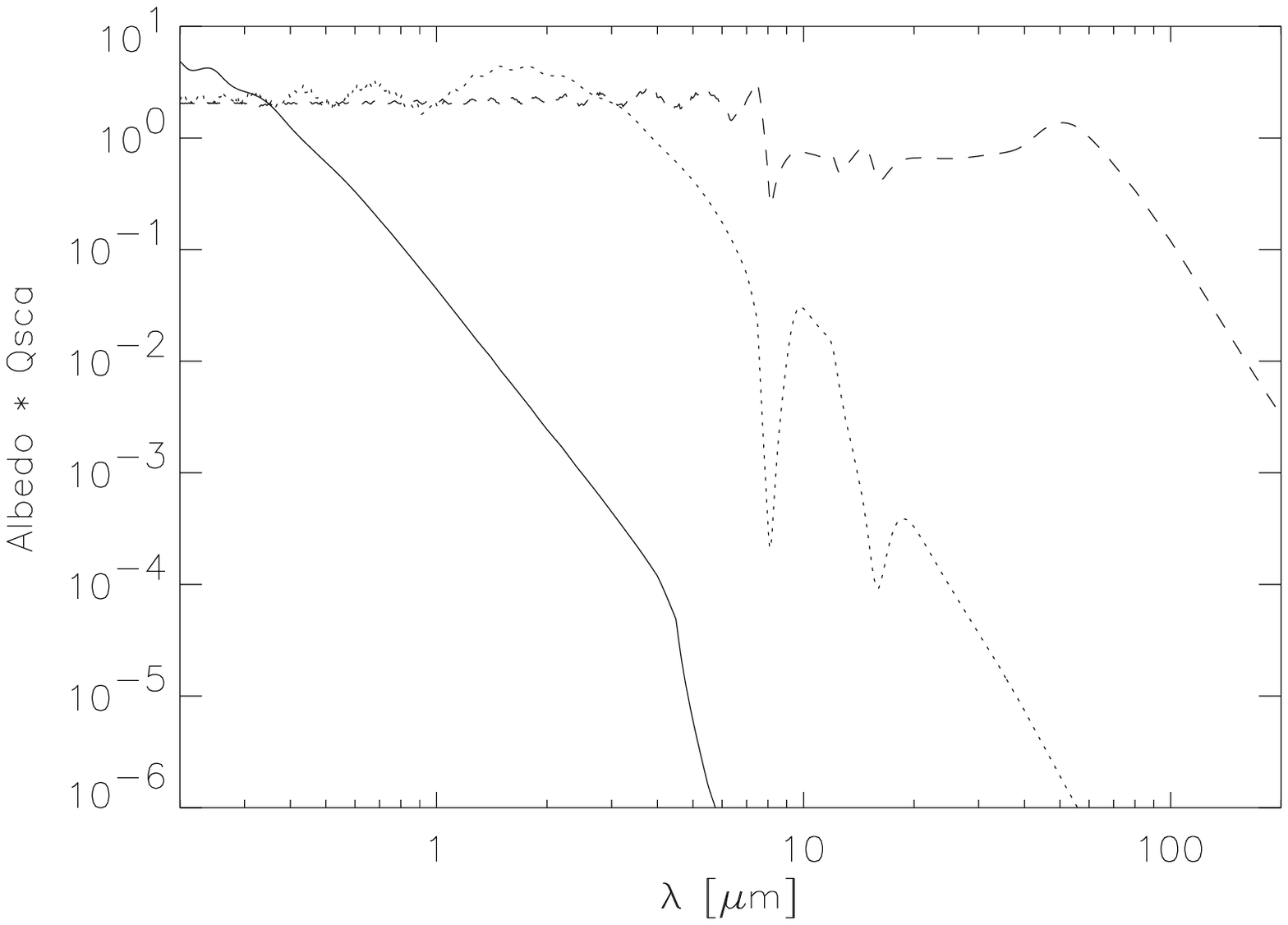}}
    \resizebox{0.44\hsize}{!}{\includegraphics{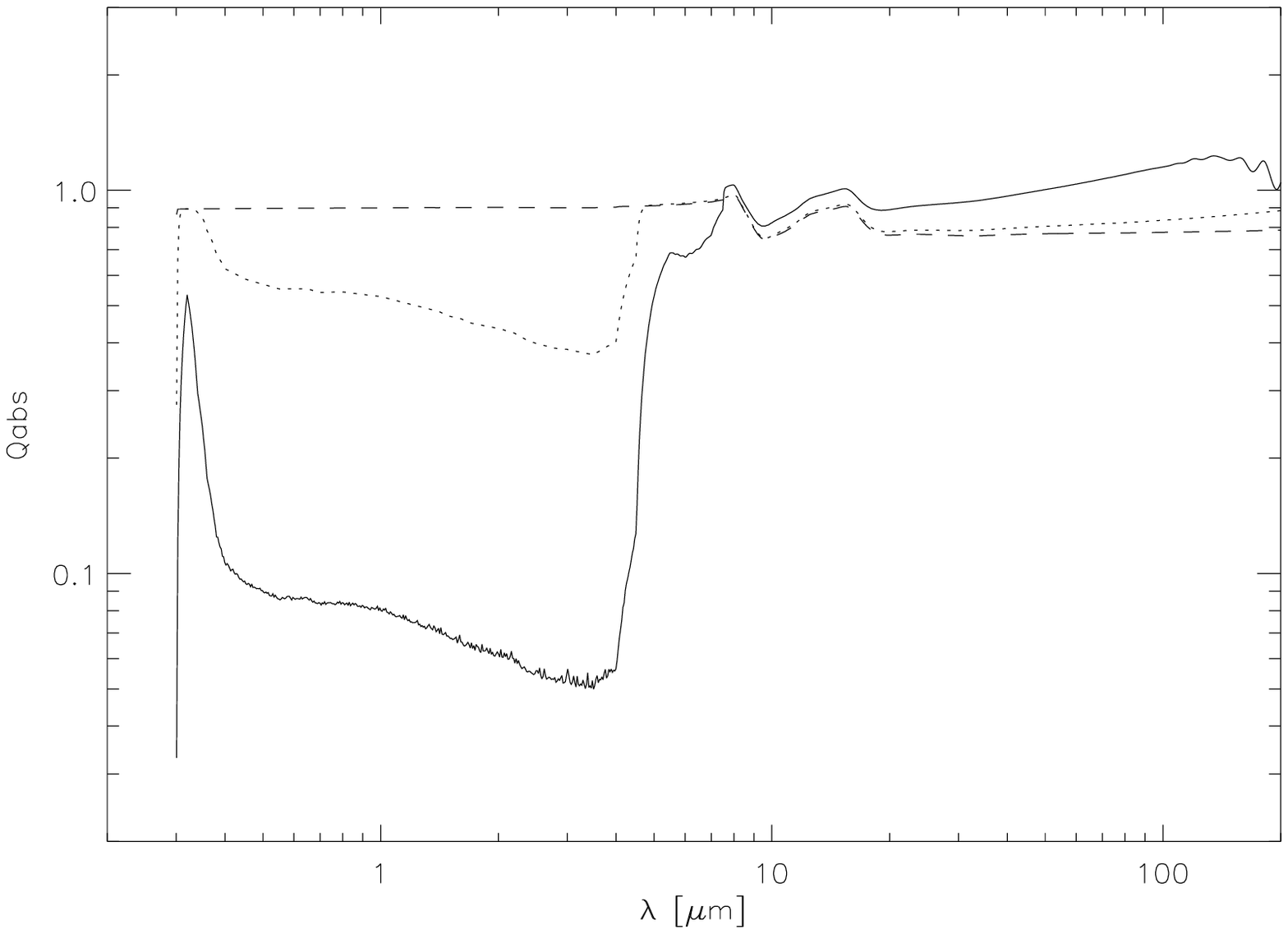}}\hspace*{10mm}
    \resizebox{0.44\hsize}{!}{\includegraphics{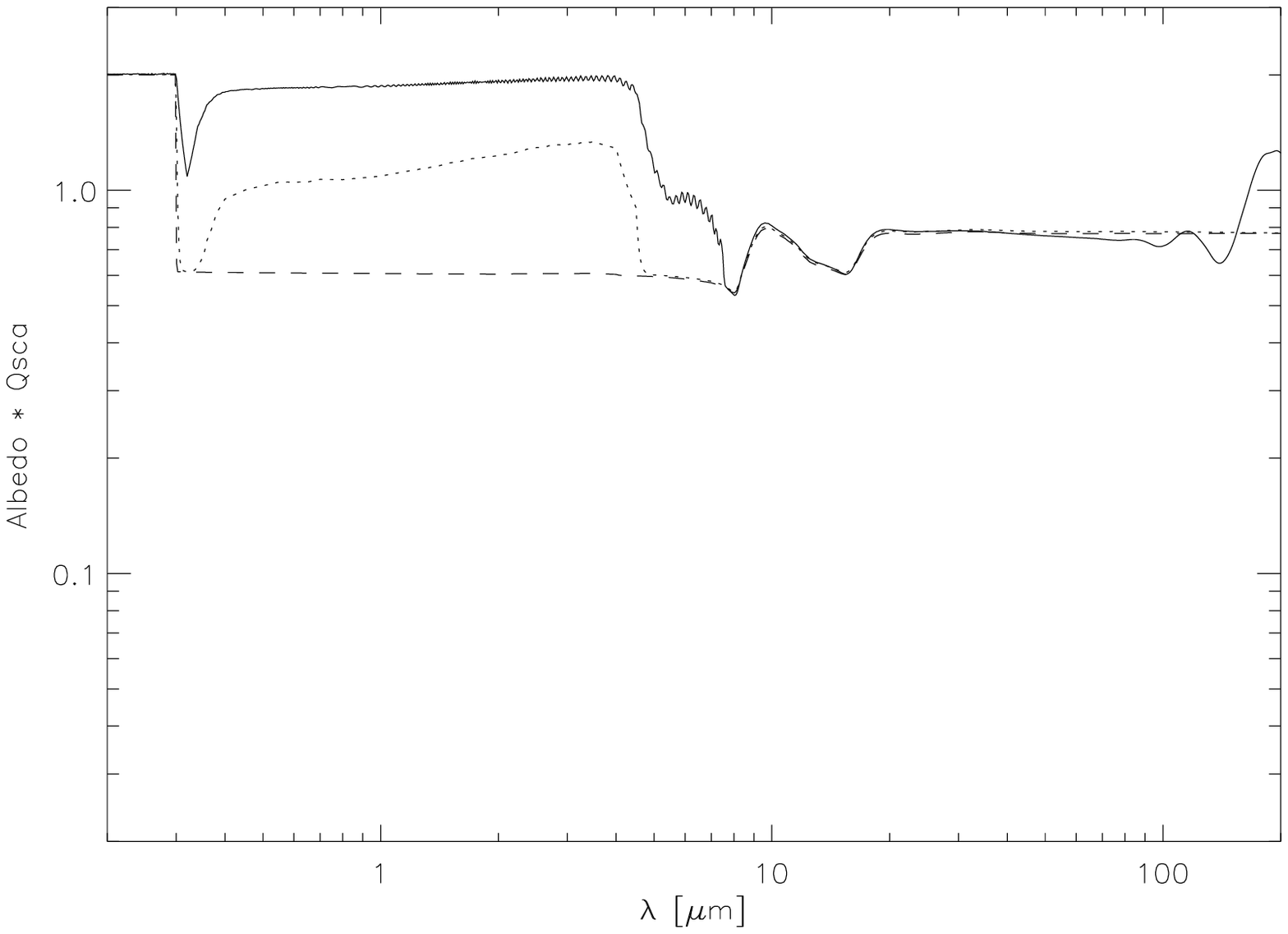}}
 \end{center}
  \caption{Absorption and scattering coefficients for Mg-rich / Fe-deficient Silicate: MgSiO$_3$.
    \newline
    {\sl (a) upper row:} grain radii
    $0.1\mu$m (solid line),
    $1\mu$m   (dotted line),
    $10\mu$m  (dashed line);
    \newline
    {\sl (b) lower row:} grain radii
    $0.1$mm (solid line),
    $1$mm   (dotted line),
    $10$mm  (dashed line).
  }
  \label{scaabs-1}
  \bigskip
\end{figure}

\begin{figure}[t]
  \begin{center}
    \resizebox{0.44\hsize}{!}{\includegraphics{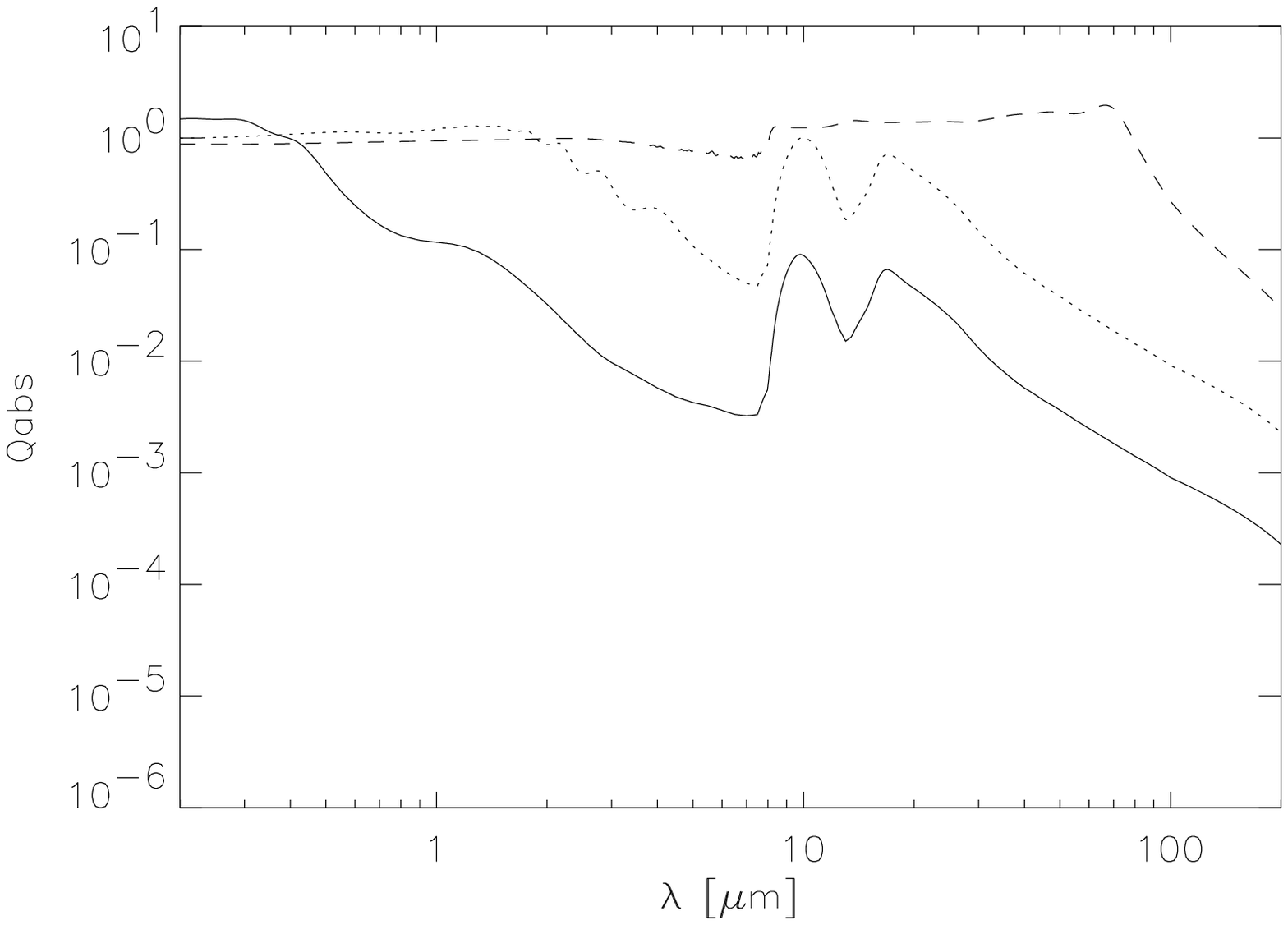}}\hspace*{10mm}
    \resizebox{0.44\hsize}{!}{\includegraphics{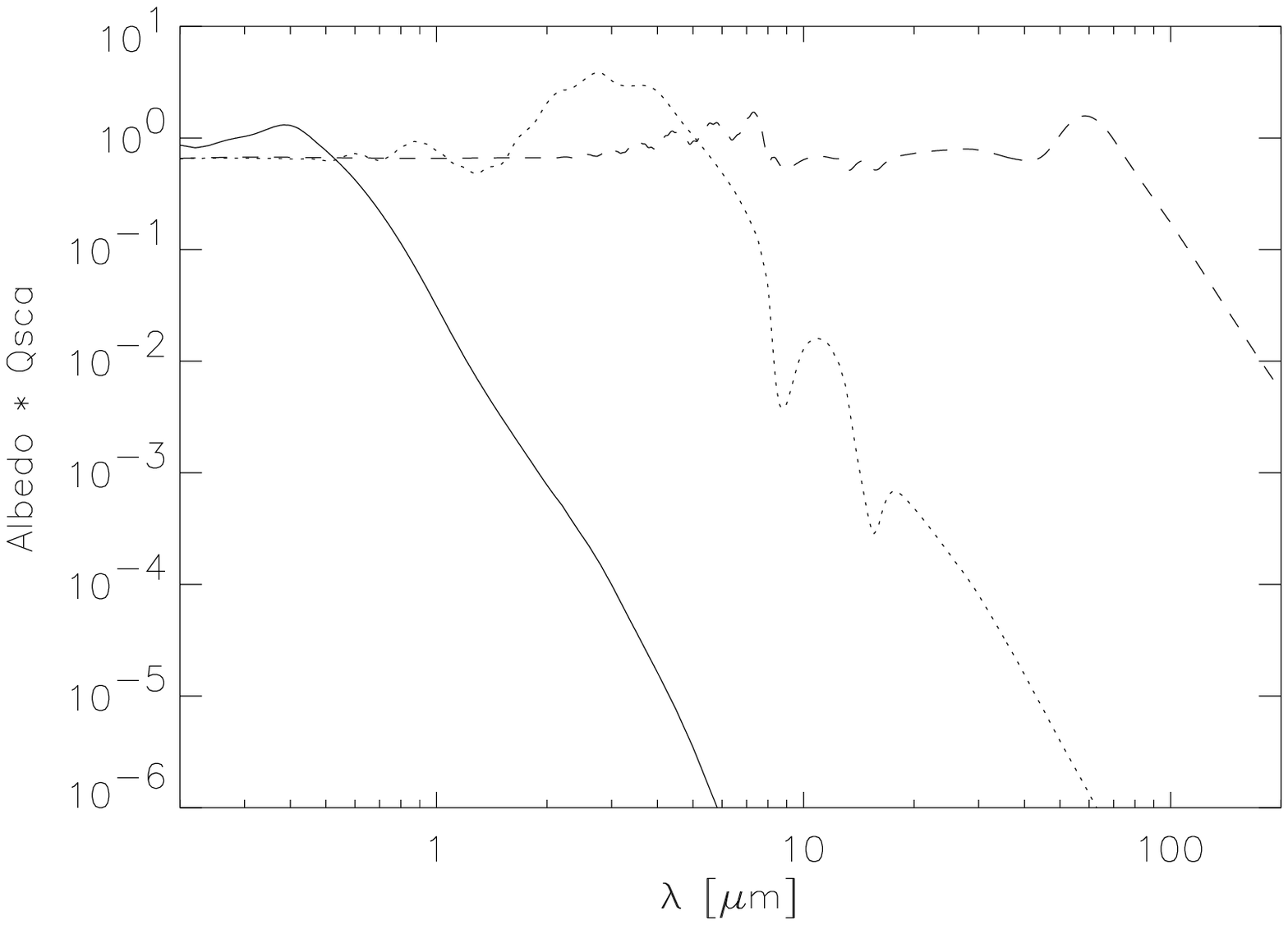}}
    \resizebox{0.44\hsize}{!}{\includegraphics{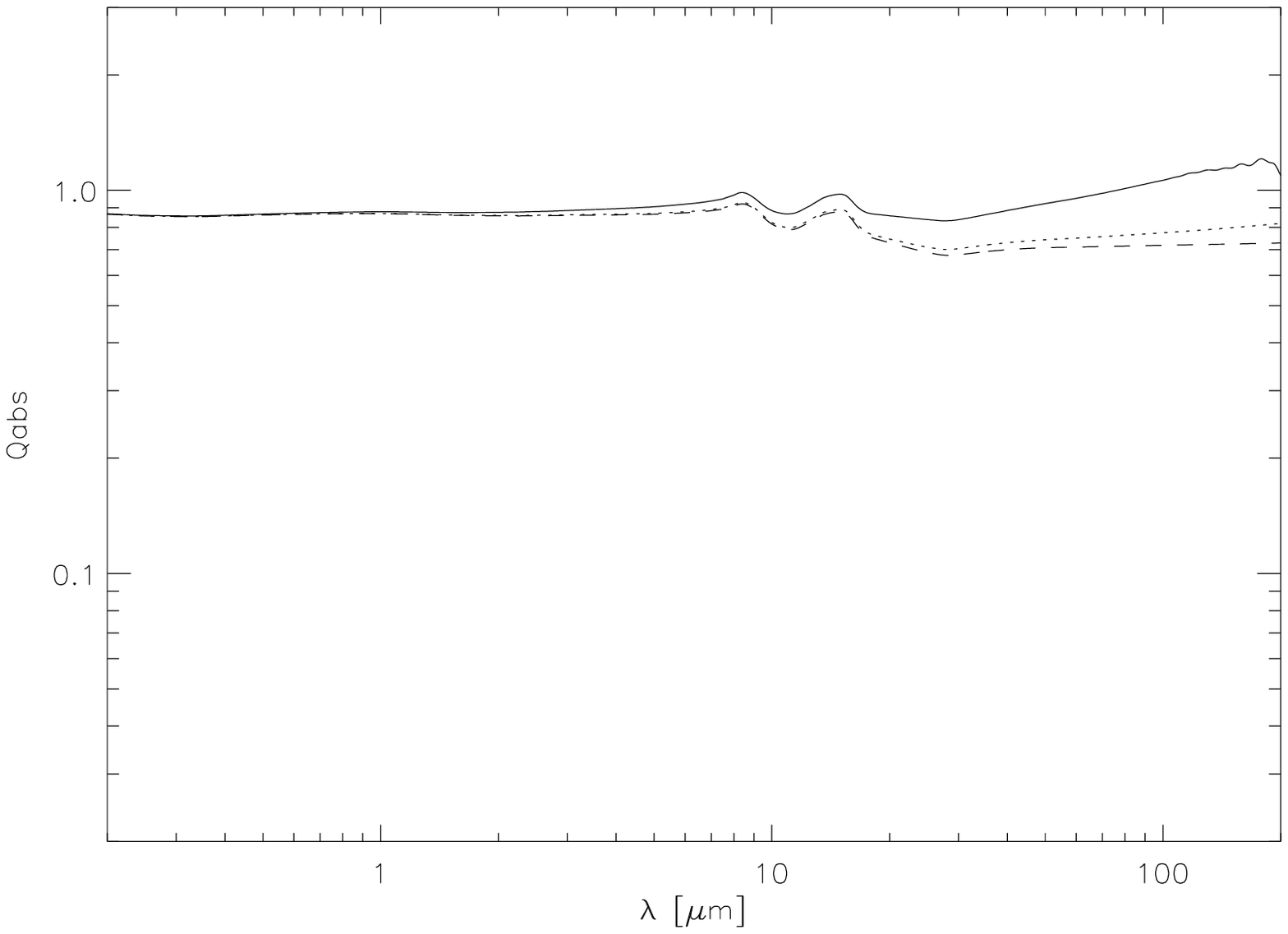}}\hspace*{10mm}
    \resizebox{0.44\hsize}{!}{\includegraphics{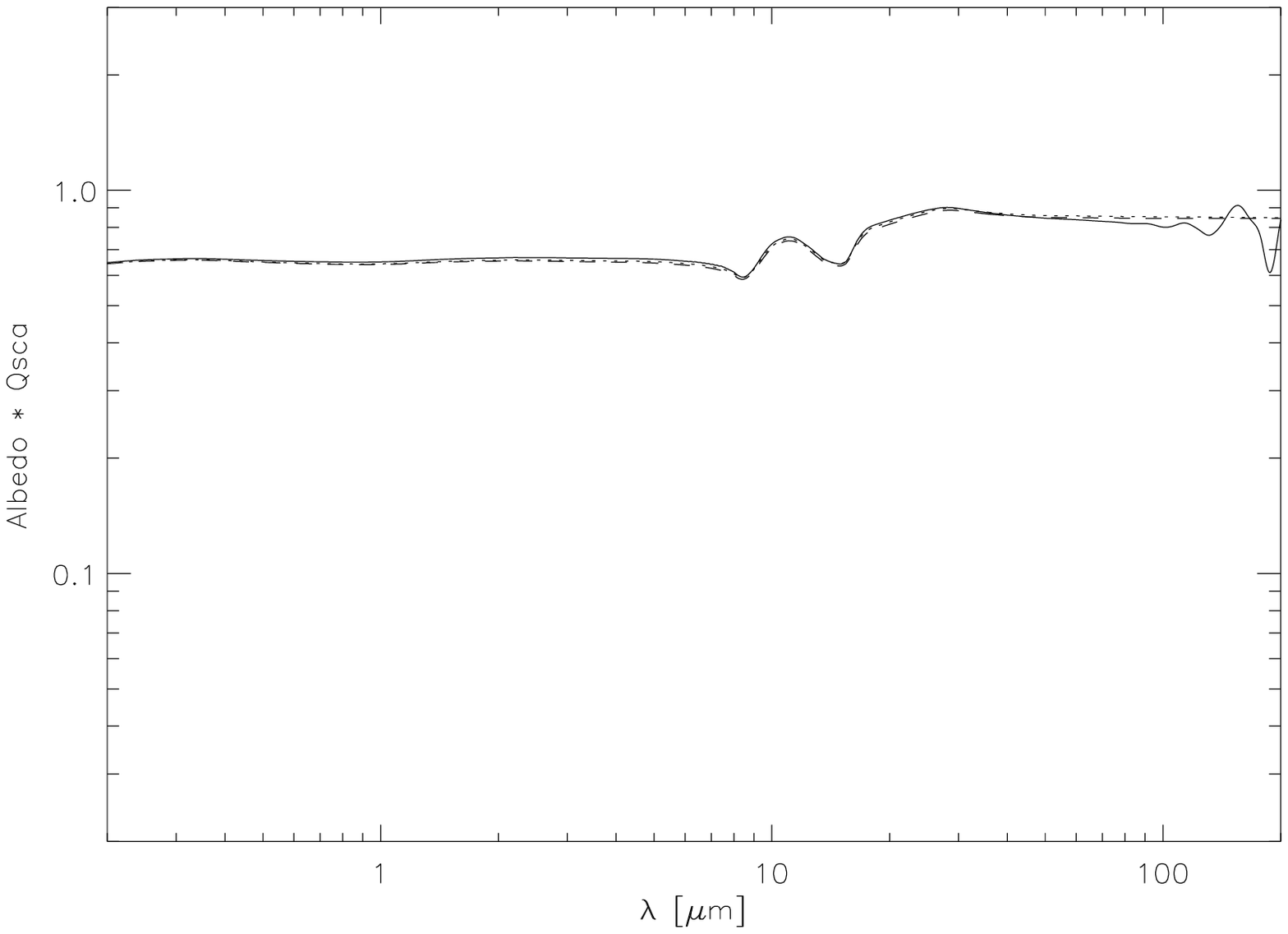}}
  \end{center}
  \caption{Absorption and scattering coefficients for Mg-poor / Fe-rich Silicate: Mg$_{0.8}$Fe$_{1.2}$SiO$_4$.
    \newline
    {\sl (a) upper row:} grain radii
    $0.1\mu$m (solid line),
    $1\mu$m   (dotted line),
    $10\mu$m  (dashed line);
    \newline
    {\sl (b) lower row:} grain radii
    $0.1$mm (solid line),
    $1$mm   (dotted line),
    $10$mm  (dashed line).
  }
  \label{scaabs-2}
  \bigskip
\end{figure}

\begin{figure}[t]
  \begin{center}
    \resizebox{0.44\hsize}{!}{\includegraphics{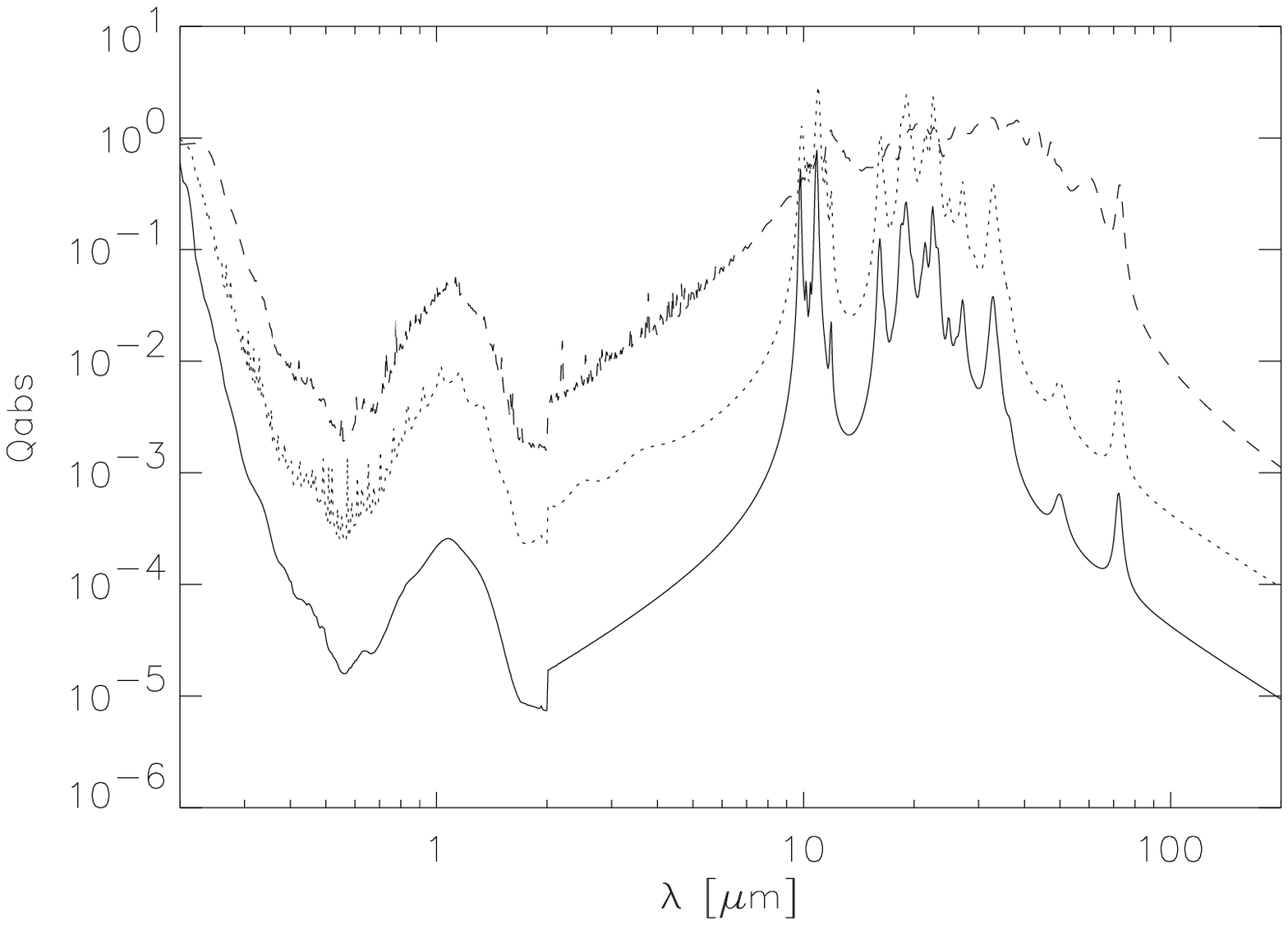}}\hspace*{10mm}
    \resizebox{0.44\hsize}{!}{\includegraphics{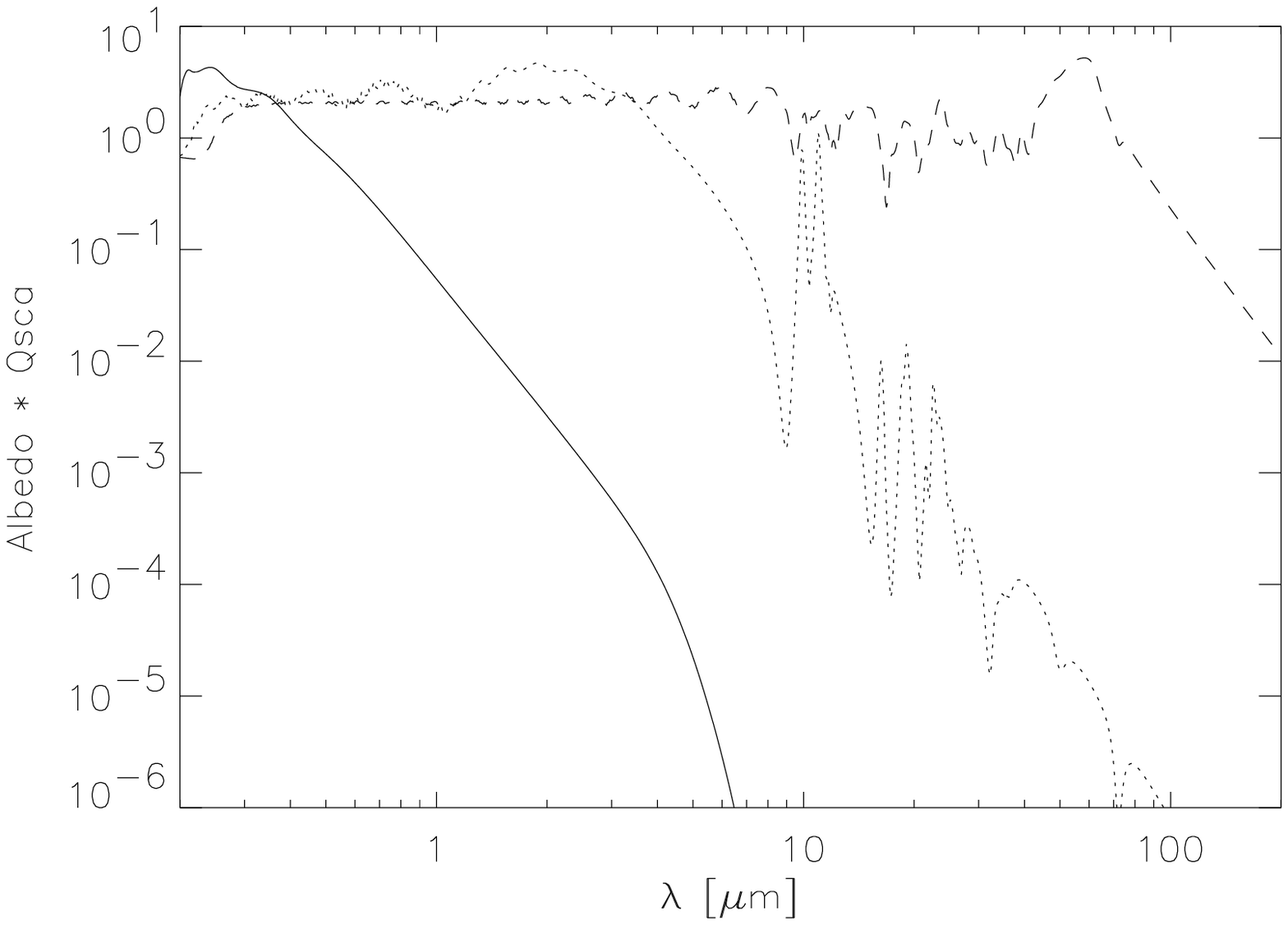}}
    \resizebox{0.44\hsize}{!}{\includegraphics{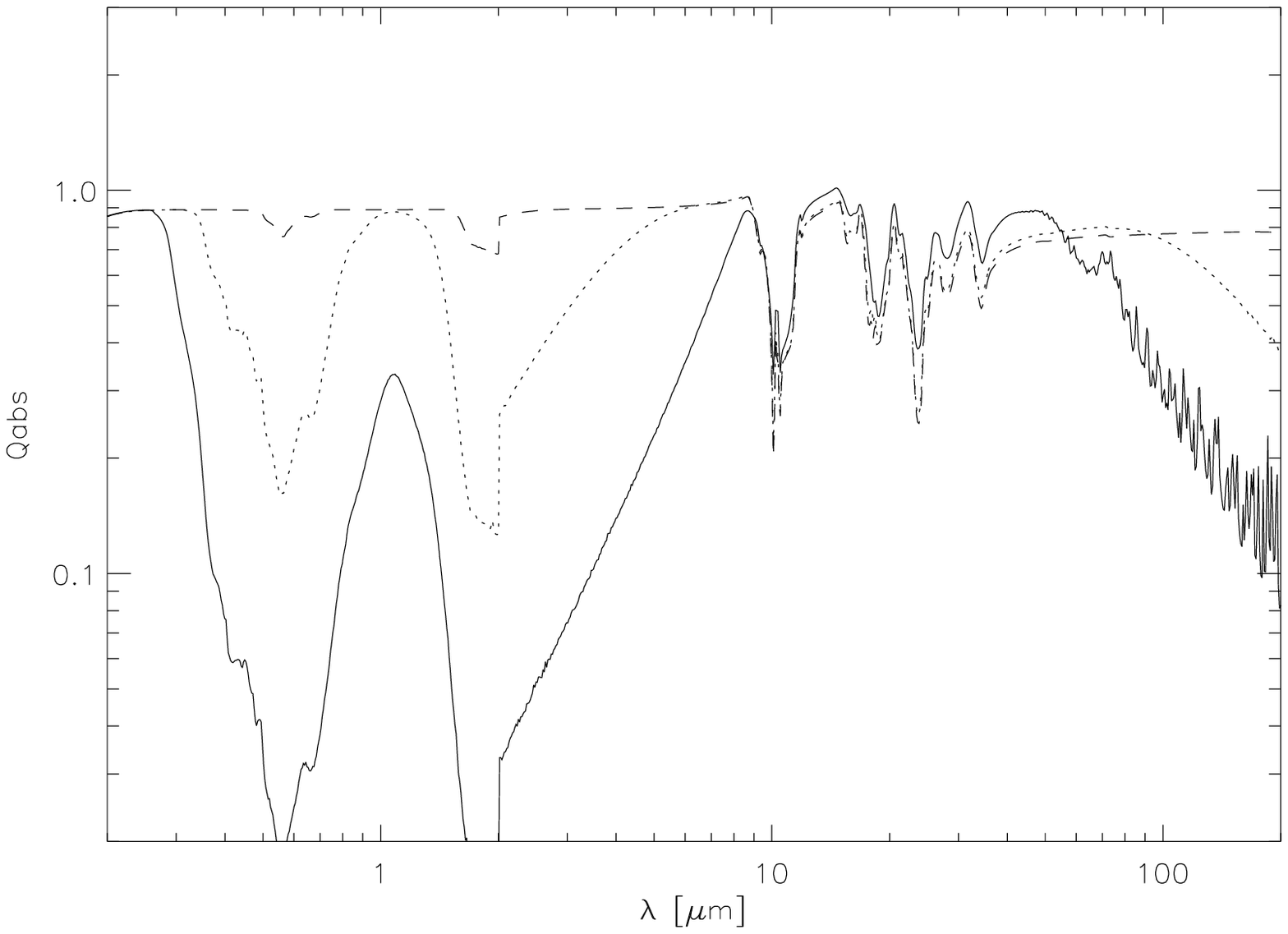}}\hspace*{10mm}
    \resizebox{0.44\hsize}{!}{\includegraphics{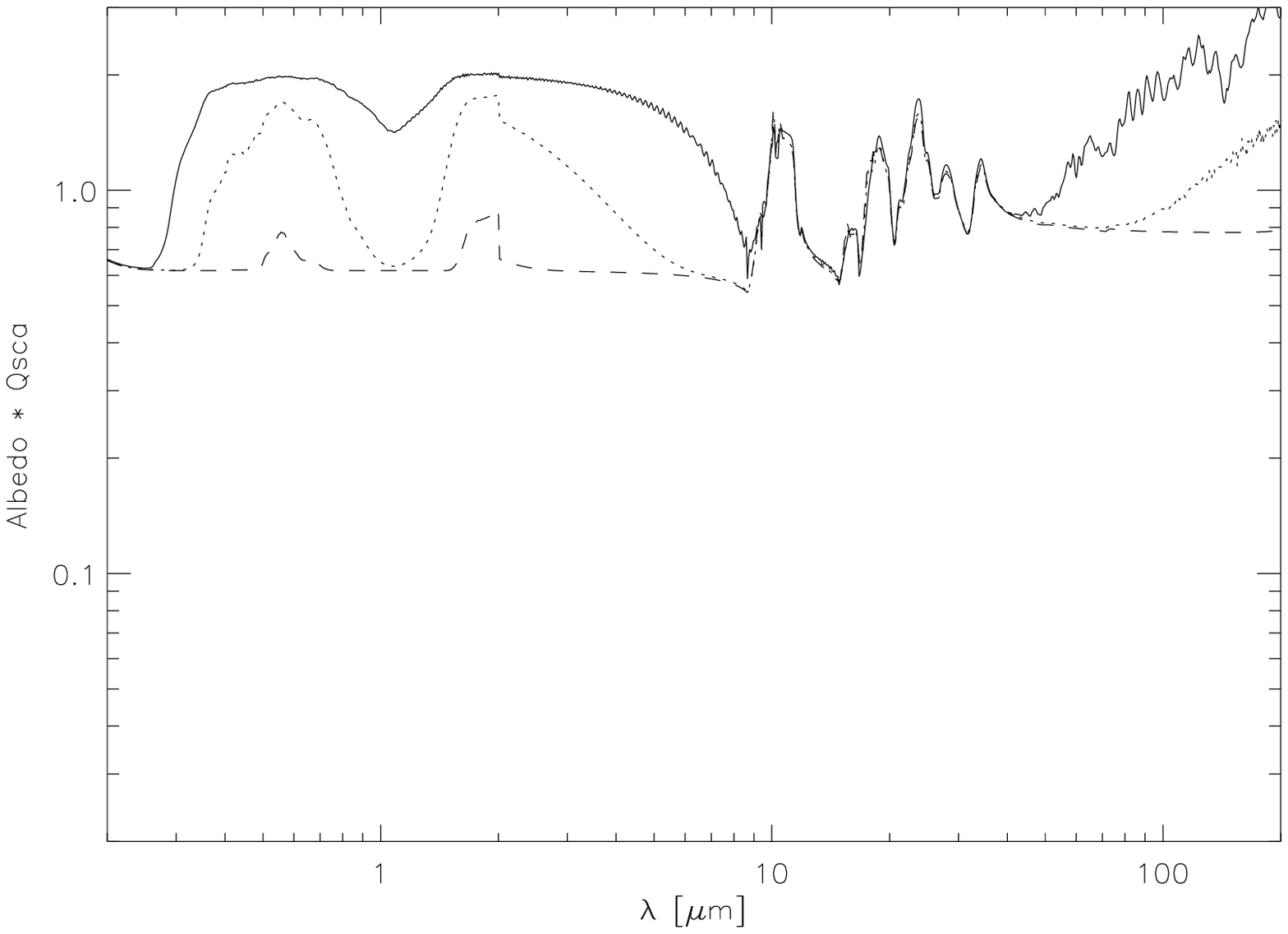}}
  \end{center}
  \caption{Absorption and scattering coefficients for Olivine (crystalline, but not oriented): Mg$_{1.9}$Fe$_{0.1}$SiO$_4$.
    \newline
    {\sl (a) upper row:} grain radii
    $0.1\mu$m (solid line),
    $1\mu$m   (dotted line),
    $10\mu$m  (dashed line);
    \newline
    {\sl (b) lower row:} grain radii
    $0.1$mm (solid line),
    $1$mm   (dotted line),
    $10$mm  (dashed line).
  }
  \label{scaabs-5}
  \bigskip
\end{figure}

\begin{figure}[t]
  \begin{center}
    \resizebox{0.44\hsize}{!}{\includegraphics{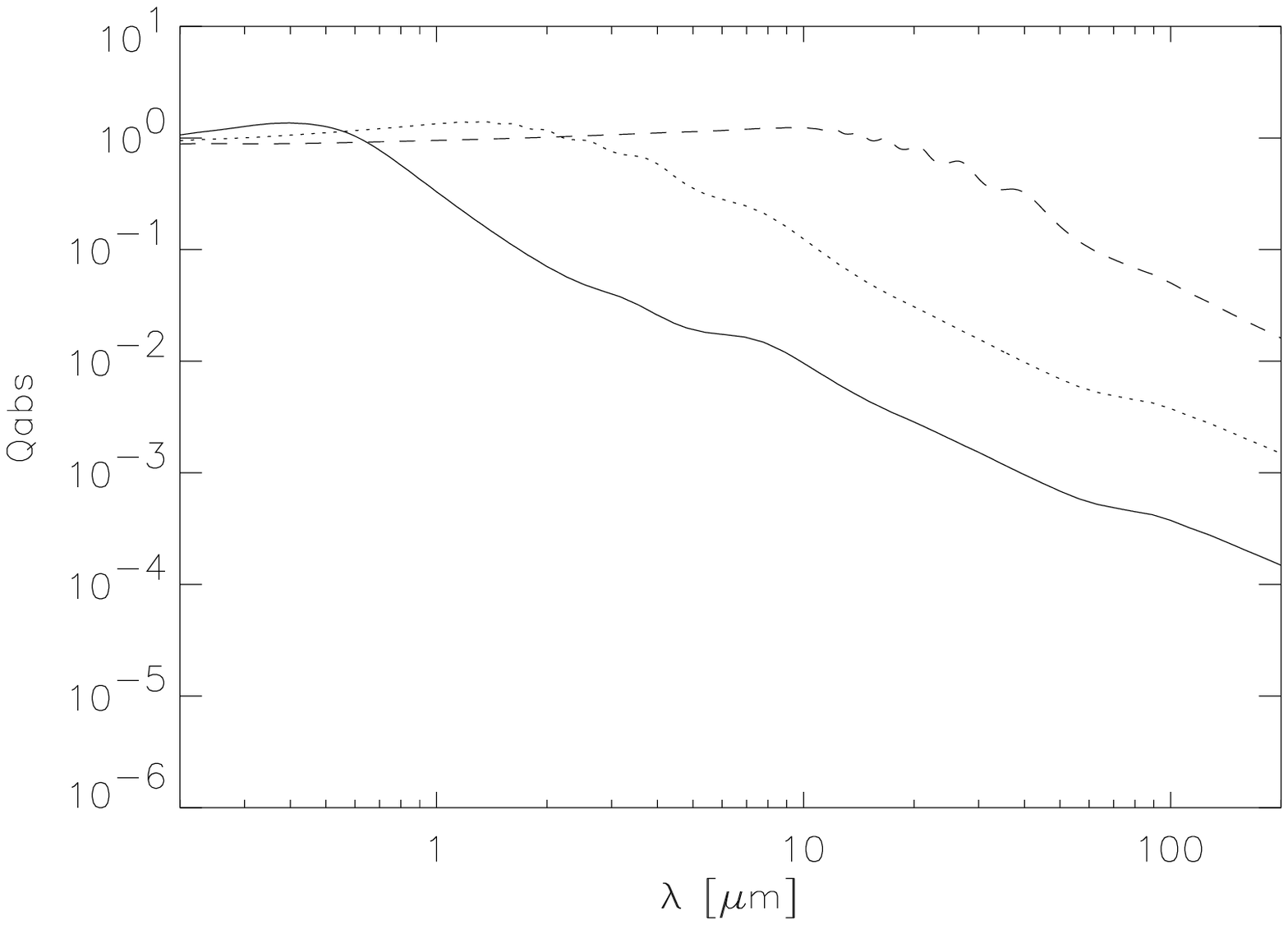}}\hspace*{10mm}
    \resizebox{0.44\hsize}{!}{\includegraphics{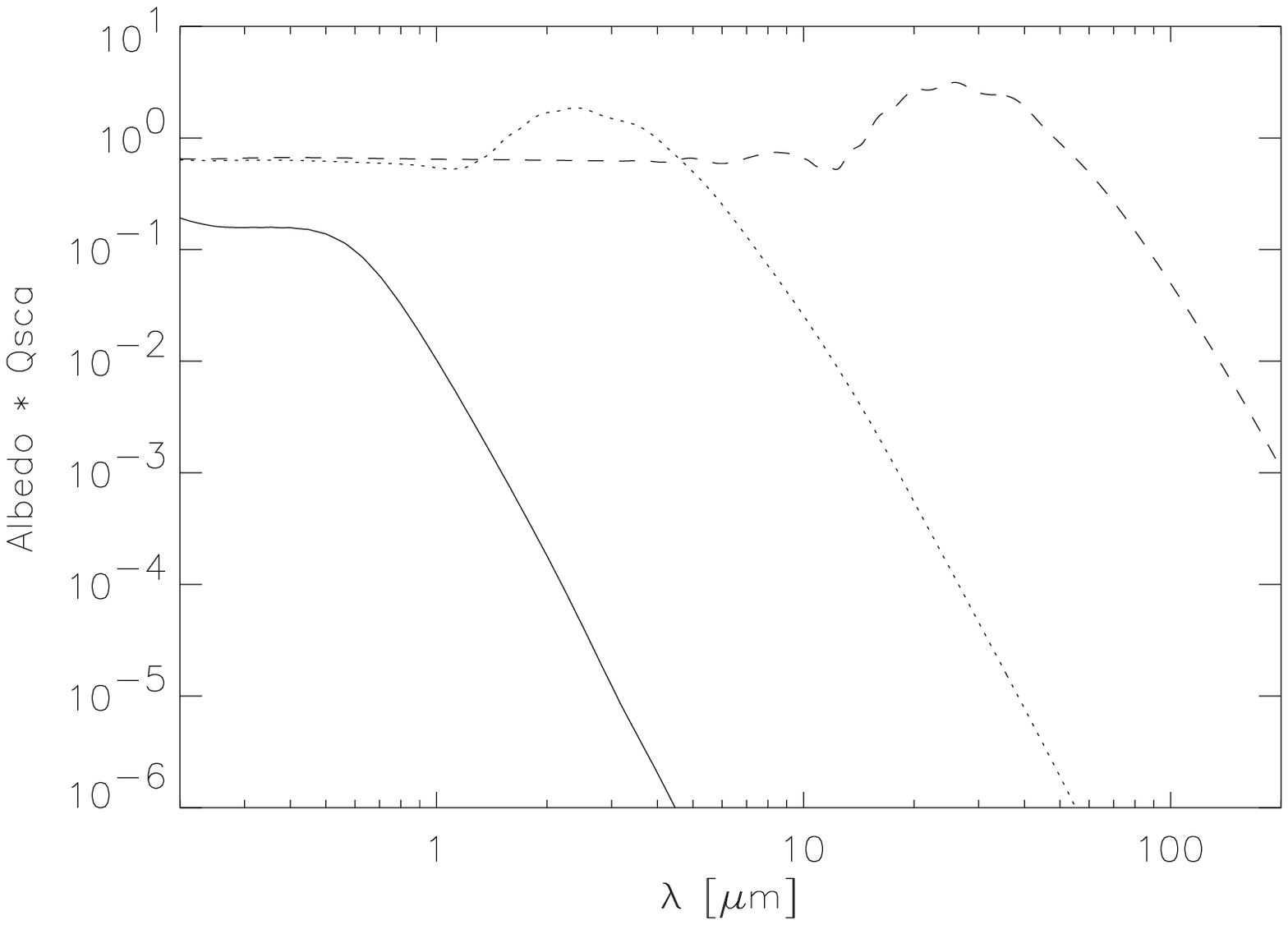}}
    \resizebox{0.44\hsize}{!}{\includegraphics{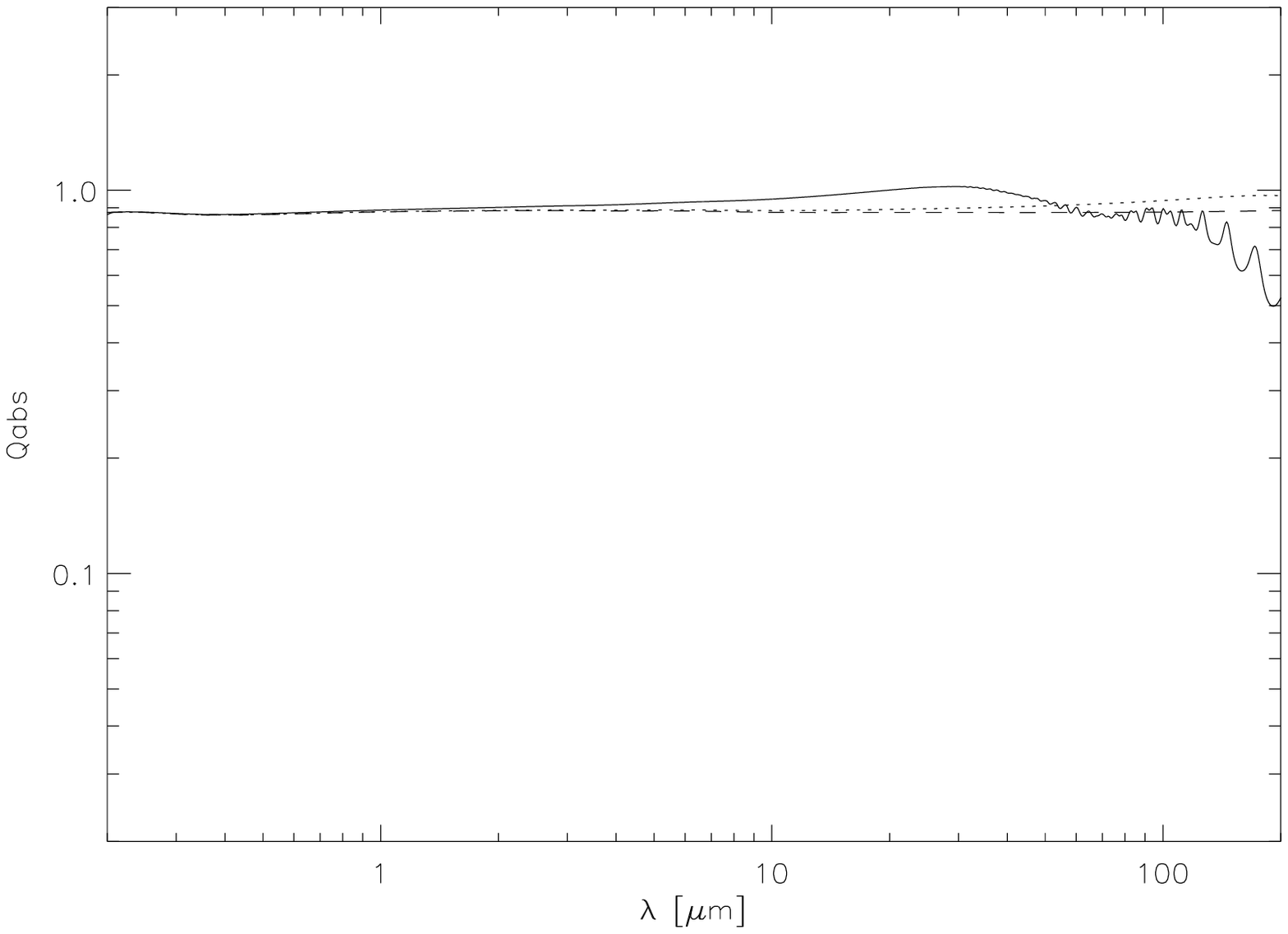}}\hspace*{10mm}
    \resizebox{0.44\hsize}{!}{\includegraphics{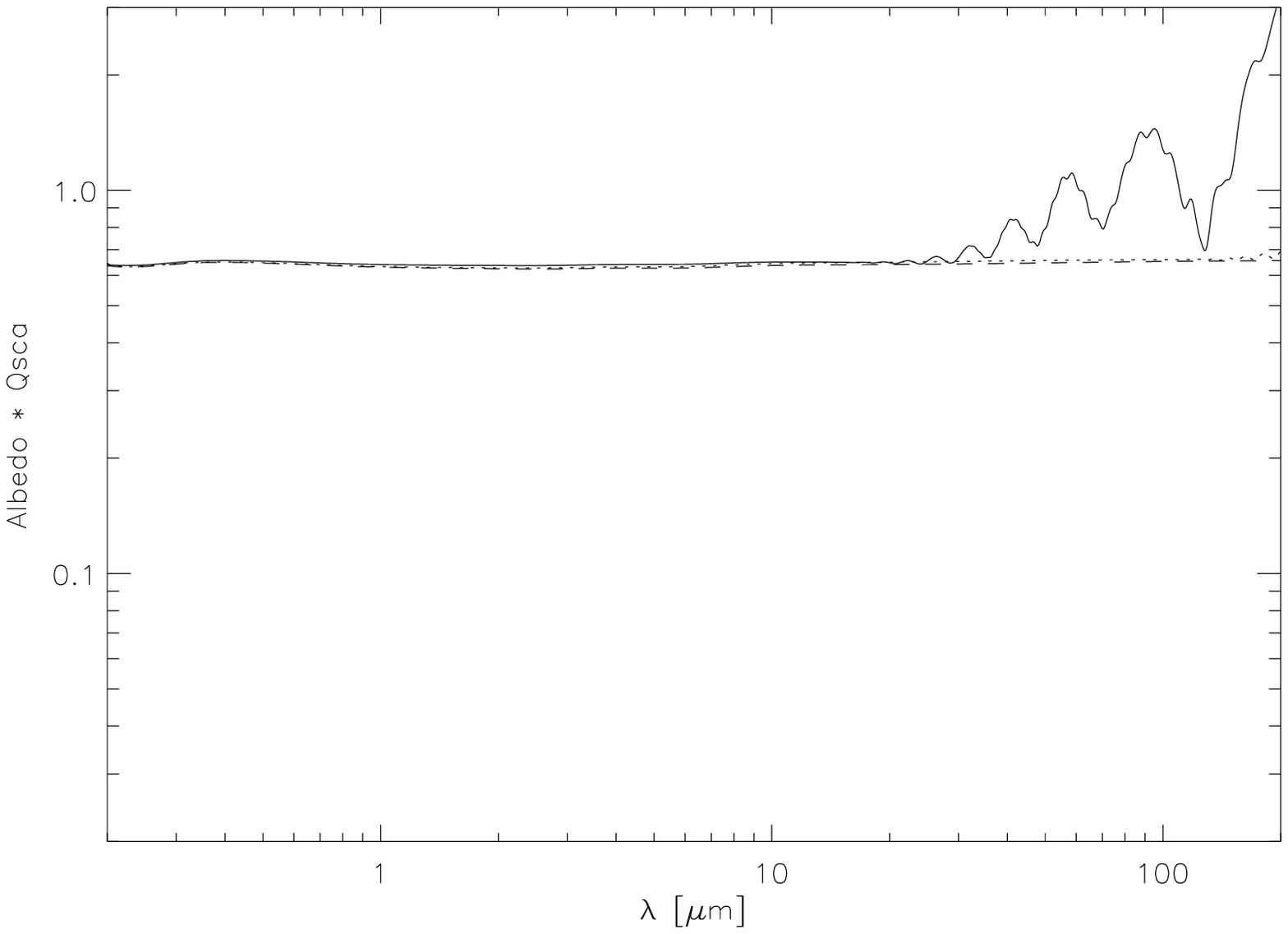}}
  \end{center}
  \caption{Absorption and scattering coefficients for Carbon: 400\,K configuration (``graphite poor'').
    \newline
    {\sl (a) upper row:} grain radii
    $0.1\mu$m (solid line),
    $1\mu$m   (dotted line),
    $10\mu$m  (dashed line);
    \newline
    {\sl (b) lower row:} grain radii
    $0.1$mm (solid line),
    $1$mm   (dotted line),
    $10$mm  (dashed line).
  }
  \label{scaabs-3}
  \bigskip
\end{figure}

\begin{figure}[t]
  \begin{center}
    \resizebox{0.4\hsize}{!}{\includegraphics{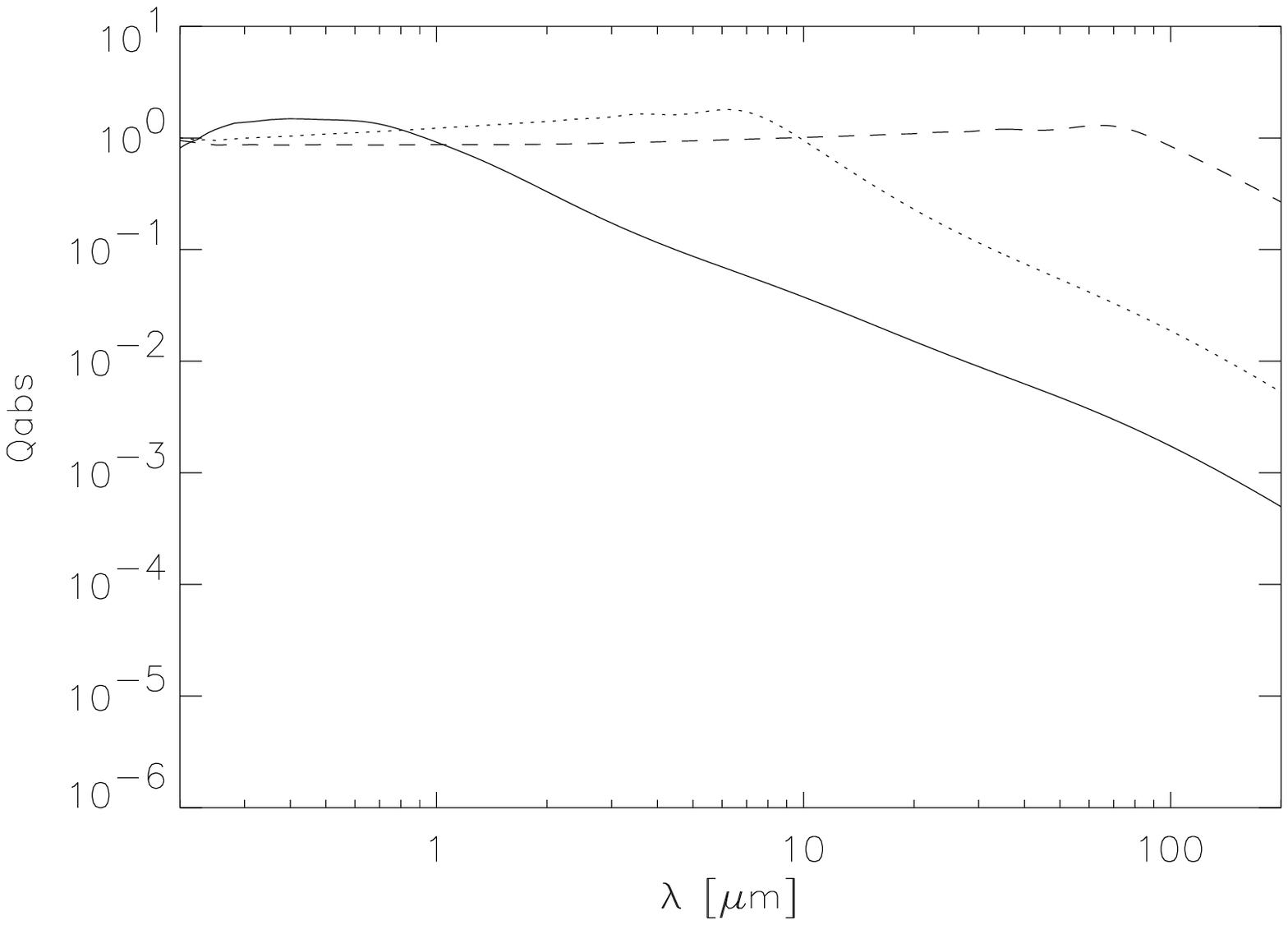}}\hspace*{10mm}
    \resizebox{0.4\hsize}{!}{\includegraphics{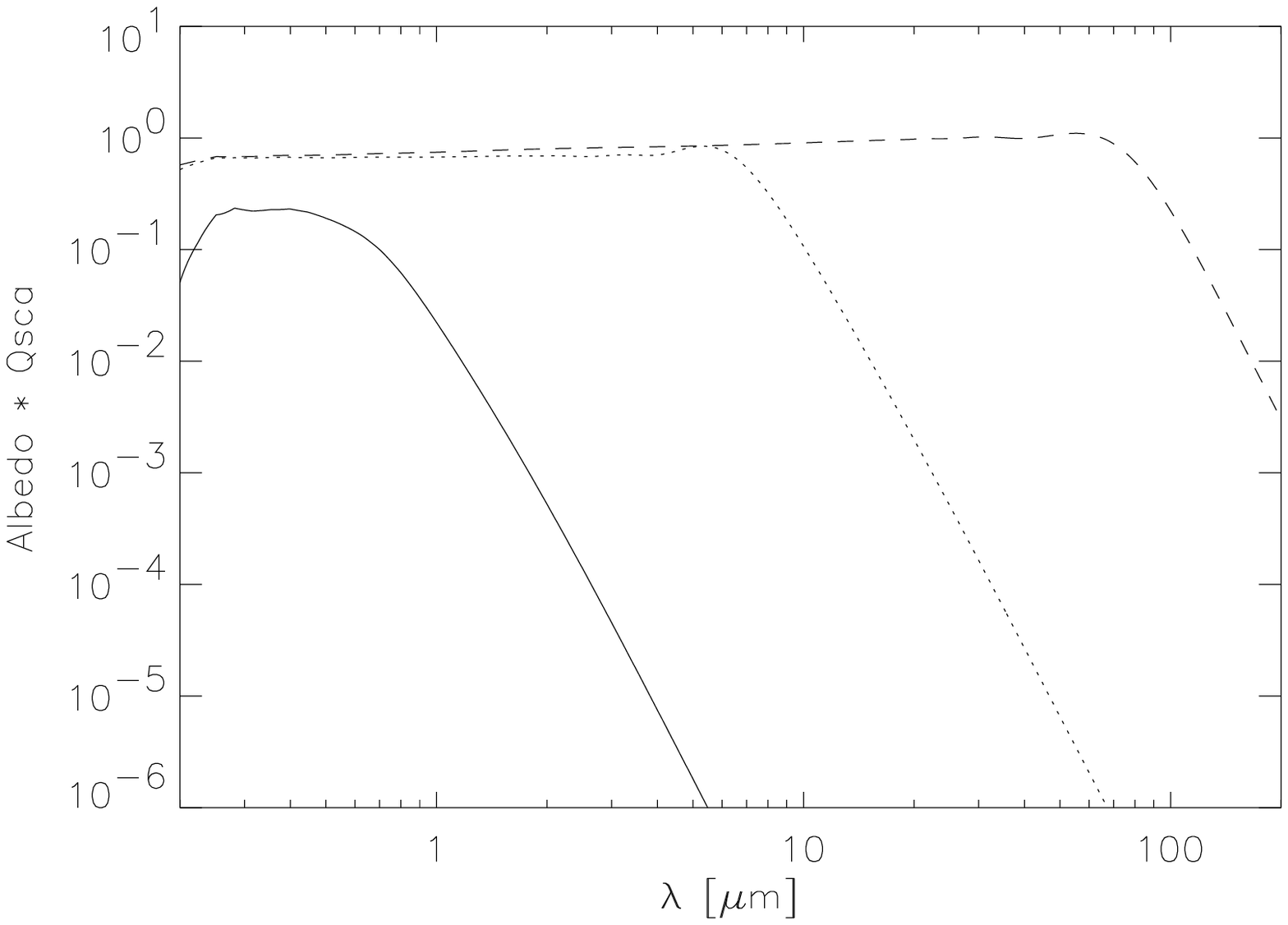}}
    \resizebox{0.4\hsize}{!}{\includegraphics{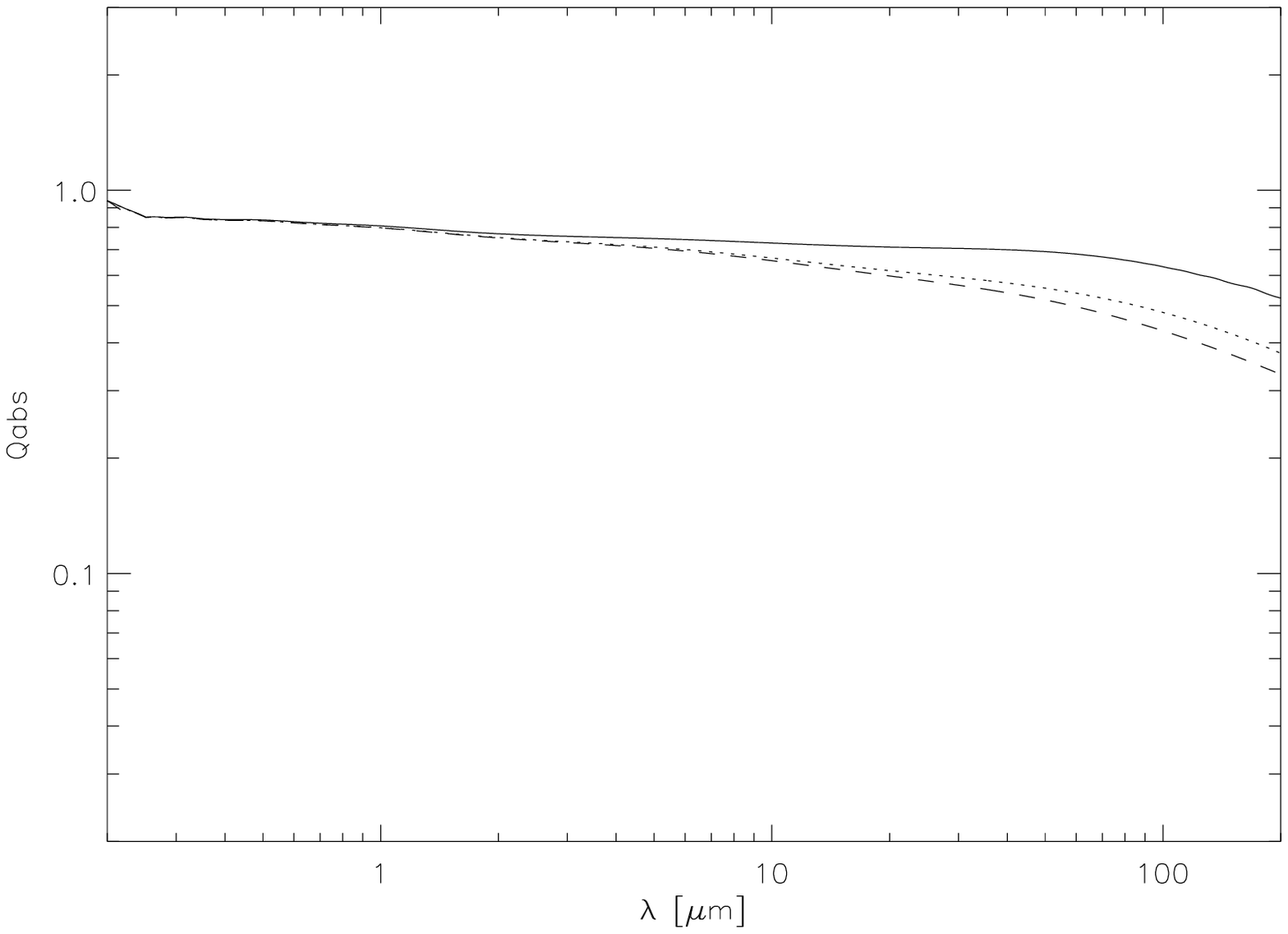}}\hspace*{10mm}
    \resizebox{0.4\hsize}{!}{\includegraphics{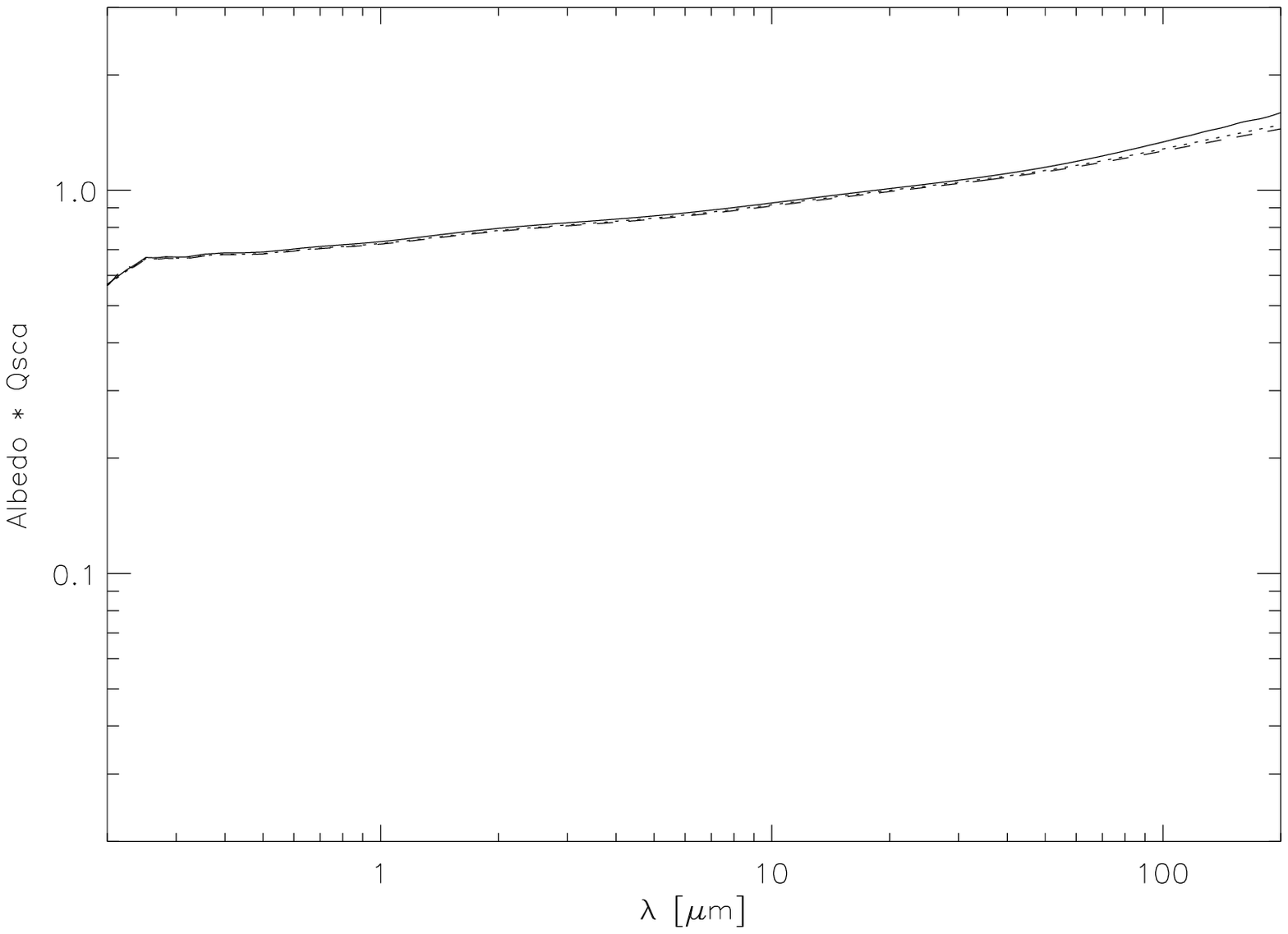}}
  \end{center}
  \caption{Absorption and scattering coefficients for Carbon: 1000\,K configuration (``graphite rich'').
    \newline
    {\sl (a) upper row:} grain radii
    $0.1\mu$m (solid line),
    $1\mu$m   (dotted line),
    $10\mu$m  (dashed line);
    \newline
    {\sl (b) lower row:} grain radii
    $0.1$mm (solid line),
    $1$mm   (dotted line),
    $10$mm  (dashed line).
  }
  \label{scaabs-4}
  \bigskip
\end{figure}


We consider the following chemical components:
\begin{enumerate}
\item Amorphous silicates with varying iron content from Fe-poor through Fe-rich,
\item Crystalline silicates (in particular Olivine), and
\item Amorphous carbon species with different graphite/diamond structure ratios.
\end{enumerate}
{\em ISO} spectroscopy of Herbig~Ae/Be stars revealed several further abundant chemical components, such as
Fe         (broad, weak emission feature at $\approx 2-4\mu$m; see Henning \& Stognienko~1996),
FeO        (broad emission features at      $\approx 21-25\mu$m; see Henning et al.~1995) and/or 
FeS        (see Henning \& Stognienko~1996)
and
H$_2$O ice (broad emission features between $\approx 40-80\mu$m; Warren~1984).
For model SEDs taking into account these additional components see, e.g.,
Malfait et al.~(1999), Bouwman et al.~(2000), and Meeus et al.~(2001).
However, the SEDs of debris disks measured so far do not allow one to perform a 
comparably detailed chemical analysis; SIRTF is expected to be revolutionary in this regard.

We consider the dust grains to be homogeneous spheres. Although dust grains are expected to have a fractal structure,
the scattering behaviour is similar to that of spheres (see Lumme \& Rahola~1994 for porous dust particle 
light scattering). Furthermore, dust grains are expected to have a non-spherical shape 
(see, e.g., 
Elvius \& Hall~1967;
Scarrott, Draper, \& Warren-Smith~1989;
Hajjar \& Bastien~1996;
Kastner \& Weintraub~1996;
Dollfus \& Suchail~1987;
Johnson \& Jones~1991;
Chrysostomou et al.~2000). 
However, we do not consider the grains to be aligned on a large scale by magnetic fields and thus, 
the assumption of spherical grains is a valid approximation (Wolf, Voshchinnikov, \& Henning~2002).
The optical parameters of the grains required for the estimation of the dust absorption, reemission and
scattering, are derived from the complex refractive index $m = n + ik$ and the grain radius.
We use measurements of the complex refractive index $m$ published by 
Dorschner et al.~(1995; Silicates)
and J\"ager et al.~(1997; Carbon)\footnote{The complex refractive indices are available at\\
{\tt http://www.astro.uni-jena.de/Laboratory/Database/odata.html}.}.
Furthermore, we compare selected dust properties with the optical properties of ``astronomical silicate''
and graphite provided by Weingartner \& Draine~(2001; see also Draine \& Lee~1984) since these data are used 
in a large variety of simulations of circumstellar dust configurations.
According to Eqs.~\ref{eq_dist5} and \ref{eq_scatt}, the absorption and scattering efficiencies
determine the particular contribution of each chemical component to the net SED.
We consider absorption/re-emission and scattering for individual grains of given composition
and size over the wavelength range 0.2 - 200$\mu$m.

The interaction of the stellar radiation field with the dust grains is described by Mie scattering theory.
We calculate the Mie scattering function using the numerical solution for the estimation of 
the Mie scattering coefficients published by Voshchinnikov~(2003), which achieves accurate results 
both in the small as well as in the --~arbitrarily~-- large size parameter 
regime\footnote{{The code for calculation of the Mie scattering coefficients is available at\hfill\break}
{\tt http://mc.caltech.edu/$\sim$swolf/miex-web/miex.htm.}}

In the following we briefly describe the silicate and carbon species considered:

{\bf Silicates:} 
Previous studies concerning the mid-infrared SED of evolved stars and the $\beta$~Pic circumstellar disk revealed
that a large variety of silicate species is present in these different environments.
For example, Pantin, Lagage, \& Artymowicz~(1997) 
find amorphous olivine (MgFeSiO$_4$) to be the dominant chemical dust
component in the $\beta$~Pic disk (55\% amorphous olivine, 35\% amorphous pyroxene, and
10\% crystalline Olivine). In contrast to this, Molster, Waters, \& Tielens~(2002a,b) find narrow-band emission
features in the SED of dust around evolved stars, clearly pointing to the existence 
of crystalline silicates.
In order to cover this large range of different silicates, we investigate the main differences between
a number of different amorphous pyroxene species (Mg$_x$Fe$_{1-x}$SiO$_3$ 
with $x$=0.4, 0.5, 0.6, 0.7, 0.8, 0.95 and 1.0),
and amorphous Olivine species (Mg$_{2y}$Fe$_{2-2y}$SiO$_4$ with $x$=0.4 and 0.5).
The efficiency factors for 1$\mu$m grains are shown in Fig.~\ref{si-reihe-q}.
Despite the characteristic 10$\mu$m and 19$\mu$m maxima of the absorption efficiency, 
the most significant property is the strong dependence of the ultraviolet (UV) to near-infrared (NIR)
absorption efficiency on the Fe content
(see J\"ager et al.~1994 for a detailed compilation and discussion of absorption bands of amorphous pyroxene).
With increasing relative Fe content the absorption efficiency increases by about three orders of magnitude
in the case of a micron-sized grain in this wavelength range from Fe-deficient MgSiO$_3$ 
to Fe-rich Mg$_{0.8}$Fe$_{1.2}$SiO$_4$.
Thus, the temperature and reemitted flux of the grains at a given radial distance from the star increases.
These effects are demonstrated in Fig.~\ref{si-reihe-sed} where the resulting density distribution for 
our standard disk model (see below) is shown. 
Furthermore, the comparison with ``astronomical silicate'' (Weingartner \& Draine~2001) shows that
it is much more similar to Fe-rich than to Fe-poor silicates.
For comparison, the scattering/reemission SED of crystalline Olivine (MgFeSiO$_4$) is shown. 
Similar to Fe-poor silicates, crystalline Olivine has a very low absorption efficieny in the 
wavelength range of stellar emission (see Fig.~\ref{scaabs-5}) and can therefore only be traced (by its pronounced
numerous narrow emission features) either based on highly accurate photometric measurements 
or if a high relative amount of this material is present in the disk.
Furthermore, the dependence of the efficieny factors on the grain size is illustrated in 
Fig.~\ref{scaabs-1}, \ref{scaabs-2}, and \ref{scaabs-5} for Fe-deficient and Fe-rich amorphous silicate
and crystalline Olivine.

{\bf Carbon:}
It is commonly assumed that the carbonaceous dust component occurs in the form of graphite
(see, e.g., Draine \& Lee~1984, Li \& Greenberg~1997, Weingartner \& Draine 2001).
In contrast to this common assumption, the dust emissivity described by a power-law with a spectral index 
of $\approx 1$ (Campbell et al.~1976; 
Sopka et al.~1985; Martin \& Rogers~1987; G\"urtler, K\"ompe, \& Henning~1996) 
can be explained by the occurence of very disordered two-dimensional material like amorphous carbon (Kittel~1963).
J\"ager, Mutschke, \& Henning~(1998) synthesized structurally different carbon material 
with an increasing sp$^2$/sp$^3$ ratio by pyrolizing cellulose materials at 400-1000$^{\rm o}$C.
The different carbon structures therefore represent different graphite (sp$^2$ hybridization) to diamond structure
(sp$^3$ hybridization) abundance ratios.
In Fig.~\ref{scaabs-3} and \ref{scaabs-4} we show the efficiency factors calculated for the two extreme cases:
the 400\,K carbon modification (graphite-poor) and 1000\,K carbon modification (graphite-rich).
As these figures show, carbon has a high absorption efficiency in the UV/visual wavelength range, 
but - in contrast to silicate species - no prominent absorption features (see also Fig.~\ref{si-reihe-sed}). 
Thus, it mainly adds a strong continuum to the SED. The absorption efficiency in the UV/visual wavelength range
is comparable to Fe-rich silicates. With increasing sublimation temperature the range of almost
constant, high absorption efficiency is extended towards longer wavelengths. The particular turn-over point
beyond which the absorption drops continiously, increases as the grain size is increased.
It is noteworthy that the absorption efficiency of graphite as published by Weingartner \& Draine~(2001)
is smaller than that of the other carbon configurations in the mid-infrared wavelength range.


\begin{figure}[t]
  \begin{center}
    \resizebox{0.47\hsize}{!}{\includegraphics{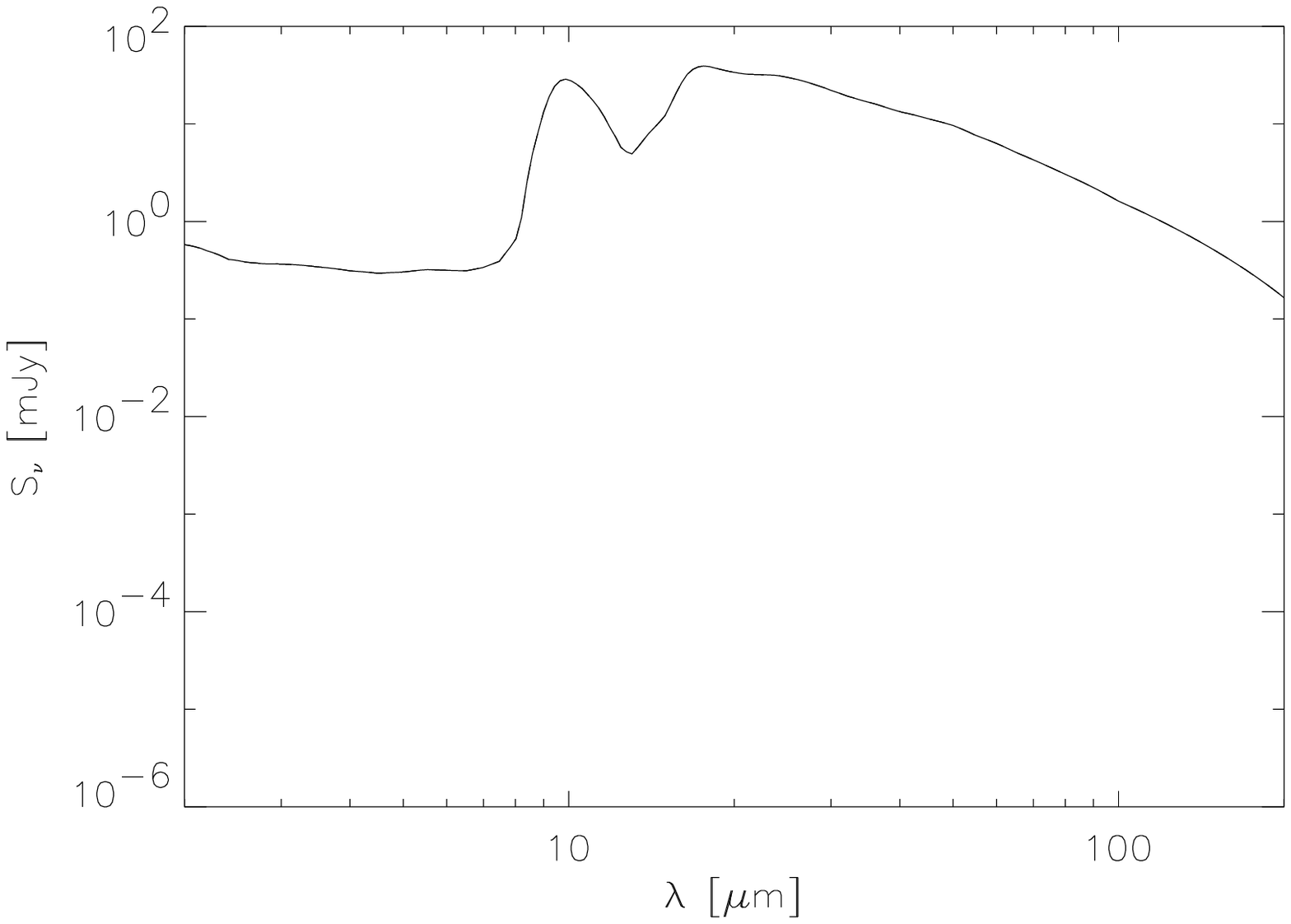}}
    \resizebox{0.47\hsize}{!}{\includegraphics{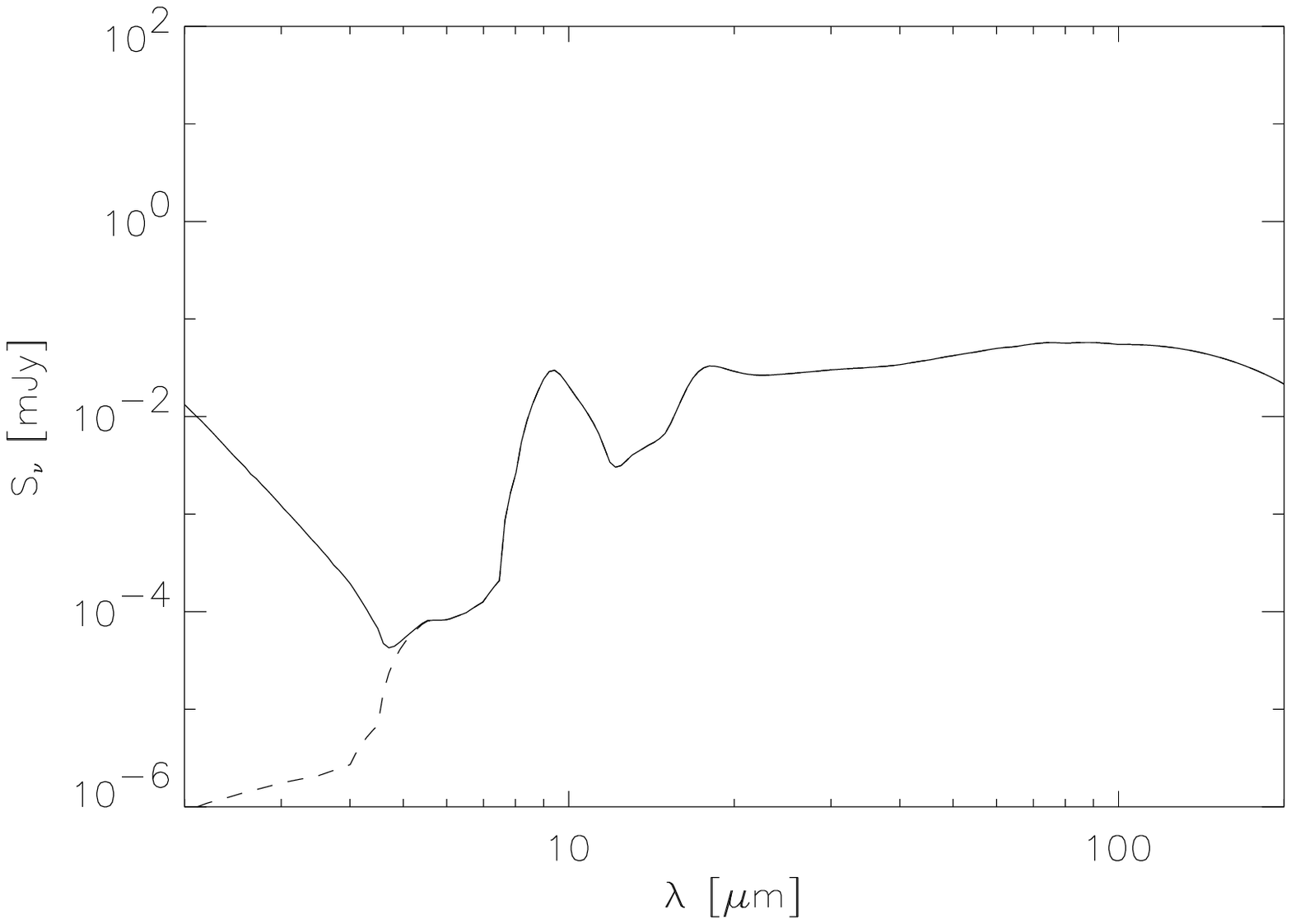}}

    \resizebox{0.47\hsize}{!}{\includegraphics{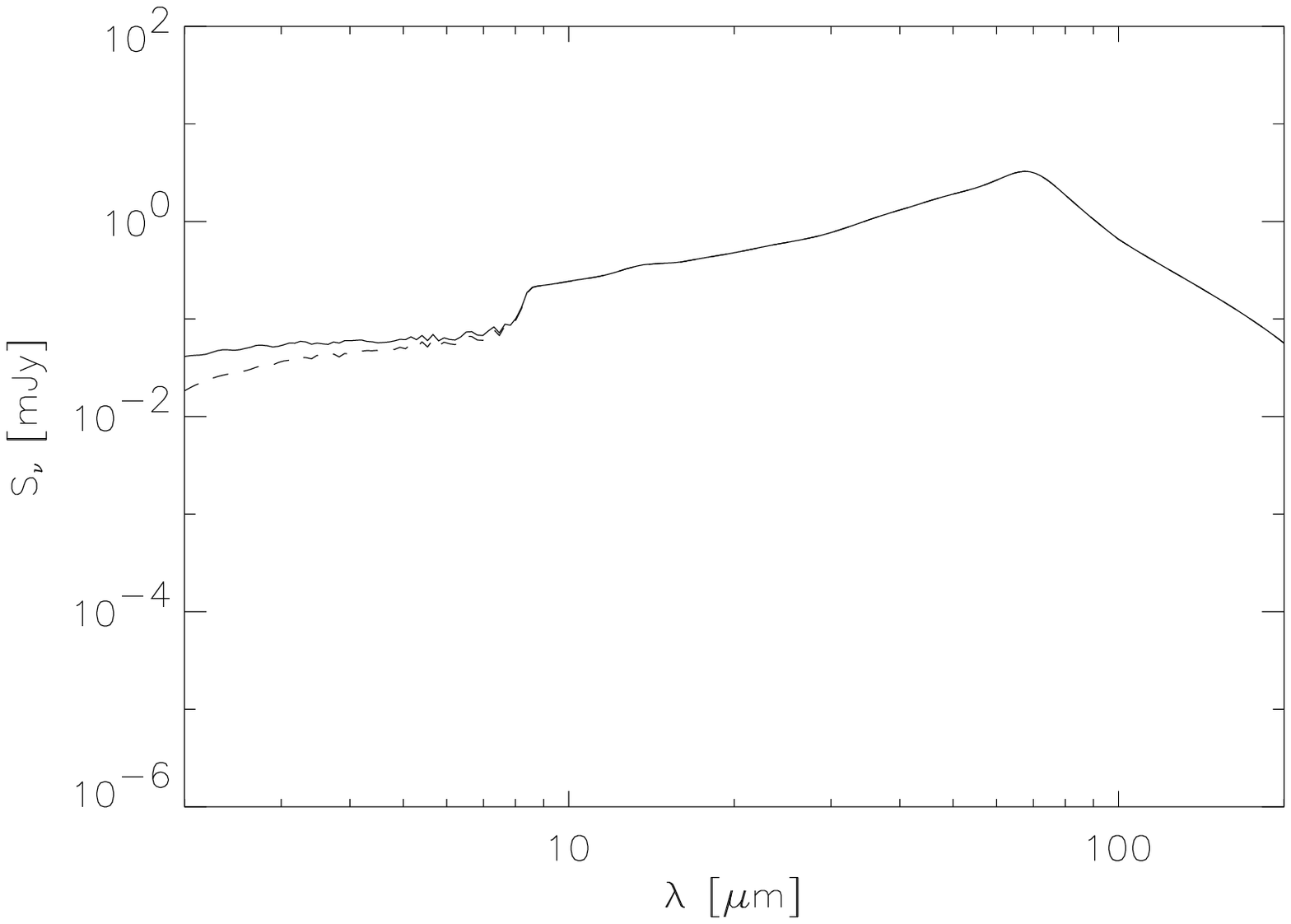}}
    \resizebox{0.47\hsize}{!}{\includegraphics{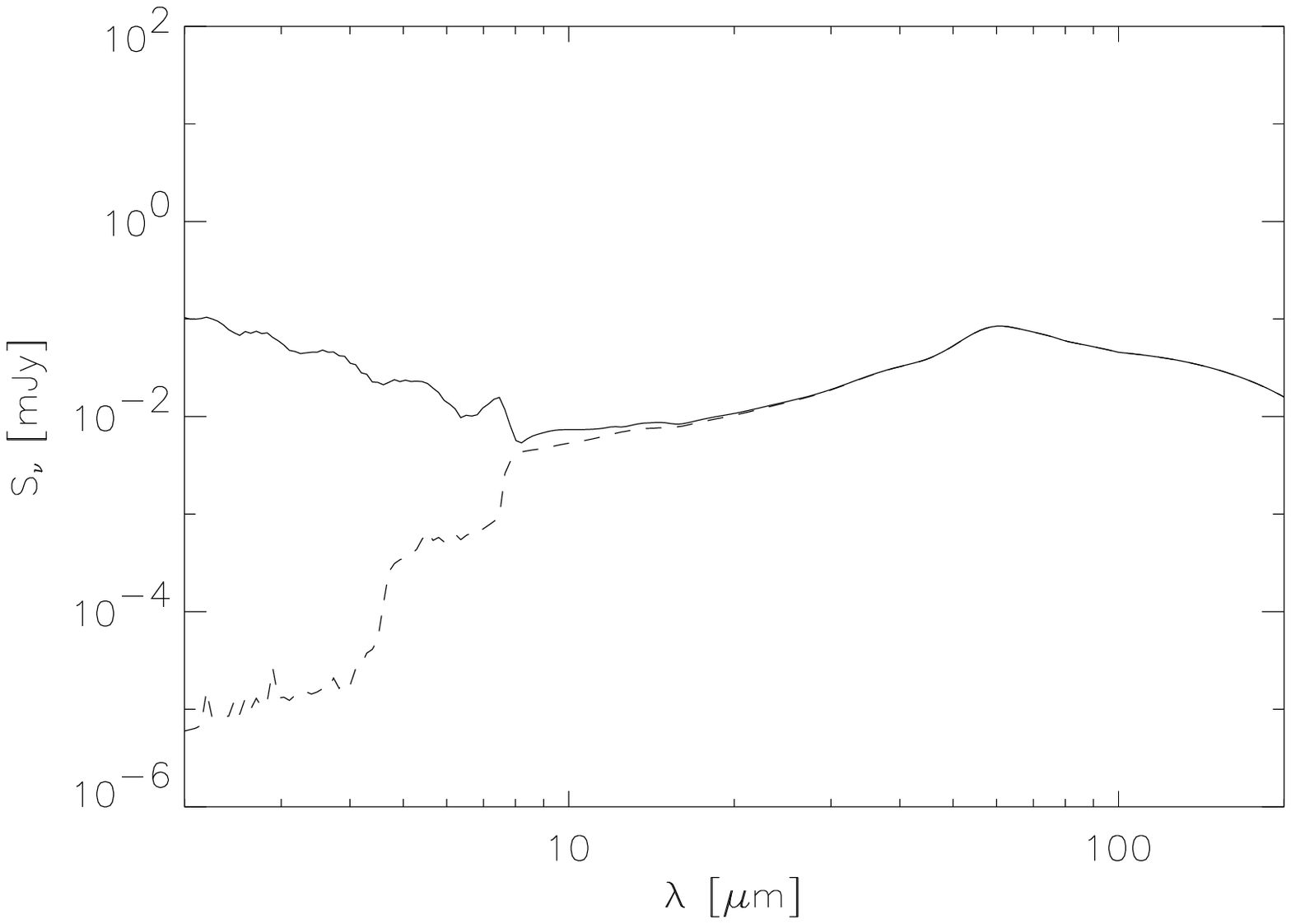}}

    \resizebox{0.47\hsize}{!}{\includegraphics{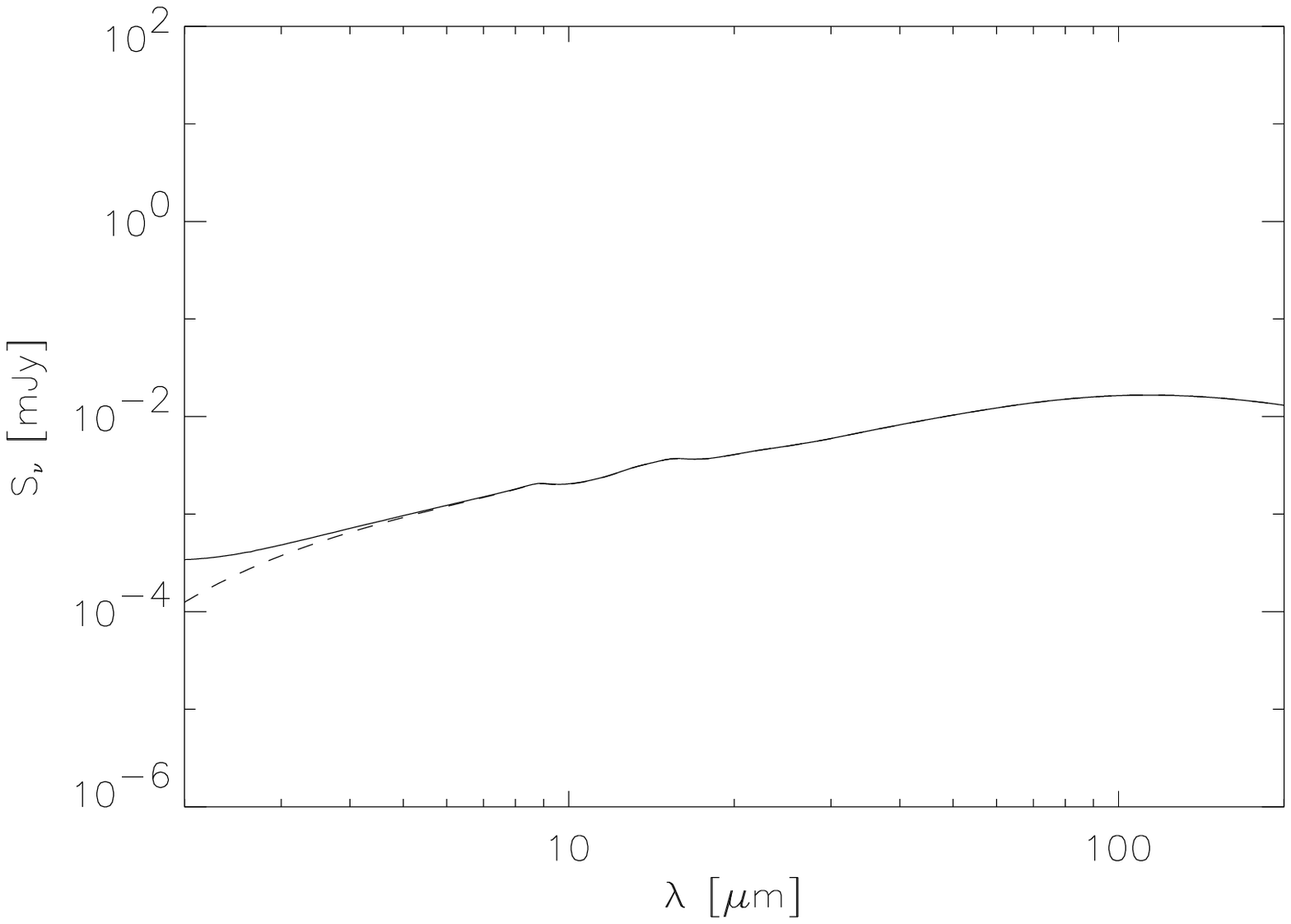}}
    \resizebox{0.47\hsize}{!}{\includegraphics{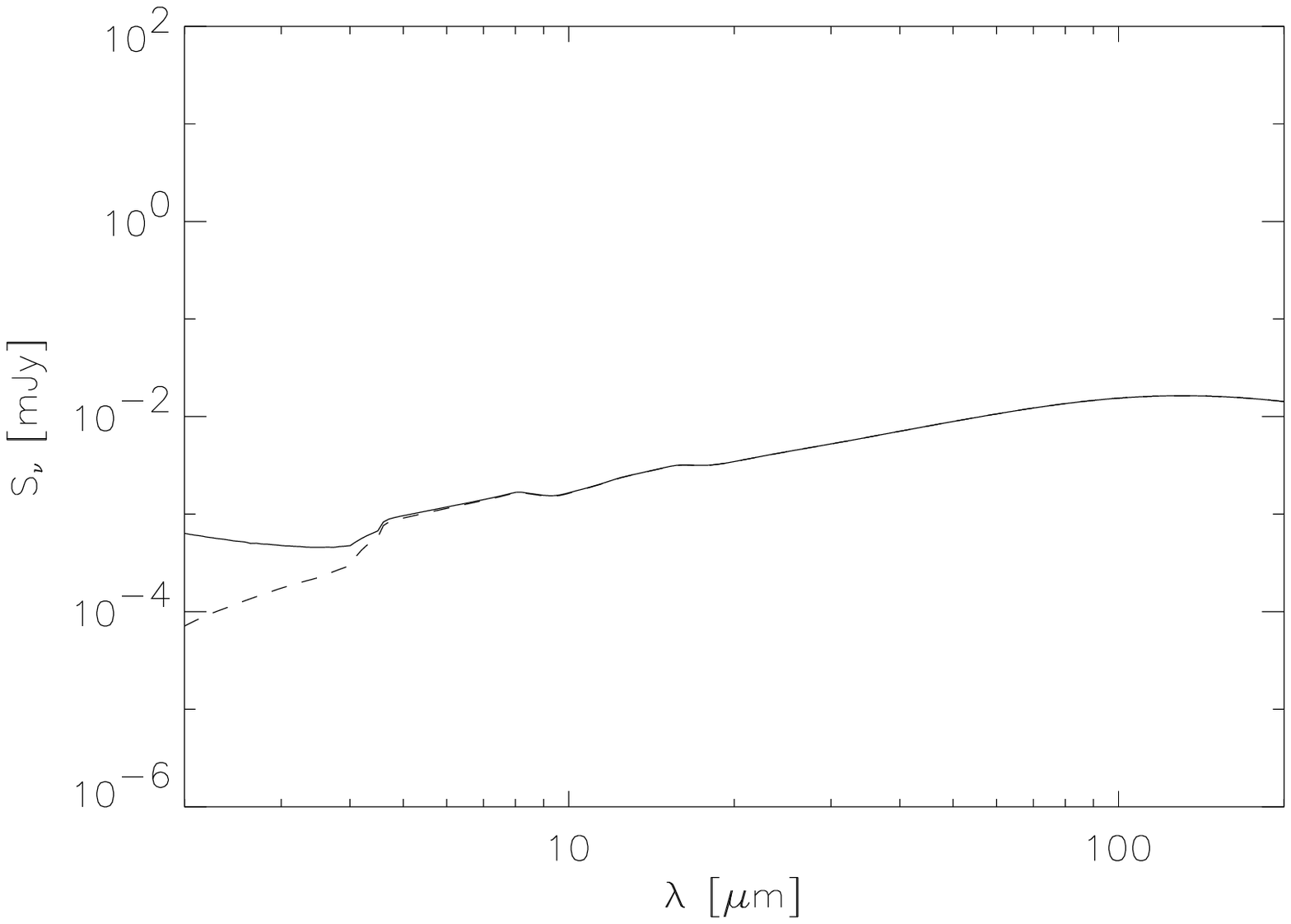}}
  \end{center}
  \caption{Illustration of the relative contributions of light scattering and thermal equilibrium emission
  in the emergent SED.
    \newline
    Solid line: Scattered + Reemitted radiation.
    Dashed line: Reemitted radiation only.
    \newline
    Left column:   Fe-rich Silicate (amorphous Olivine, MgFeSiO$_4$).
    \newline
    Right  column: Fe-poor Silicate (MgSiO$_3$).
    \newline
    {\sl Top:}    $a$ = 0.1$\mu$m,
    {\sl Middle:} $a$ = 10$\mu$m,
    {\sl Bottom:} $a$ = 1mm.
    Disk mass: $10^{-10}$M$_{\sun}$.
  }
  \label{scattlight}
  \bigskip
\end{figure}


Finally, we illustrate the importance of scattering of stellar radiation by the dust in the case 
of iron-deficient and iron-rich silicate (see Fig.~\ref{scattlight}). Mainly determined by the solar-type SED
of the star, the contribution of scattering to the resulting SED is negligible beyond $\approx 10\mu$m.
At shorter wavelengths, however, the lower absorption efficiency  in the case of an iron-deficient silicate
results in a larger relative constribution of the scattered light, while the iron-rich amorphous
Olivine dust reemission dominates at wavelengths of a few microns. Thus, if iron-poor silicates
or crystalline Olivine are the dominating dust species, spectral features of the star will be present
in the near- to mid-infrared wavelength range even after correct substraction of the stellar SED
(e.g., resulting from independent stellar models).

\section{Parameter Study}\label{sec_parstudy}

In this section we explore the parameter space allowed by our adopted analytic disk model.
We are strongly guided by observations and theoretical constraints in considering
viable values for the disk density distribution, inner and outer disk radius, and the grain size distribution.

\subsection{Disk density distribution}\label{diskdendis}


\begin{figure}[t]
  \begin{center}
    \resizebox{0.55\hsize}{!}{\includegraphics{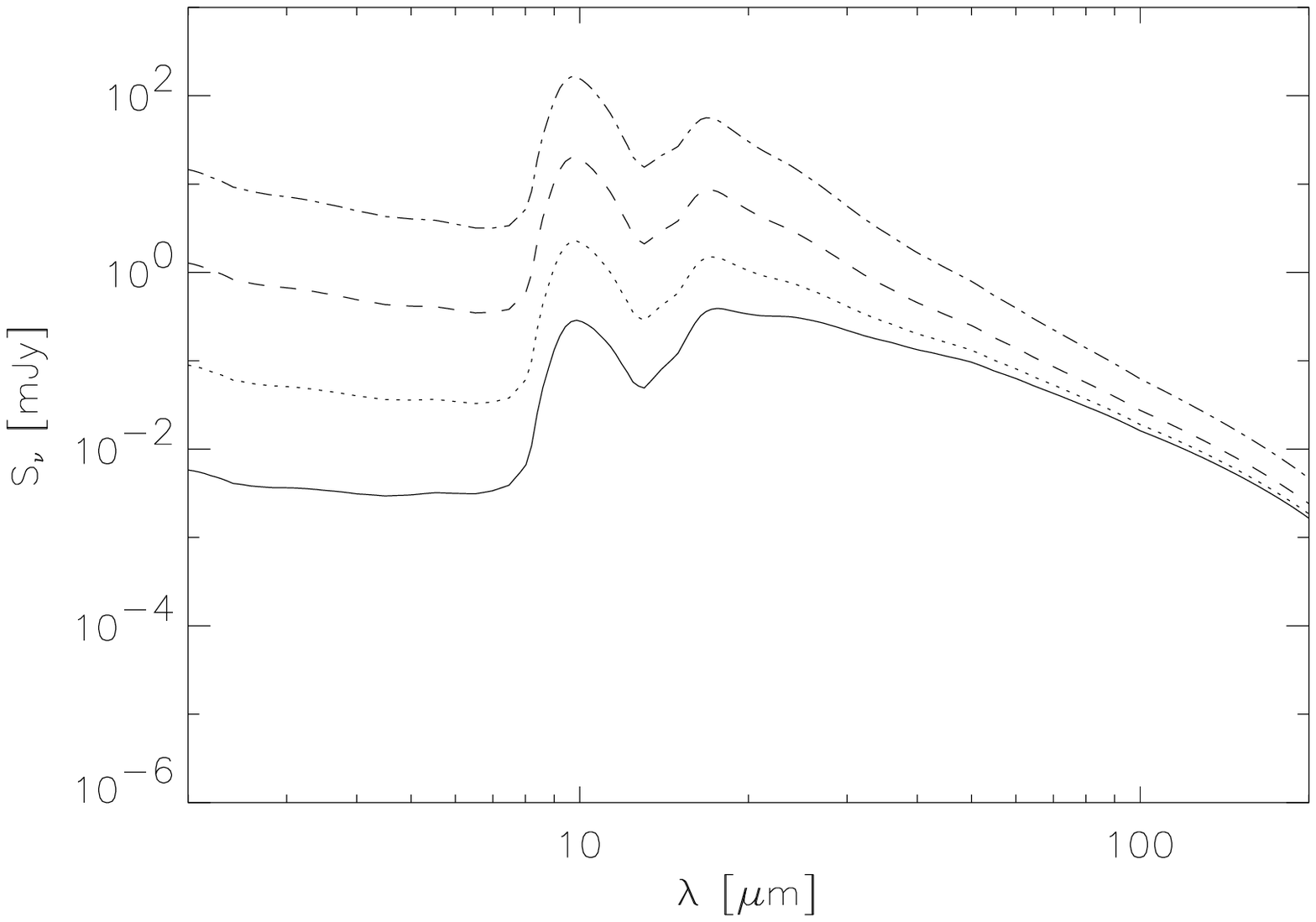}}
    \resizebox{0.55\hsize}{!}{\includegraphics{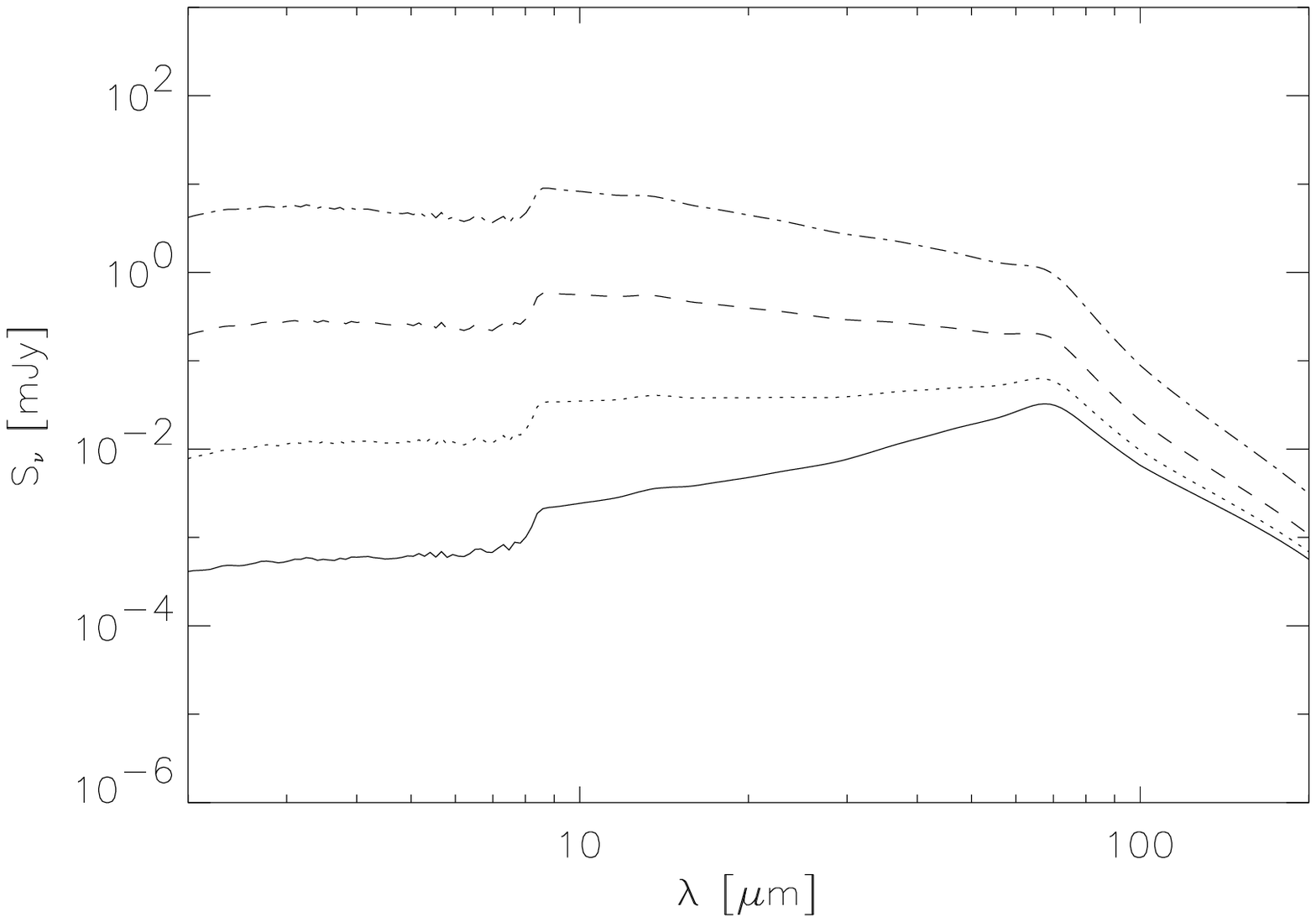}}
    \resizebox{0.55\hsize}{!}{\includegraphics{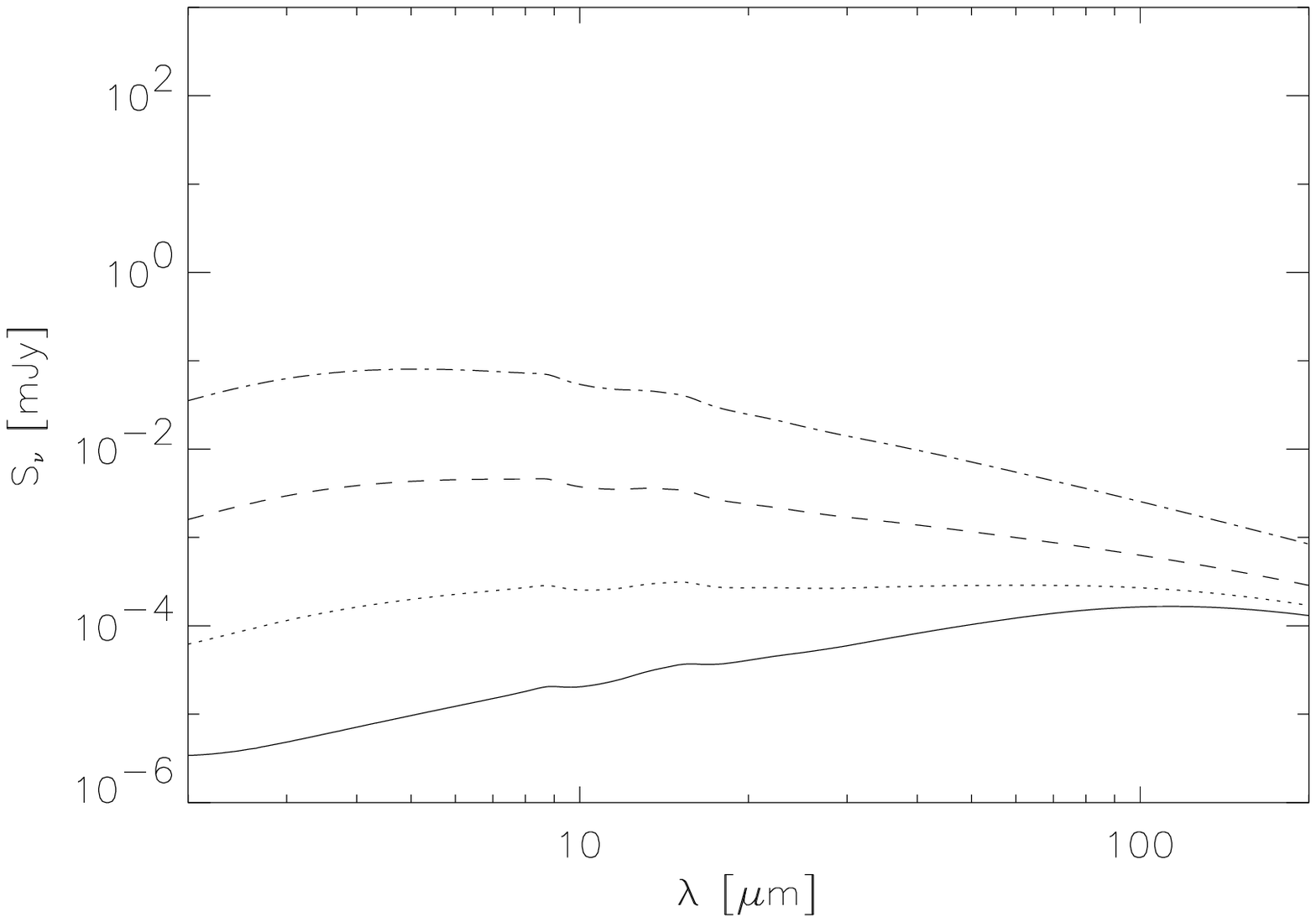}}
  \end{center}
  \caption{Dependence of the emergent SED on the disk density distribution.
    Disk density profile:
    $q$=1.0 (solid line),
    1.5 (dotted),
    2.0 (dashed), and
    2.5 (dash-dotted).
    {\sl Top:}    $a$ = 0.1$\mu$m,
    {\sl Middle:} $a$ = 10$\mu$m,
    {\sl Bottom:} $a$ = 1mm.
    All calculations use amorphous olivine (MgFeSiO$_4$) composition and a 
    disk mass of $10^{-12}$M$_{\sun}$.
  }
  \label{q-dendis}
  \bigskip
\end{figure}


Based on hydrostatic models, the radial density distribution in T\,Tauri disks is proposed to be in the range
\begin{equation}
n(r) \propto r^{-(1.9 - 2.4)}
\end{equation}
(see, e.g., 
Kenyon \& Hartmann~1987, 
D'Alessio et al.~1999, 
Chiang \& Goldreich~1997).
Detailed modelling of the near-infrared - to - millimeter appearance of several 
spatially resolved T\,Tauri disks has confirmed these predictions
(see, e.g., 
Burrows et al.~1996; 
Stapelfeldt et al.~1998;
Cotera et al.~2001;
Wood et al.~2002;
Wolf et al.~2003).
Optically thin debris disks which - in contrast to T\,Tauri disks - are not dominated by gas, 
are assumed to cover this range of radial density profiles as well but also to manifest values
outside of this range.

In the most simple model of debris disks, structure is dominated exclusively by gravitation,
radiation pressure, and Poynting-Robertson drag. In a collisionless system without sources or sinks and
grains in circular orbits, the exponent $q$ amounts to 1 (see, e.g., Briggs~1962).
This ``classical solution'' approximately represents the overall distribution of dust in the solar system
(see, e.g., Ishimoto~2000).
However, as for instance Gor'kavyi et al.~(1997) pointed out, there exist several effects that may change this
distribution remarkably, including
resonance effects with planets,
gravitational encounters with planets which occur in the form of elastic gravitational scattering,
mutual collisions of particles, evaporation of dust grains, and
existence of sources of dust with highly eccentric orbits (such as the Encke comet in the case of the Solar system;
	(see, e.g., Whipple~1976, Sykes~1988, Epifani et al.~2001).
The resulting density distribution is in most cases no longer described by a single power-law as stated in 
Eq.~\ref{eq_radprof}, but depends strongly on the properties (mainly the orbits and masses of planets 
and dust sources) of each particular disk/planetary system.
The exponents of the radial density distribution derived for different radial zones in
the model of the Solar system reach from $q$=1.0 to 2.4 (Gor'kavyi et al.~1997).
Another simply-structured debris disk would be a cloud of grains moving outward from the centre in hyperbolic
trajectories. This solution, which applies to so-called $\beta$-meteorites in the interplanetary space
is described by $q$=2 (see, e.g.,
Lecalvier des Etangs, Vidal-Madjar, \& Ferlet~1998;
Ishimoto \& Mann~1999).
In the case of $\beta$~Pic, visible observations of the scattered starlight and mid-infrared images show that
the outer disk (at radii $>100$AU) can be described by a power-law in the range
[1.5,2.3] (Artymowicz, Burrows, \& Paresce~1989; Kalas \& Jewitt~1995; Smith \& Terrile~1984; 
Pantin et al.~1997)\footnote{We want to remark that Trilling et al.~(2000; see also Trilling \& Brown~1998) 
reported the detection of circumstellar disks around
55~Cnr, $\rho$~CrB, and HD~210277
which are planet-harbouring systems (Butler et al.~1997, Noyes et al.~1997, Marcy et al.~1999).
From scattered-light images these authors derive a volume density profile described by $q=4$.
However, since these measurements could not be confirmed so far by later, independent 
(see, e.g., Jayawardhana et al.~2002)
measurements, we do not take so steep density profiles into account.}.
In the following we consider the cases $q$=1.0, 1.5, 2.0, and 2.5.

We find that for a given grain size and chemistry, the radial density profile, described by the exponent $q$ 
(see Eq.~\ref{eq_radprof}), determines the overall slope of the SED (see Fig.~\ref{q-dendis}). 
The relative strength of the spectral features
is not affected by $q$. The steeper the density profile, i.e., the larger the exponent $q$, the larger
the relative amount of dust located in the inner regions of the disk. 
Since the dust temperature increases towards the star, this results in an increase of the
near/mid-infrared-to-(sub)mm luminosity ratio.
But also the net reemission of the disk increases with increasing steepness of the density profile
since the relative number of hotter and therefore more luminous grains increases.

\subsection{Inner Gaps in the Disk}\label{sec_gaps}


\begin{figure}[t]
  \begin{center}
    \resizebox{0.47\hsize}{!}{\includegraphics{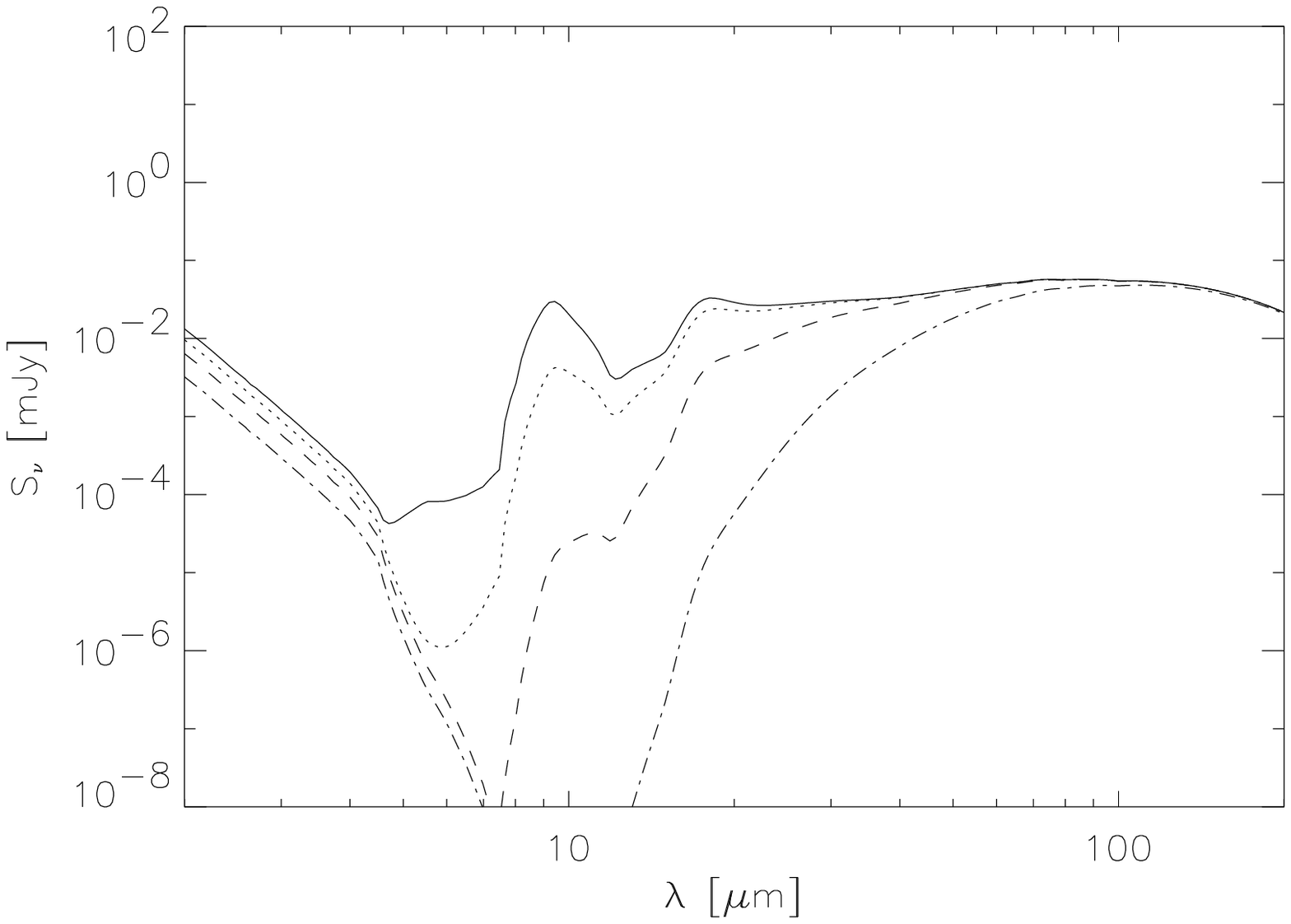}}
    \resizebox{0.47\hsize}{!}{\includegraphics{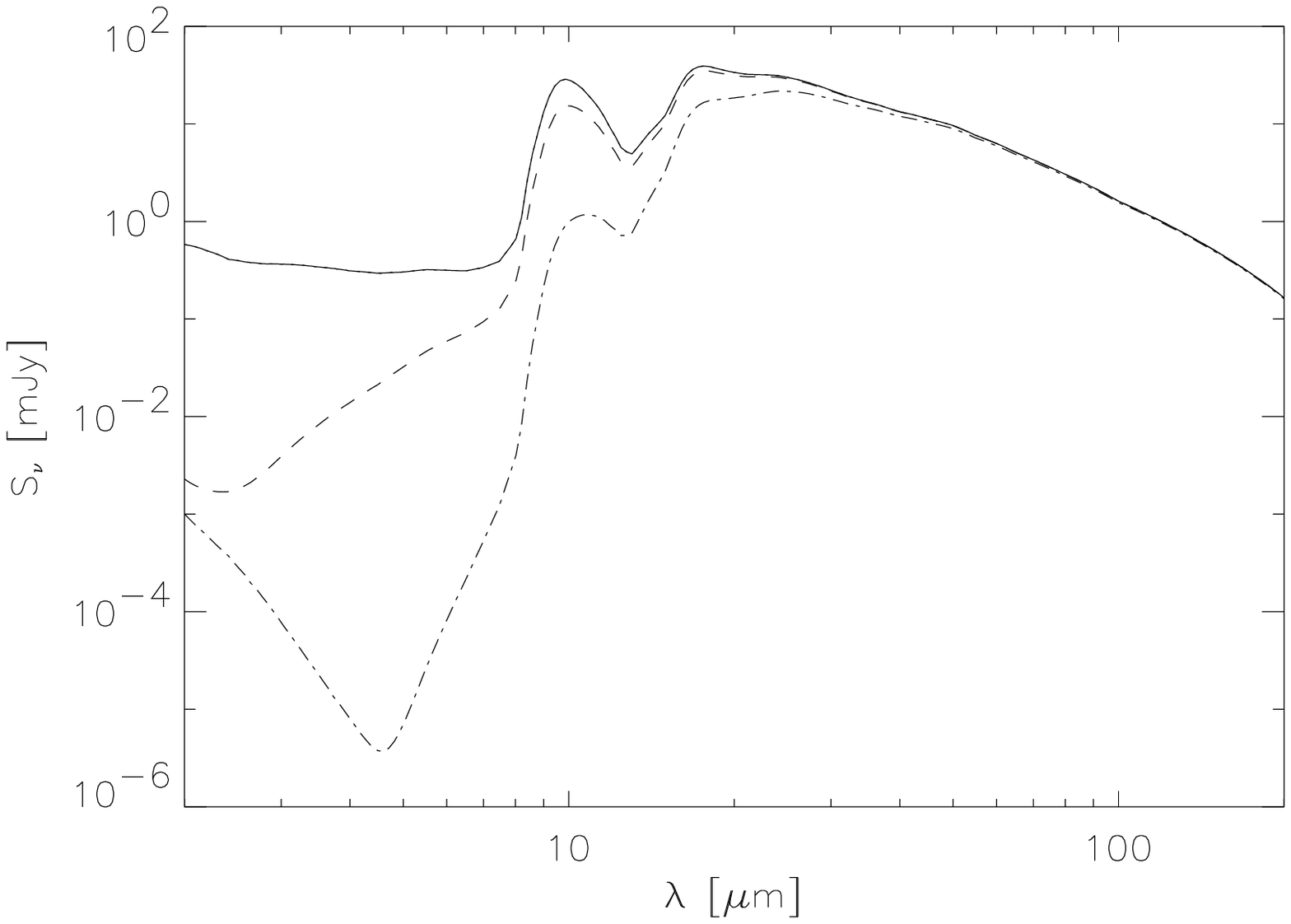}}

    \resizebox{0.47\hsize}{!}{\includegraphics{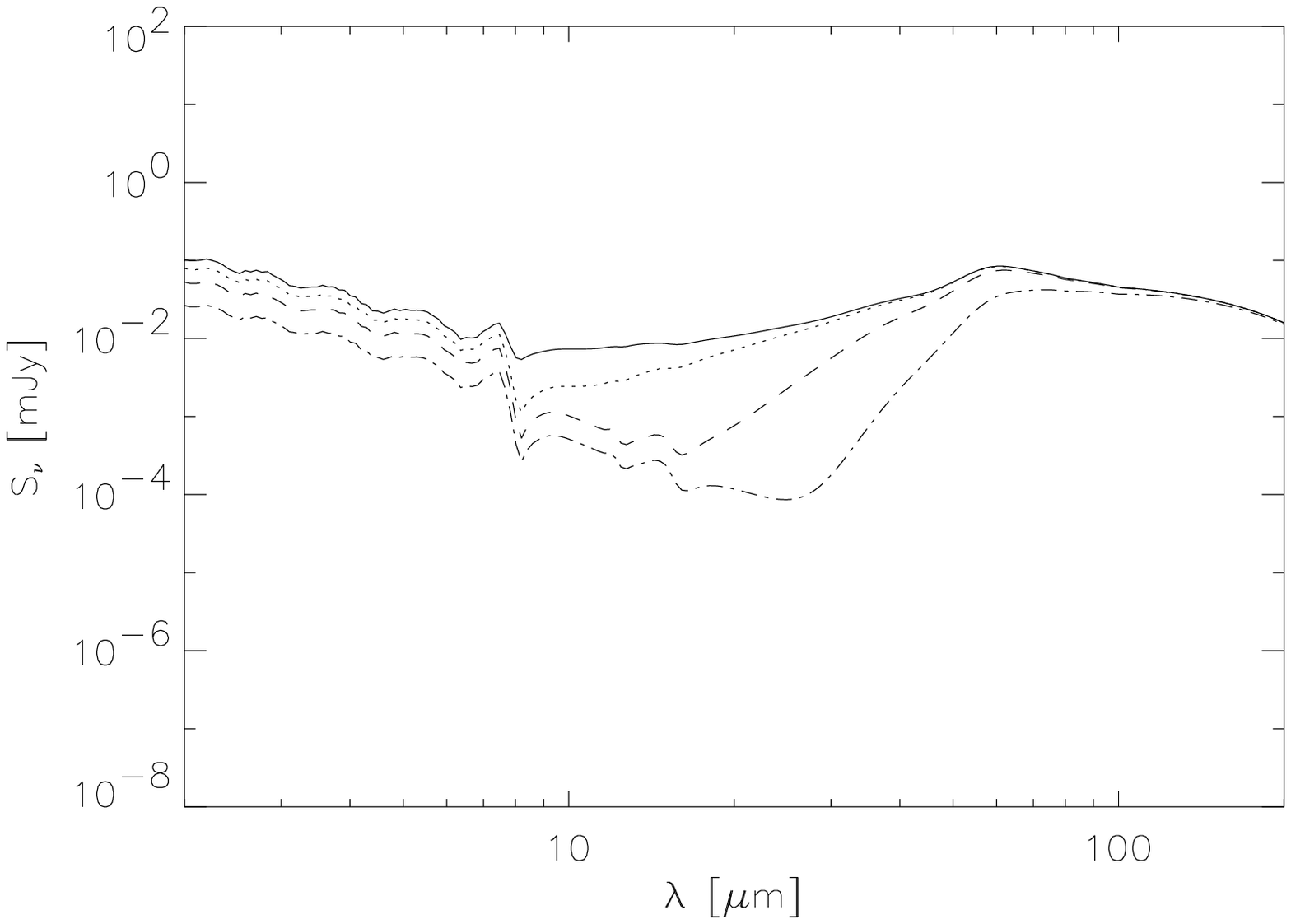}}
    \resizebox{0.47\hsize}{!}{\includegraphics{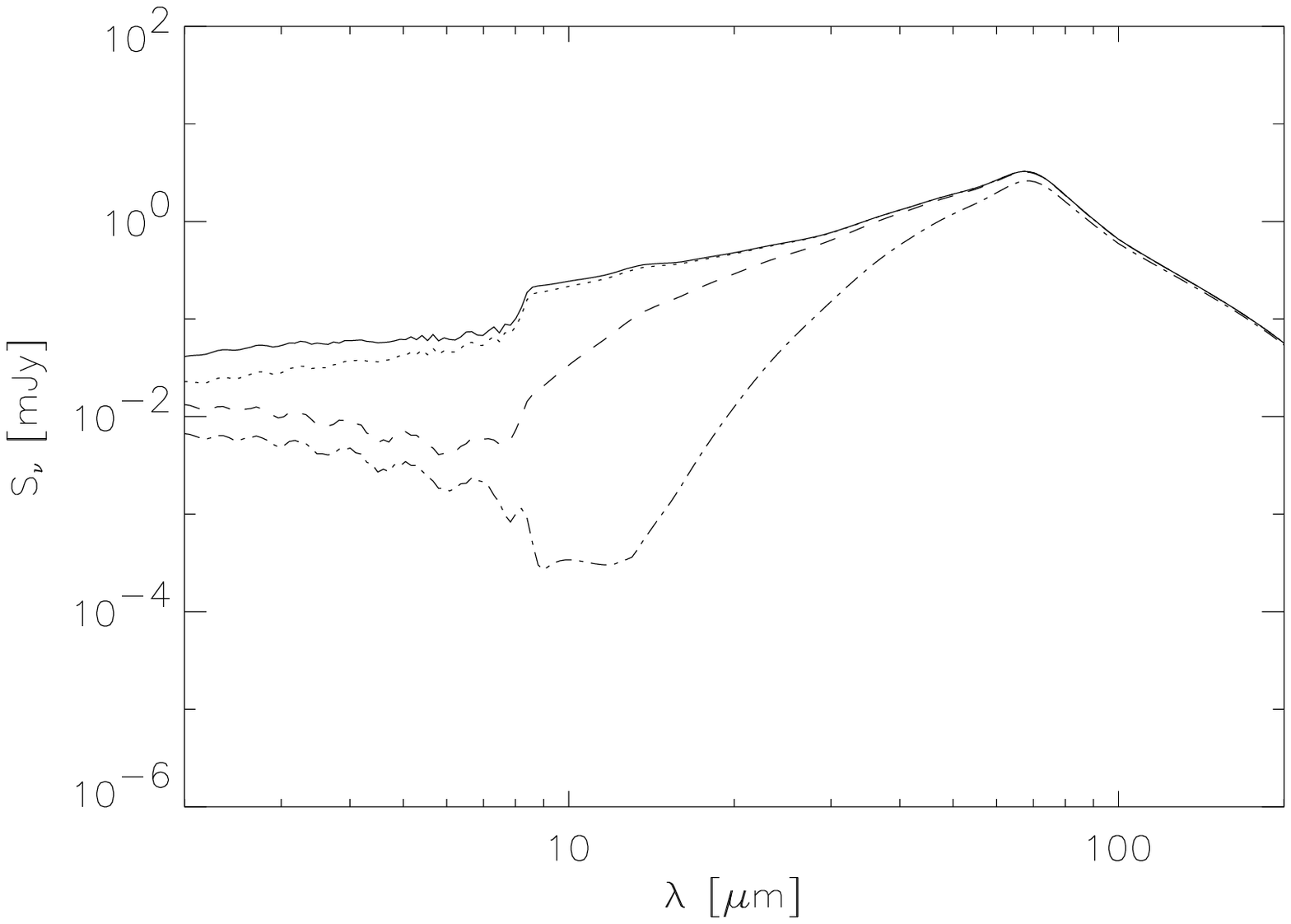}}

    \resizebox{0.47\hsize}{!}{\includegraphics{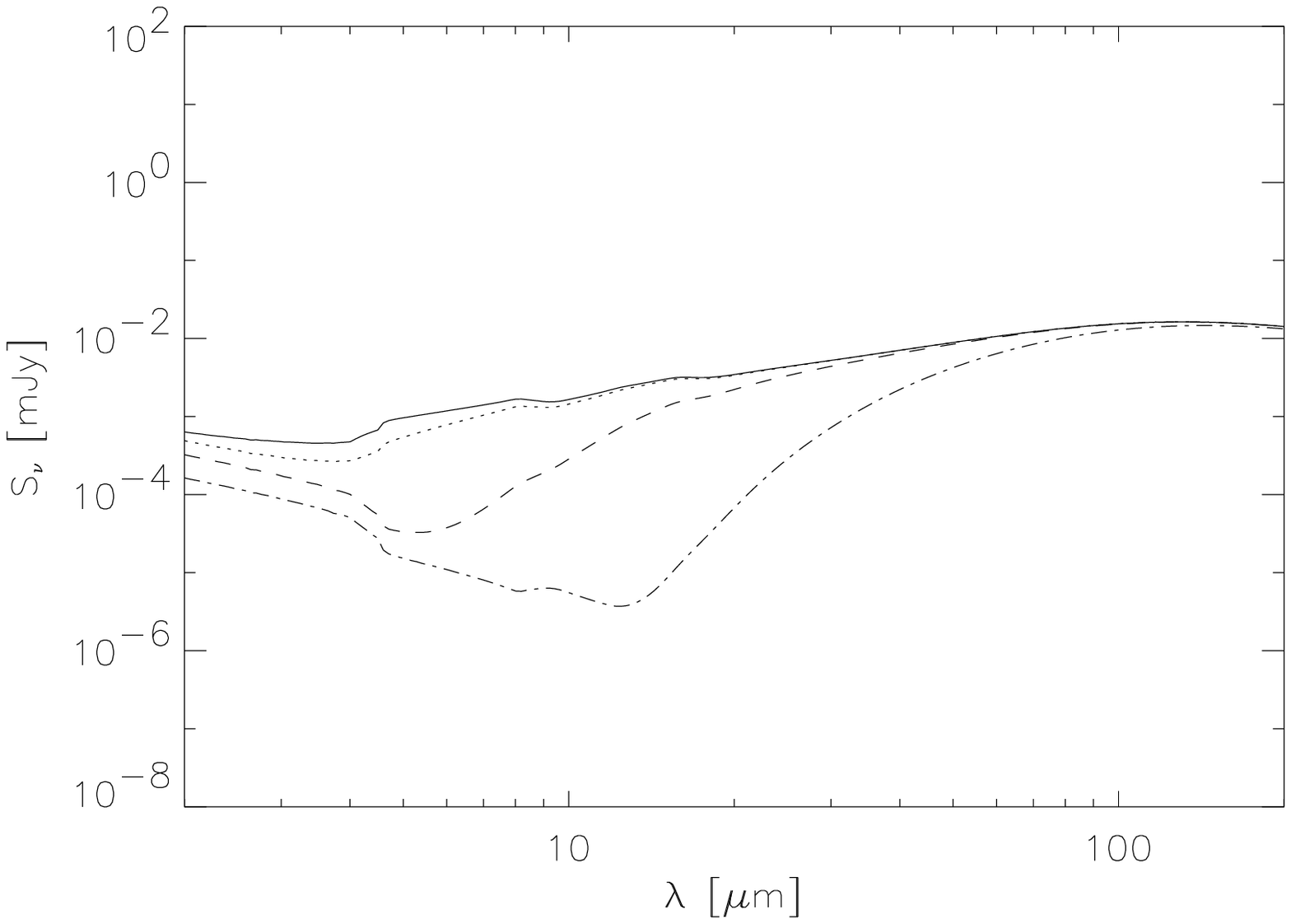}}
    \resizebox{0.47\hsize}{!}{\includegraphics{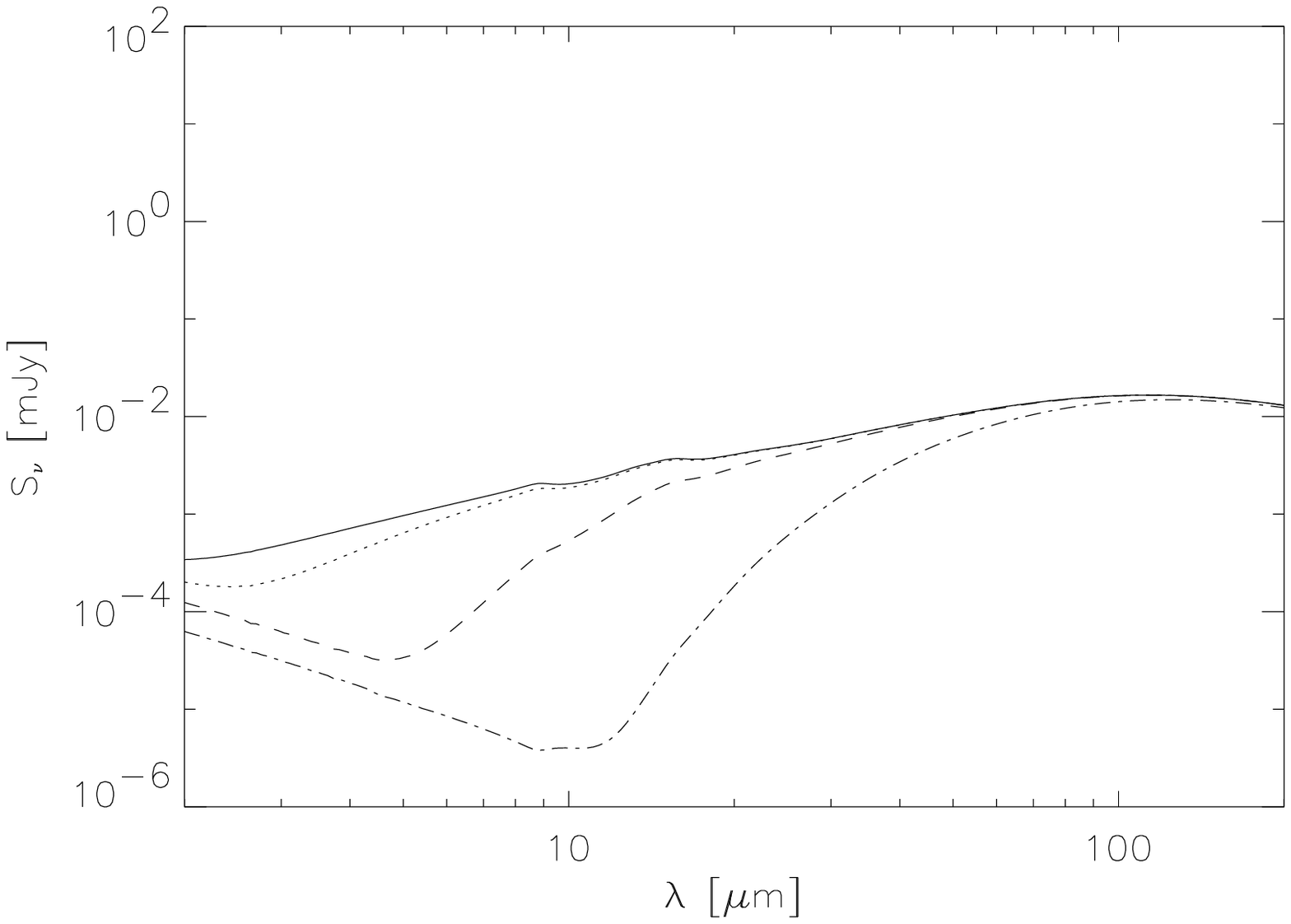}}
  \end{center}
  \caption{Influence of inner gaps on the SED.
    Inner disk radius: 
    dust sublimation radius (solid line),
    0.1\,AU (dotted),
     1\,AU (dashed), and
    10\,AU (dash-dotted).
    \newline
    Left column:   Fe-poor Silicate (MgSiO$_3$).
    \newline
    Right  column: Fe-rich (amorphous Olivine, MgFeSiO$_4$).
    \newline
    {\sl Top:}    $a$ = 0.1$\mu$m,
    {\sl Middle:} $a$ = 10$\mu$m,
    {\sl Bottom:} $a$ = 1mm.
    Assuming a dust sublimation temperature of 1550\,K, the sublimation radius 
    of MgSiO$_3$ / MgFeSiO$_4$
    amounts to 
    $7.3\times10^{-3}$\,AU / $1.3\times10^{-1}$\,AU ($a$ = 0.1$\mu$m),
    $1.2\times10^{-2}$\,AU / $3.5\times10^{-2}$\,AU ($a$ = 10$\mu$m), and
    $3.3\times10^{-2}$\,AU / $3.2\times10^{-2}$\,AU ($a$ = 1mm).
    Disk mass: $10^{-10}$M$_{\sun}$.
  }
  \label{innergap}
  \bigskip
\end{figure}


Observed SEDs of selected disks around T\,Tauri and debris-type disks have been found to show hints of 
inner cavities such that the inner radius of the disk is apparently much larger than the
sublimation radius of possible dust species.
Examples among T\,Tauri disks are GM~Aurigae (Koerner et al.~1993, Rice et al.~2003)
and TW~Hya (Calvet et al.~2002).
In several prominent debris disks, inner cavities have been found as well:
$\beta$~Pic (inner radius: 20\,AU\footnote{We want to remark that this value derived by Dent et al.~(2000)
from optical-to-submillimeter spectroscopy is in contrast to the result given by Weinberger et al.~(2003)
who assume a substantial amount of dust grains within that radius based on spatially resolved
mid-infrared spectroscopy (see \S~\ref{sizedis}).}), 
HR~4796A (30-50\,AU),
$\epsilon$~Eri (50\,AU), 
Vega (80\,AU), and
Fomalhaut (125\,AU) - 
see, e.g., Dent et al.~2000; 
Greaves, Mannings, \& Holland~2000b; Wilner et al.~2002; Holland et al.~2003.
Taking into account the physical processes responsible for the spatial dust density distribution
in debris disks, scattering of particles by massive planets is assumed to be the major effect
in explaining these large inner holes:
Dust grains drifting inwards due to the Poynting-Robertson effect are likely to be scattered into
larger orbits resulting in a lower dust number density within the planet's orbit.
Based on the assumed cavity sizes of the debris disks quoted above (relative to the total sizes of these disks), 
we consider the following cases: 0.1\,AU, 1.0\,AU, and 10\,AU and compare the resulting SED to the
case of a disk without any gap, that is, extending inward to the dust sublimation radius.

Clearing the inner disk results in a loss of warm dust which is mainly responsible for the near/mid-infrared
shape of the SED. With increasing gap size the minimum value of the mid-infrared flux becomes smaller and
is shifted towards longer wavelengths (see Fig.~\ref{innergap}).
Because of the higher temperature and therefore more significant contribution to the near/mid-infrared
spectrum by small grains, the decrease of the flux in this wavelength range is more pronounced when 
the relative number of small grains in the system goes up.
Keeping the disk mass constant, the flux in the millimeter region (not accessible by \SIRTF)
increases slightly, but the net flux is smaller compared to a disk without an inner gap.
This is because the fraction of the stellar flux absorbed by a single dust grain
decreases with increasing radial distance from the star (see Eq.~\ref{eq_dist2}).
Applying the same explanation to the scattering cross section (see Eq.~\ref{eq_scatt}), the slight decrease
of the scattered stellar light with increasing gap size can be explained by analogy.

\subsection{Outer disk radius}\label{outerradius}


\begin{figure}[t]
  \begin{center}
    \resizebox{0.55\hsize}{!}{\includegraphics{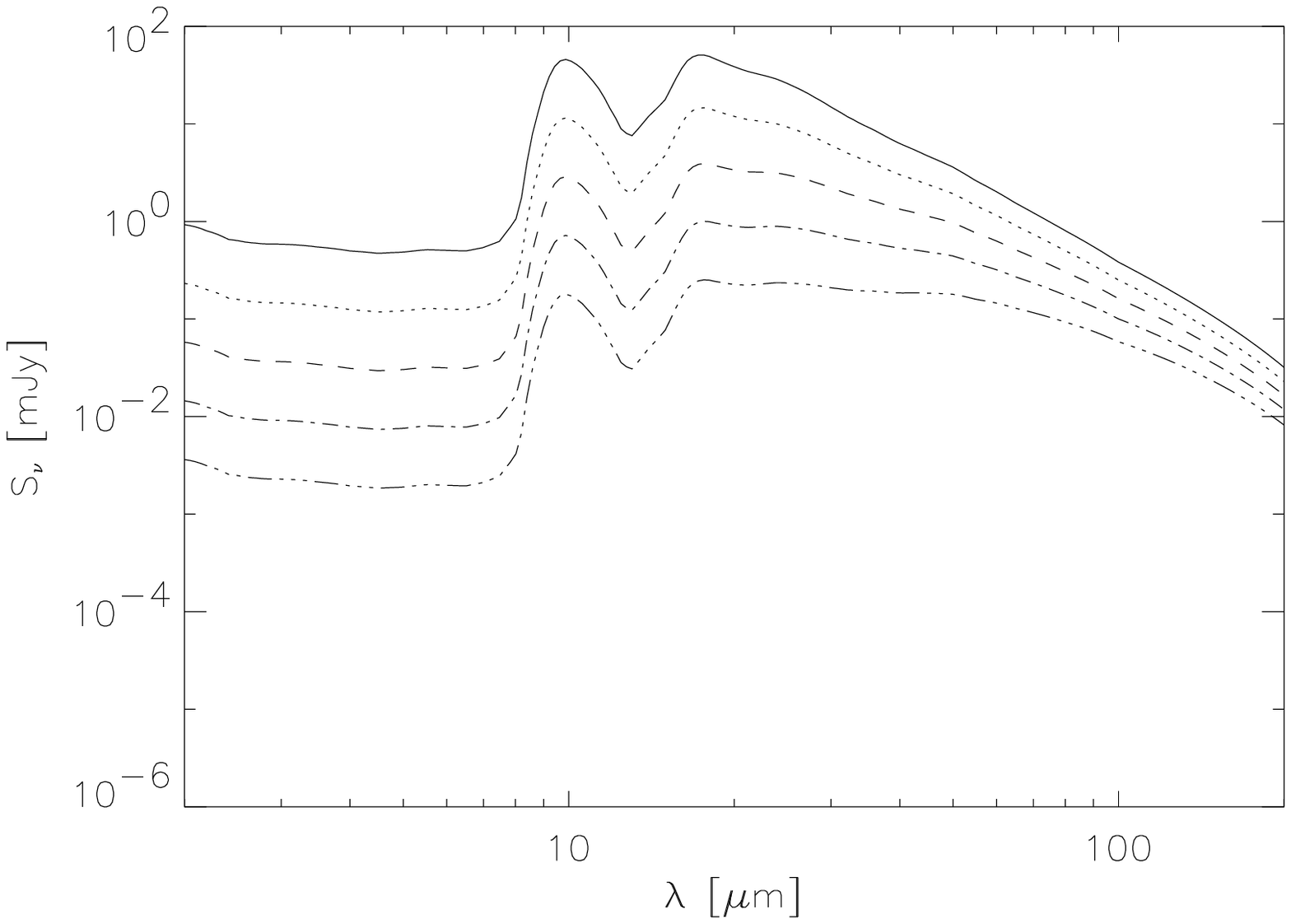}}
    \resizebox{0.55\hsize}{!}{\includegraphics{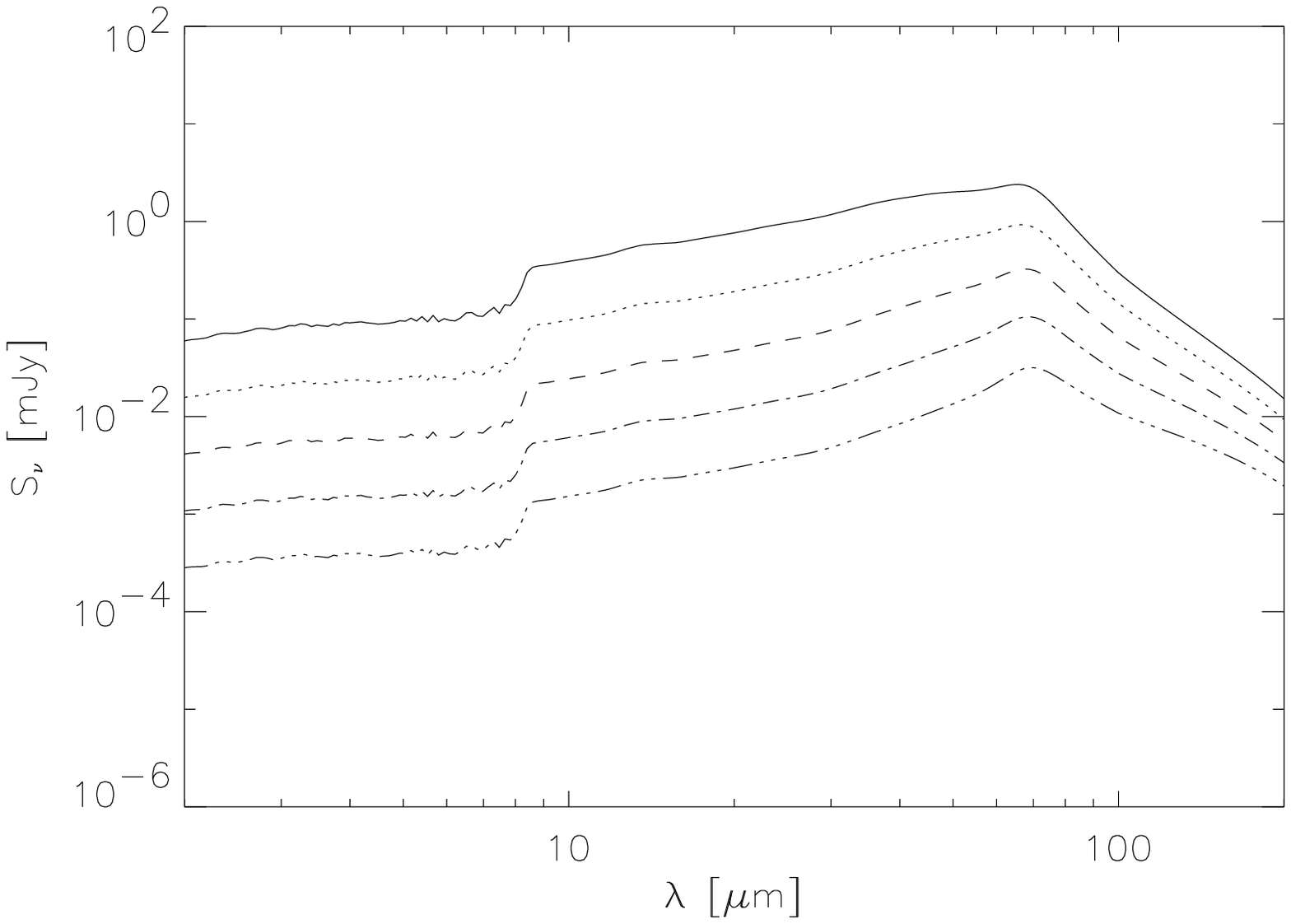}}
    \resizebox{0.55\hsize}{!}{\includegraphics{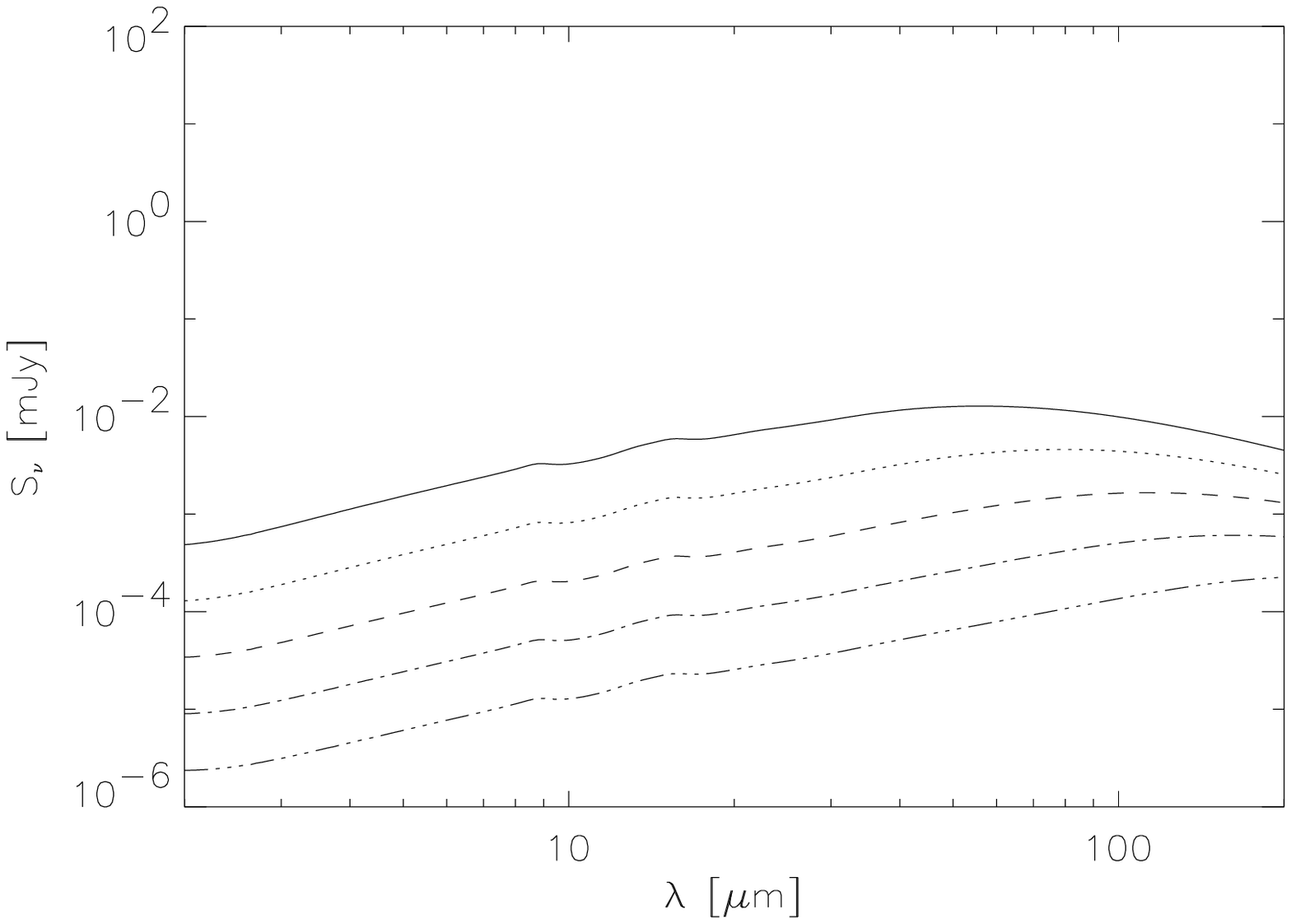}}
  \end{center}
  \caption{Influence of the outer disk radius on the grain size distribution.
    Outer radius of the disk:
    25\,AU (solid),
    50\,AU (dot),
    100\,AU (dash),
    200\,AU (dash-dot),
    400\,AU (dash-dot-dot-dot).
    Grain radius: 0.1$\mu$m (top), 10$\mu$m (middle), 1mm (bottom).
    Amorphous olivine (MgFeSiO$_4$) composition and
    disk mass $10^{-11}$M$_{\sun}$ are assumed.
  }
  \label{rou}
  \bigskip
\end{figure}


Similar to T\,Tauri disks, circumstellar disks at the late stage of their evolution appear
to show a large range of possible radii, typically in the range of about 100\,AU to several hundered AU.
So far, scattered light images and SEDs revealed radii of 
80\,AU ($\epsilon$Eri), 
120\,AU (Vega), 
125\,AU (HR~4796),
185\,AU (Fomalhaut)
and 1000\,AU in case of the disk around $\beta$~Pic (see Vidal-Madjar et al.~1998;
Dent et al.~2000; Greaves, Mannings, \& Holland~2000b; Holland et al.~2003). 
However, scattered light images as well as near/mid-infrared SEDs of debris disks are
mainly determined by the pronounced spectral features and the total flux
in the small grain component in the disk (see \S~\ref{diskdendis}, Fig.~\ref{sidi}),
while larger grains with radii of several tens of micron and larger can be more efficiently found by
(sub)millimeter measurements.
One has to take into account that the abundance and spatial distribution of the small grain component, 
especially the one with $\beta \gtrsim 1$ (see Eq.~\ref{eq_beta}), depends on the location of dust sources
and the likelihood of collisional events. Since these parameters are not necessarily correlated
with the actual disk size, submillimeter/millimeter measurements are required to trace larger grains
which are less affected by the radiation pressure and Poynting-Robertson effect. 
The question to be answered here is whether
there are significant differences in the SED resulting from different disk sizes.

As result of our simulations,
Fig.~\ref{rou} shows the SED for disks with outer radii between 25 and 400\,AU.
The influence on the near/mid-infrared wavelength range is seen only as a decrease of the net flux
with increasing disk size which is due to the decrease of the mean temperature of the disk.
At several tens of microns and beyond 100$\mu$m also the shape of the SED is affected because
the smaller mean temperature results in an increase of the (sub)millimeter--to-near/mid-infrared flux ratio.

\subsection{Grain size distribution}\label{sizedis}


\begin{figure}[t]
  \begin{center}
    \resizebox{0.47\hsize}{!}{\includegraphics{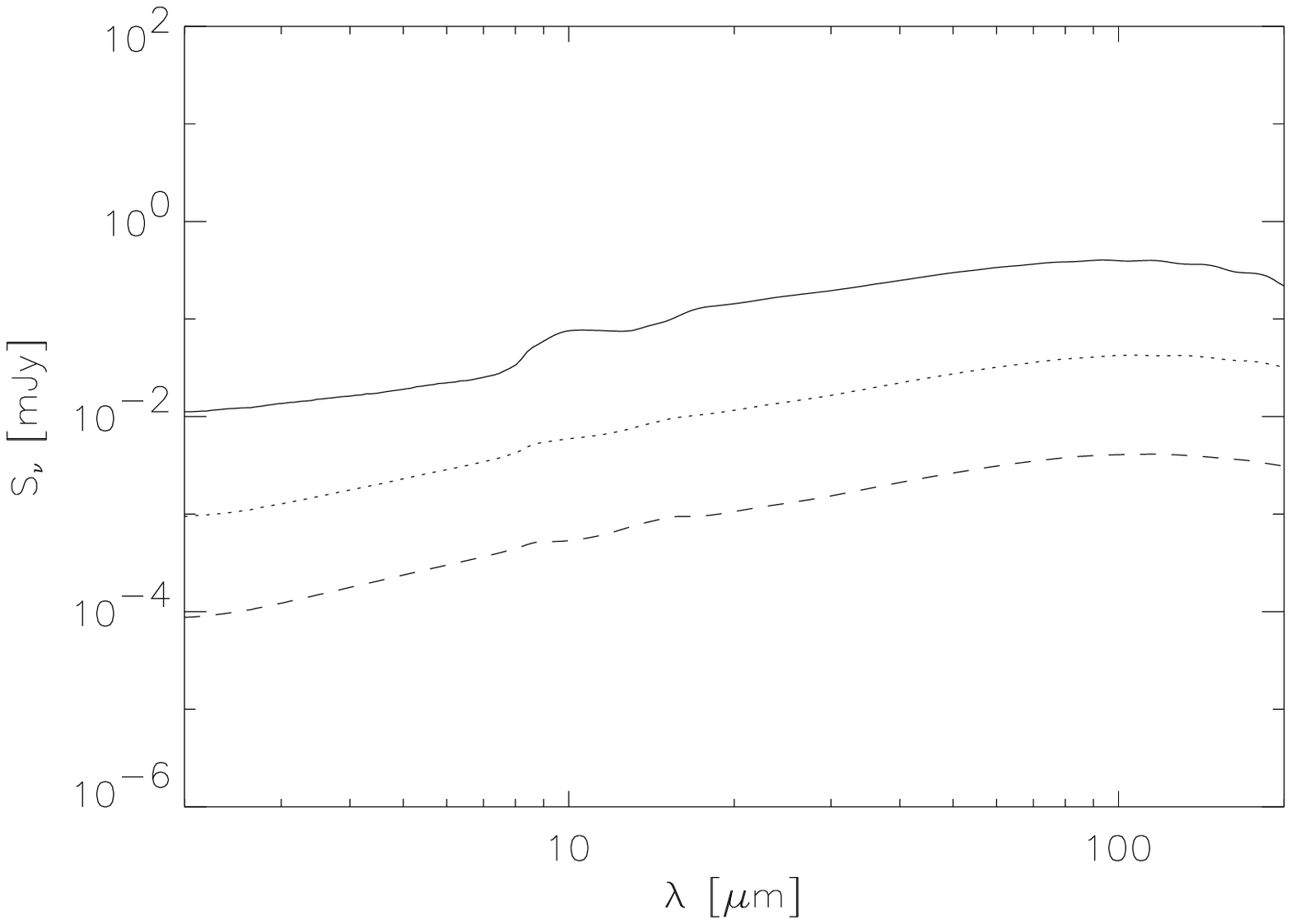}}
    \resizebox{0.47\hsize}{!}{\includegraphics{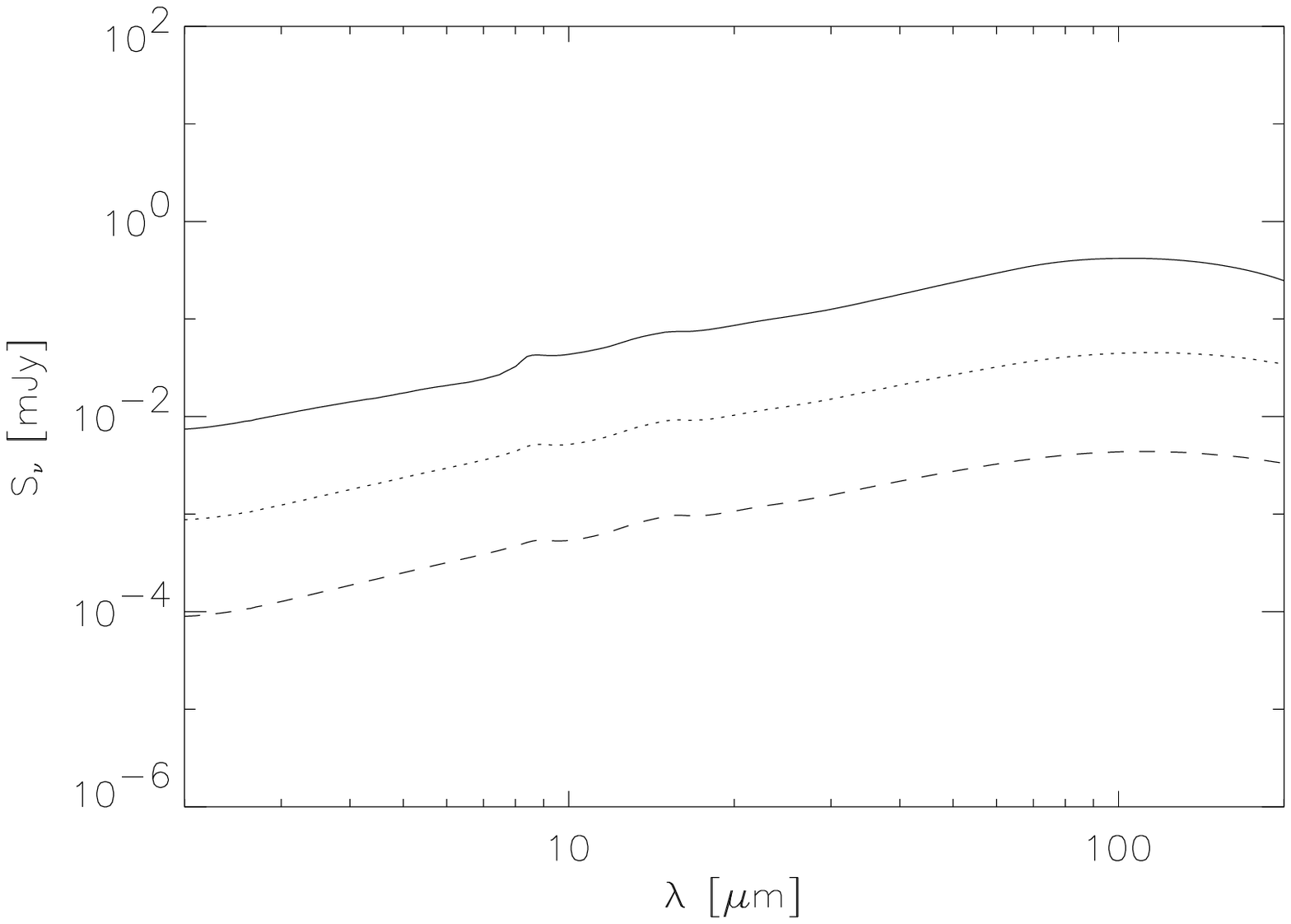}}
 
    \resizebox{0.47\hsize}{!}{\includegraphics{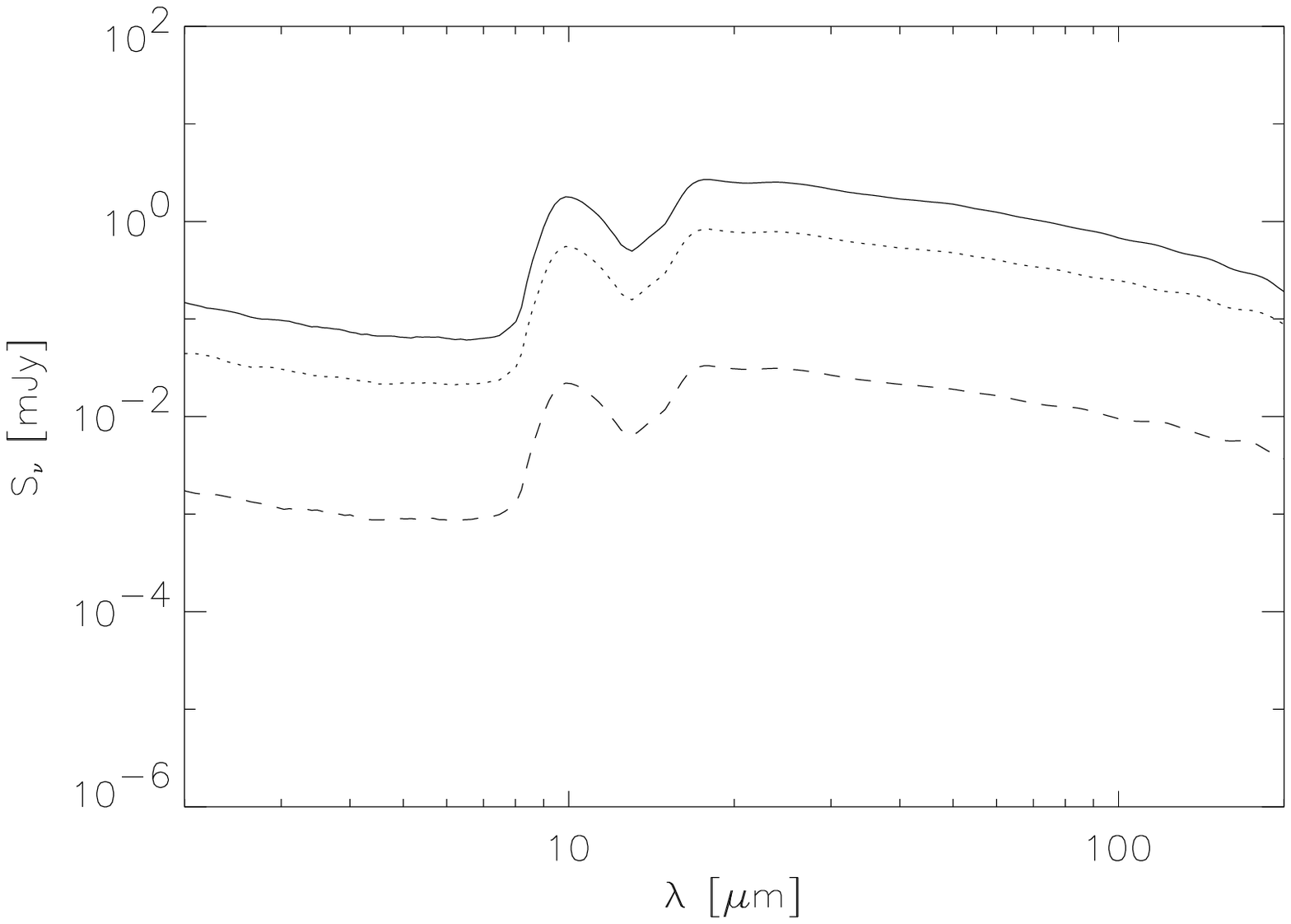}}
    \resizebox{0.47\hsize}{!}{\includegraphics{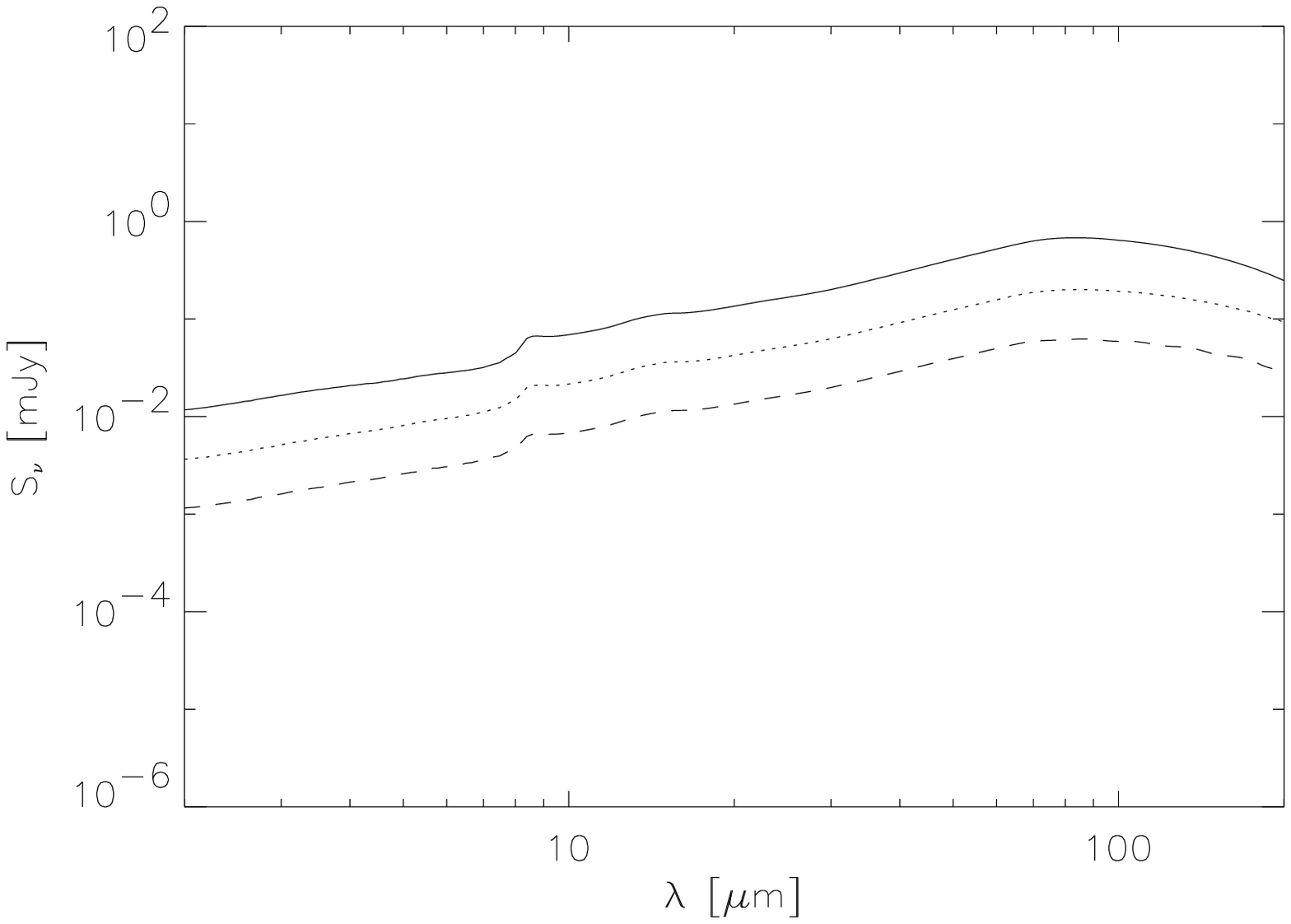}}
  \end{center}
  \caption{Influence of the grain size distribution on the SED.\newline
    Grain size distribution exponent:
     $p$ = 2.5 (upper row) and 3.5 (lower row).
    \newline
    Minimum grain size:
     $a_{\rm min}$ = 0.1$\mu$m (left column),
     and 10$\mu$m (right column).
    \newline
    Maximum grain size:
     $a_{\rm max}$ = 0.1mm (solid line),
     1mm   (dotted), and
     10mm  (dashed).
     \newline
    Amorphous olivine (MgFeSiO$_4$) composition and
     disk mass $10^{-10}$M$_{\sun}$ are assumed.
  }
  \label{sidi}
  \bigskip
\end{figure}


In this section we consider dust models with a grain size distribution which is characterized by
a (minimal) set of three parameters:
the minimum and maximum grain radius ($a_{\rm min}$, $a_{\rm max}$) and a power-law exponent $p$ of the 
size distribution:
\begin{equation}\label{eq_grsi}
  n(a) \propto a^{-p}.
\end{equation}
Since debris disks are assumed to be a place of continuous dust production due to collisional processes between
larger particles (rather than being nurtured from a hostile interstellar medium - see, e.g., 
Artymowicz \& Clampin~1997), all three parameters depend on the particular debris disk system. In contrast to
the interstellar medium or disks around very young stellar objects, all three parameters may therefore
vary in a wide range even between different radial zones within one and the same disk (Gor'kavyi et al.~1997).
Furthermore, the collision rate is assumed to decrease as the disk ages since the fraction
of disk material present in larger bodies increases with time (see, e.g., Kokubo \& Ida~2002).
This might allow us to classify debris disks based on their grain size distribution,
at least on a statistical basis.

\noindent
{\em Constraints on the minimum grain size}
$a_{\rm min}$:
The smallest grain size expected to be present is determined on the one hand by 
the likelihood of collisional processes in the disk,
and on the other hand by the optical properties of the dust and the strength of the
stellar radiation field.
If no collisions are occuring, the minimum grain size can be estimated from the ratio
of the radiation pressure to gravity, also known as $\beta$, (Burns et al.~1979)
\begin{equation}\label{eq_beta}
\beta = \frac{3}{16 \pi} \frac{L_{\rm s} Q_{\rm pr}}{G M_{\rm s} c a \rho_{\rm g}},
\end{equation}
where
$L_{\rm s}/M_{\rm s}$ is the central stellar luminosity/mass,
$G$ is the gravitational constant,
$c$ is the speed of light,
$\rho_{\rm g}$ is the material density of the grain,
and
$Q_{\rm pr}$ is the radiation pressure coefficient averaged over the entire stellar spectrum.
Those grains for which $\beta>1$ are ``blown away'' by radiation pressure on a timescale
much shorter than the age of the star (e.g., Aumann et al.~1984).
Mid-infrared observations of the $\beta$~Pic disk by Weinberger et al.~(2003) show that there must exist
a substantial amount of grains with radii $\le 10\mu$m within 20\,AU, while Pantin et al.~(1997)
even constrain the minimum grain size to be as small as $\approx 0.1\mu$m.
In case of Fomalhaut's debris disk, Wyatt \& Dent~(2002) derive a minimum grain size of $\approx 7\mu$m
from submillimeter observations.
In our parameter study we consider $a_{\rm min}$ = 0.1$\mu$m and 10$\mu$m since having the bulk of
the mass in larger grains does not produce a substantial mid-infrared contribution to the SED.

\noindent
{\em Constraints on the maximum grain size}
$a_{\rm max}$:
The upper grain size is a very uncertain parameter since debris disks are expected to harbour dust ``particles''
ranging from (sub)micron-sized grains to planetesimals/planets as the result of an (almost) finished
process of dust grain growth, planetesimal formation, and simultaneous dust production
due to collision events (see, e.g., Beckwith, Henning, \& Nakagawa~2000).
However, according to the wavelength range considered, grains larger than $\approx 1$mm will not give a
remarkable contribution to the SED. The existence of these grains can therefore not be proven or ruled out
by \SIRTF\ measurements.
Complementary ground-based (sub)millimeter observations can in principal increase this upper grain size
limit by about one order of magnitude.
Based on previous observations of the very limited number of debris disks only very weak constraints
on upper grain sizes could be derived. Paresce \& Burrows (1987), for example, found an absence of significant
colour excess in visible observations at the outer part of the $\beta$~Pic disk, which implies that
the dominant particles for the scattering are larger than a few $\mu$m.
Furthermore, observations of main-sequence and pre-main-sequence (post-Herbig Ae/Be, post-T~Tauri)
stars show that the distribution is broad and extends from below 1\,$\mu$m to at least $\approx$1\,mm
(see, e.g., 
Skinner, Barlow, \& Justtanont~1992;
Sylvester et al.~1996;
Sylvester \& Skinner~1996;
Li \& Greenberg~1998).
In our parameter study we consider $a_{\rm max}$ = 0.1mm, 1mm, and 10mm.

\noindent
{\em Constraints on the grain size distribution exponent}
$p$:
The equilibrium size distribution resulting from a collisional 
cascade can be described by a power-law (Eq.~\ref{eq_grsi}) with the exponent $p=3p_{\rm d}-2$ (Dohnanyi~1969).
Furthermore, Tanaka, Inaba, \& Nakazawa~1996 could show 
that $p_{\rm d}=1.833$ ($p=3.499$) describes the case of a self-similar
(and therefore infinite) collisional cascade. Wyatt \& Dent~(2002) find in fact a very good agreement 
of the dust SED from Fomalhaut's debris disk with the assumption of $p=3.499$ (see also Holland et al.~2003).
However, in general, real dust particle/planetesimal ensembles are expected
to be characterized by size-dependent particle strengths, therefore resulting in a distribution of
exponents $p$ as a function of grain size. The ``mean'' size distribution exponent could then be slightly
shifted from 3.499 (Durda \& Dermott~1997).
For example, Patin et al.~(1997) derived two different exponents to fit mid-infrared images of the $\beta$~Pic
disk: p=3.0 and 3.3 for the smallest grains (with radii up to 10$\mu$m) and the larger population, respectively.
Since particle growth/collision processes are expected to be mainly dependent on the radial distance from 
the star (Weidenschilling \& Davis~2000), 
the dependence on the (mean) grain size translates into a dependence 
of the exponent $p$ on the location within the disk.
Another argument for a clear deviation from $p=3.499$ comes from the fact that newly created small grains
are likely to be removed from the system faster than larger ones may decrease this value, at least on short
timescales, i.e., as long as a small grain population resulting from collision processes is present
(see, e.g., Moro-Mart\'{\i}n \& Malhotra~2003).
We consider two different values of the exponent, $p$ = 2.5 and 3.5.

\noindent
In Fig.~\ref{sidi} the SEDs resulting from dust disks with different grain size distributions are shown.
Based on the discussion above, we consider the following parameter combinations:
$a_{\rm min}$ = 0.1$\mu$m, 10$\mu$m;
$a_{\rm max}$ = 0.1mm, 1mm, 10mm; and
$p$ = 2.5 and 3.5.
Comparing the SEDs within these parameter ranges and the considered wavelength range
one can draw the main conclusion that the relative number of small grains in the system
(determined by $a_{\rm min}$, $a_{\rm max}$, and $p$) influences the shape of the SED.
In particular we find that: 
\begin{itemize}
\item 	Increasing the maximum grain size results in a decrease of the net flux and a decrease of dust 
  composition specific
  emission features. Since the amount of the underlying continuum can be due either to almost
  featureless emission by carbon grains of arbitrary size (distribution) or large silicate grains,
  the maxium grain size might become the most difficult parameter to be derived from \SIRTF\ measurements
  alone. However, millimeter measurements can be used to distinguish between the scenarios based
  on different slopes of the SED
  (similar to what has been done in case of T\,Tauri disks; see Beckwith \& Sargent~1991).
  
\item 	Given a fixed disk mass,
  increasing the grain size distribution exponent and decreasing the minimum grain size increases
  the relative amount of small grains and therefore pronounced dust-specific spectral features
  in the SED.
\end{itemize}

The SEDs shown in Fig.~\ref{sidi} illustrate that the relative abundance of small grains
and therefore the minimum grain size and grain size distribution exponent will be 
well-constrained disk parameters from SIRTF measurements due to the abundance of spectral features
in the 5-40$\mu$m range (see also Fig.~\ref{scaabs-1}, \ref{scaabs-2}, \ref{scaabs-5}).

\section{SED probability distribution}\label{sec-seddis}


\begin{figure}[t]
  \begin{center}
    \hspace*{20mm}
  \end{center}
  \caption{
    {\large \bf The complete paper can be downloaded from}
    \hspace*{5cm}{\tt http://mc.caltech.edu/$\sim$swolf/downloads/deb-sed.ps.gz}.
    \vspace*{1cm}
  Distribution of normalized debris disk SEDs in the range [0.2$\mu$m, 200$\mu$m].
  Ranges of variable disk parameters:
  $p=[2.5, 3.5]$,
  $q=[1.0, 2.5]$.
  The particle size ranges from which the minimum and maximum grain size are randomly selected are given
  in each figure.
  Left/Right column:
  Minimum and maximum grain sizes are selected from a logarithmically equidistant / linear distribution
  of grain sizes.
  Color coding: Dark regions represent a high SED density, while light grey regions symbolize
  low SED densities.
  The black upper and lower curves in each figure mark the range within which solutions (SEDs) were found.}
  See \S~\ref{sizedis} for a more detailed explanation and discussion.
  \label{seddis}
  \bigskip
\end{figure}


In this section we assess the likelihood of observing a particular SED given an
assumed underlying distribution of the parameters we have varied in the previous subsections.
We consider a plausible range of values for the
minimum and maximum grain size, the radial density distribution exponent $q$, and the 
grain size distribution exponent $p$. The other disk and dust parameters are fixed:
the inner radius (= sublimation radius), the outer radius (= 100\,AU) and
the chemical composition of the dust (amorphous Olivine - MgFeSiO$_4$).
According to the previous parameter range justifications for the variable quantities we 
calculate first the separate SEDs for $10^3$ grain sizes logarithmically distributed
between 0.1$\mu$m and 1000$\mu$m in disks with 31 different radial density distribution exponents
linearly distributed in the range $[1.0, 2.5]$, i.e., $3.1 \times 10^4$ separate SEDs.
In the second step we select a lower and upper grain size randomly from this grain size distribution,
assuming an equipartition of grain sizes on the prepared logarithmic grid of possible grain radii.
Furthermore, we randomly chose the grain size distribution exponent $p$ from one of 11 possible values
linearly distributed in the range $[2.5, 3.5]$ and one of the 31 radial density distribution
exponents $q$. In the third step we derive the net SED for a disk
defined by these randomly chosen parameter settings based on the linear weighted combination of the single SEDs
calculated in the first step.
In order to allow the comparison of the (shapes of) SEDs resulting from the 
$31 \times 11 \times \sum_{i=1}^{1000}i \approx 1.7\times10^8$
possible different disks, the net luminosity is normalized to ``1'' (arbitrary unit).
For this reason, the single SEDs had to be calculated over a larger parameter range than in the previous
studies in order to minimize neglected flux reemission appearing outside the considered wavelength range; 
here we chose $\lambda = [0.2\mu$m$, 4000\mu$m].
In Fig.~\ref{seddis} the distribution of $1.2\times10^7$ normalized, individual disk SEDs is shown,
whereby the density of SEDs in the flux-wavelength-space is color-coded.

This SED density distribution of course depends on the particular distribution of dust and disk parameters
within the given parameter ranges. In order to illustrate this, in Fig.~\ref{seddis} a second set
of SED density distributions is shown for which the minimum and maximum grain size have been randomly
selected from a linear distribution of grain sizes (instead of from a logarithmic one as before).
The main difference between the approaches is the much stronger pronounciation of the silicate
emission bands in case of a logarithmic distribution since in that case both the mean minimum
and mean maximum selected grain size is smaller than in case of a linear grid of possible grain size radii.
The ``real'' (unknown) distribution of possible grain size boundaries depends on the outcome of collision
processes. The comparison of observed SED density distributions with distributions as shown in Fig.~\ref{seddis}
but based on theoretically predicted collision scenarious may therefore provide a key to prove
the underlying physics of collision processes in debris disks.

The range in normalized flux density, however, over which possible debris disk SEDs are distributed does depend 
on the parameter range of the dust and disk parameters only, and is therefore the same in both apporaches.
Our simulations show that in case of a possible grain size distribution in the range of
0.1-1000$\mu$m the most narrow distribution of already normalized SEDs still spans a range over more than
one order of magnitude in possible fluxes (in the range of $\lambda \approx 40-50\mu$m), but
increases to above 3 orders of magnitude towards the lower and upper boundary of the considered
wavelength interval. 
The scattering of the SEDs slightly decreases if the range of possible grain radii is decreased
(also Fig.~\ref{seddis}). In particular, we find an additional minimum of possible flux values
around $\lambda \approx 50 - 70\mu$m for our particular dust/disk setup in case of large grain
size distributions ($a = 10 - 1000\mu$m).  This spectral regime may be the most diagnostic
for determining disk masses relatively independent of the details of the assumed disk or
grain properties.

\section{SIRTF Colors}\label{colors}

\begin{figure}[t]
  \begin{center}
    \resizebox{\hsize}{!}{\includegraphics{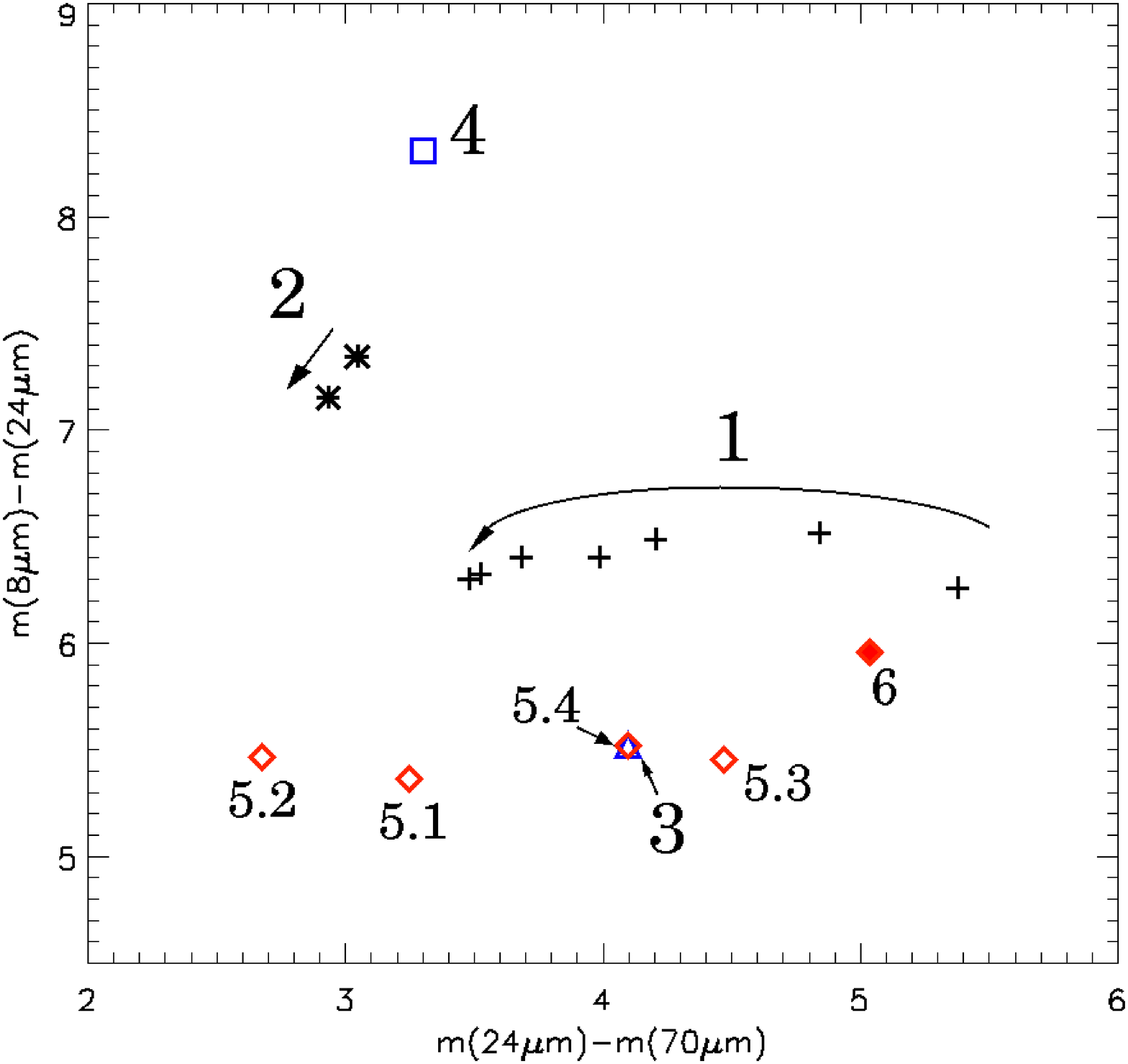}}
  \end{center}
  \caption{SIRTF color-color diagrams corresponding to the SEDs shown in Fig.~\ref{si-reihe-sed}.
    [1] 
    MgSiO$_3$ 
    $\rightarrow$
    Mg$_{0.95}$Fe$_{0.05}$SiO$_3$
    $\rightarrow$
    Mg$_{0.8}$Fe$_{0.2}$SiO$_3$
    $\rightarrow$
    Mg$_{0.7}$Fe$_{0.3}$SiO$_3$
    $\rightarrow$
    Mg$_{0.6}$Fe$_{0.4}$SiO$_3$
    $\rightarrow$
    Mg$_{0.5}$Fe$_{0.5}$SiO$_3$
    $\rightarrow$
    Mg$_{0.4}$Fe$_{0.6}$SiO$_3$;
    [2]
    MgFeSiO$_4$
    $\rightarrow$
    Mg$_{0.8}$Fe$_{1.2}$SiO$_4$;
    [3]
    ``Astronomical Silicate'';
    [4]
    Crystalline Olivine;
    [5] Carbon:
    [5.1] C (400\,K),
    [5.2] C (600\,K),
    [5.3] C (800\,K),
    [5.4] C (1000\,K);
    [6]
    Graphite.
  }
  \label{col3}
  \bigskip
\end{figure}

\begin{figure}[t]
  \begin{center}
    \resizebox{0.49\hsize}{!}{\includegraphics{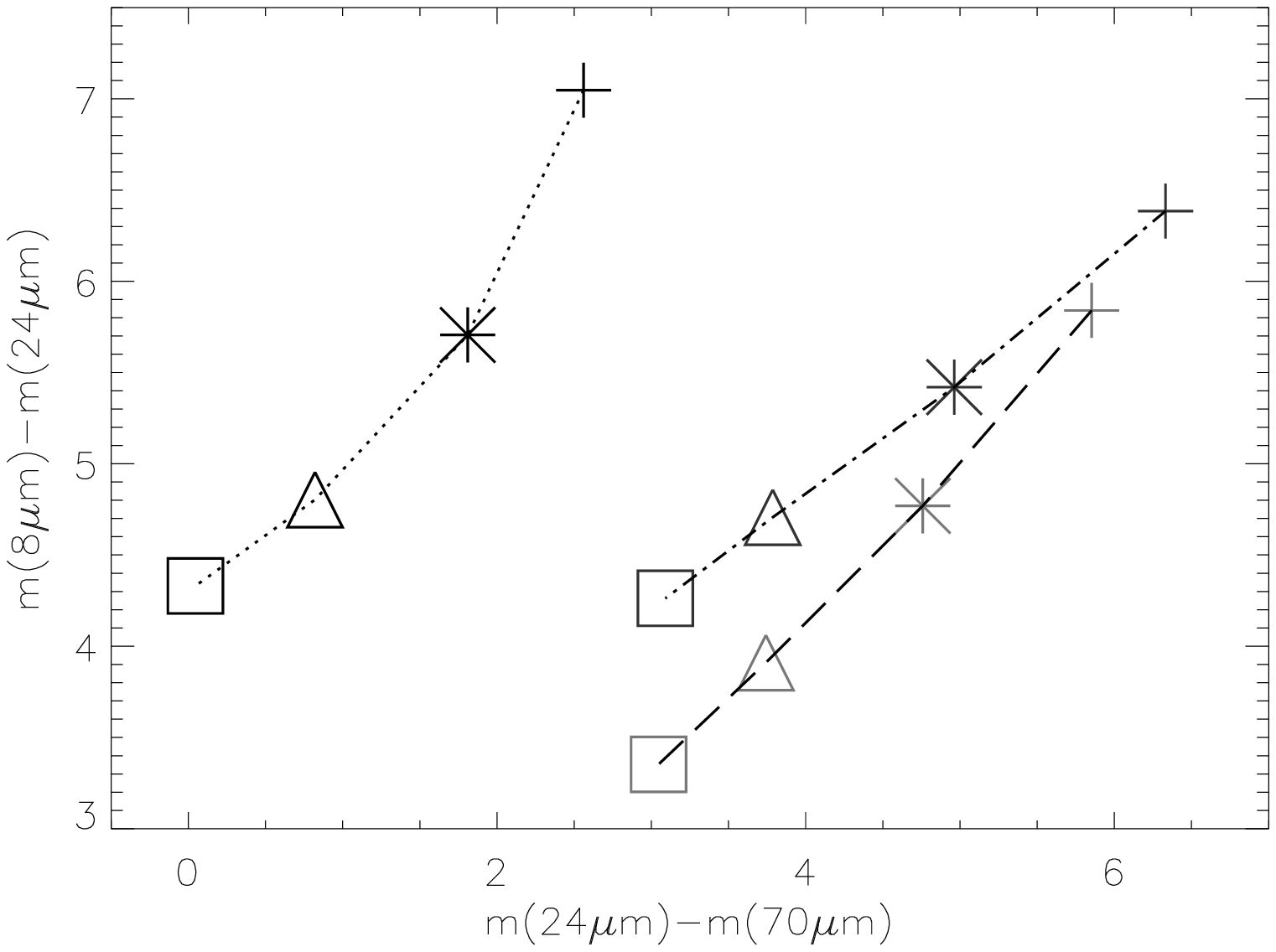}}
    \resizebox{0.49\hsize}{!}{\includegraphics{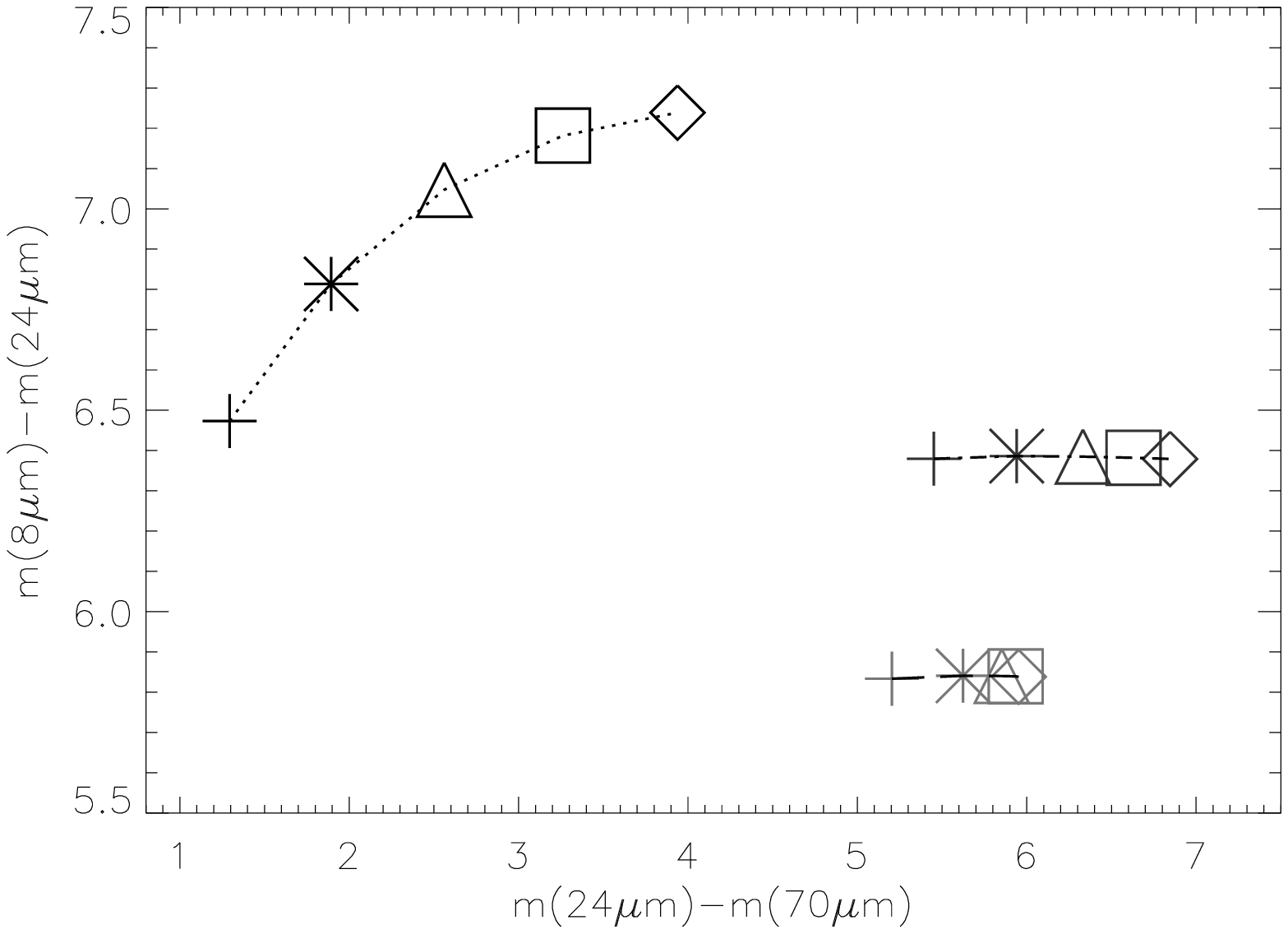}}
    \resizebox{0.49\hsize}{!}{\includegraphics{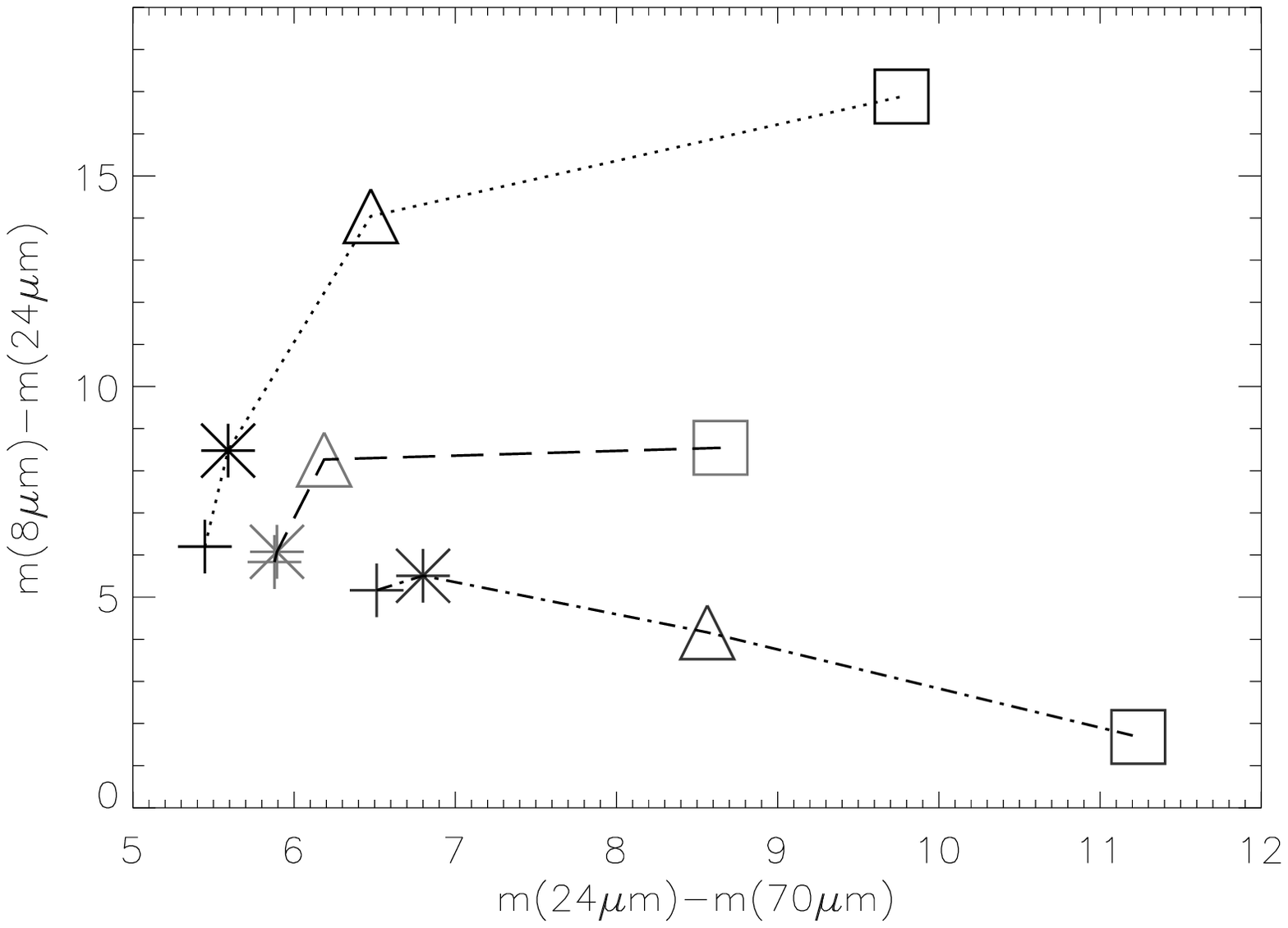}}
    \resizebox{0.49\hsize}{!}{\includegraphics{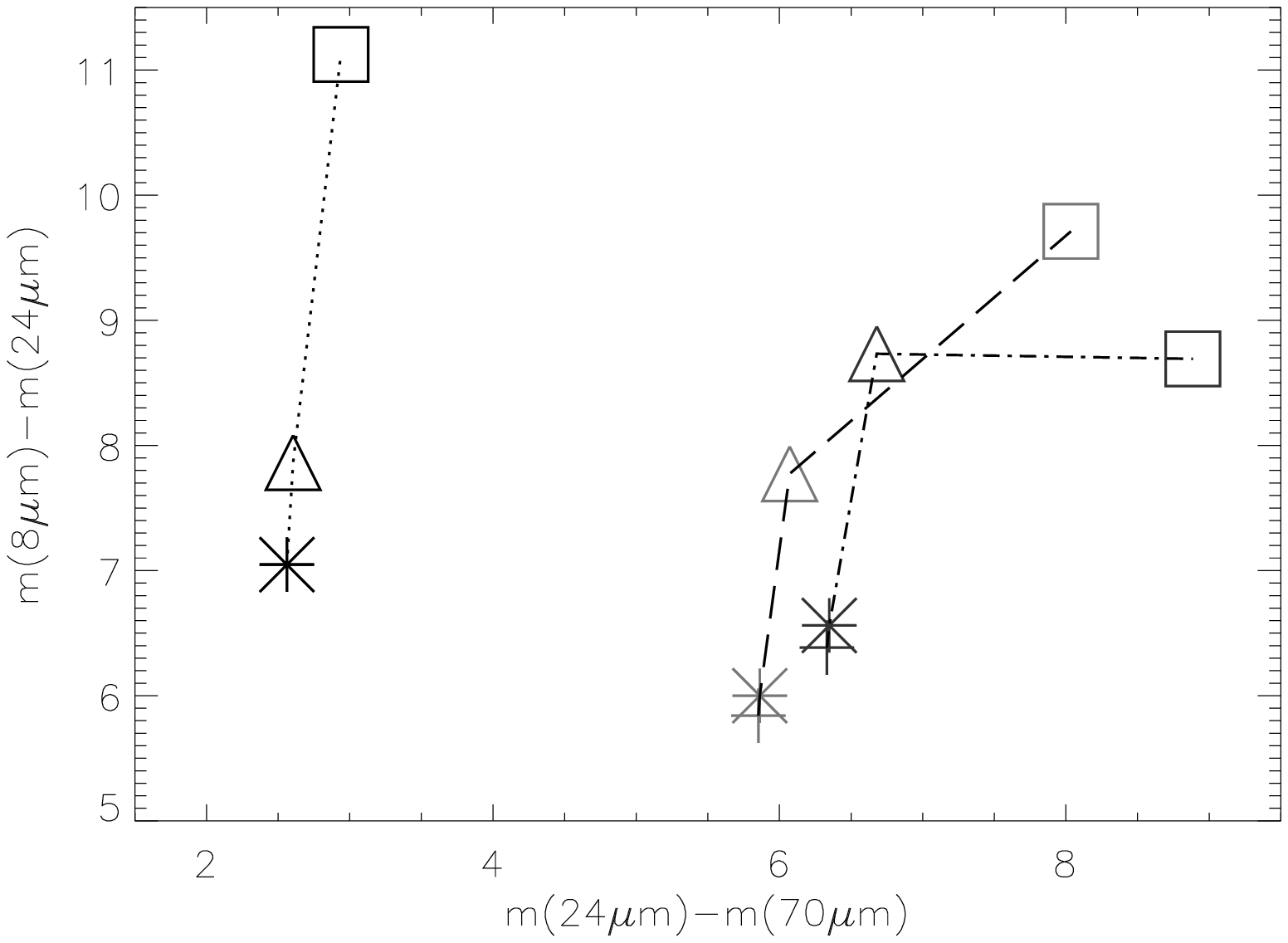}}
  \end{center}
  \caption{SIRTF color-color diagrams corresponding to the SEDs shown in 
    Fig.~\ref{q-dendis} (top left), Fig.~\ref{rou} (top right), and Fig.~\ref{innergap} (bottom).
    Different line styles represent different dust grain radii: 
    0.1$\mu$m (dotted), 
    10$\mu$m (dot-dashed), 
    1mm (dashed).
    {\em Top left:}
    Different symbols represent different power-law exponents $q$ of the disk density distribution:
    $q=1.0$ ({\sl plus}),
    $q=1.5$ ({\sl asterisk}),
    $q=2.0$ ({\sl triangle}), and
    $q=2.5$ ({\sl square}). 
    {\em Top right:}
    Different symbols represent different outer disk radii:
    25\,AU  ({\sl plus}),
    50\,AU  ({\sl asterisk}),
    100\,AU ({\sl triangle}),
    200\,AU ({\sl square}), and   
    400\,AU ({\sl rhombus}).     
    {\em Bottom left/right:}
    Different symbols represent different inner gap radii for the case
    of MgSiO$_3$ / MgFeSiO$_4$ grains:
     sublimation radius ({\sl plus}),
     0.1\,AU ({\sl asterisk}),
     1.0\,AU ({\sl triangle}), and
    10\,AU  ({\sl square}). 
  }
  \label{col1}
  \bigskip
\end{figure}

\begin{figure}[t]
  \begin{center}
    \resizebox{0.49\hsize}{!}{\includegraphics{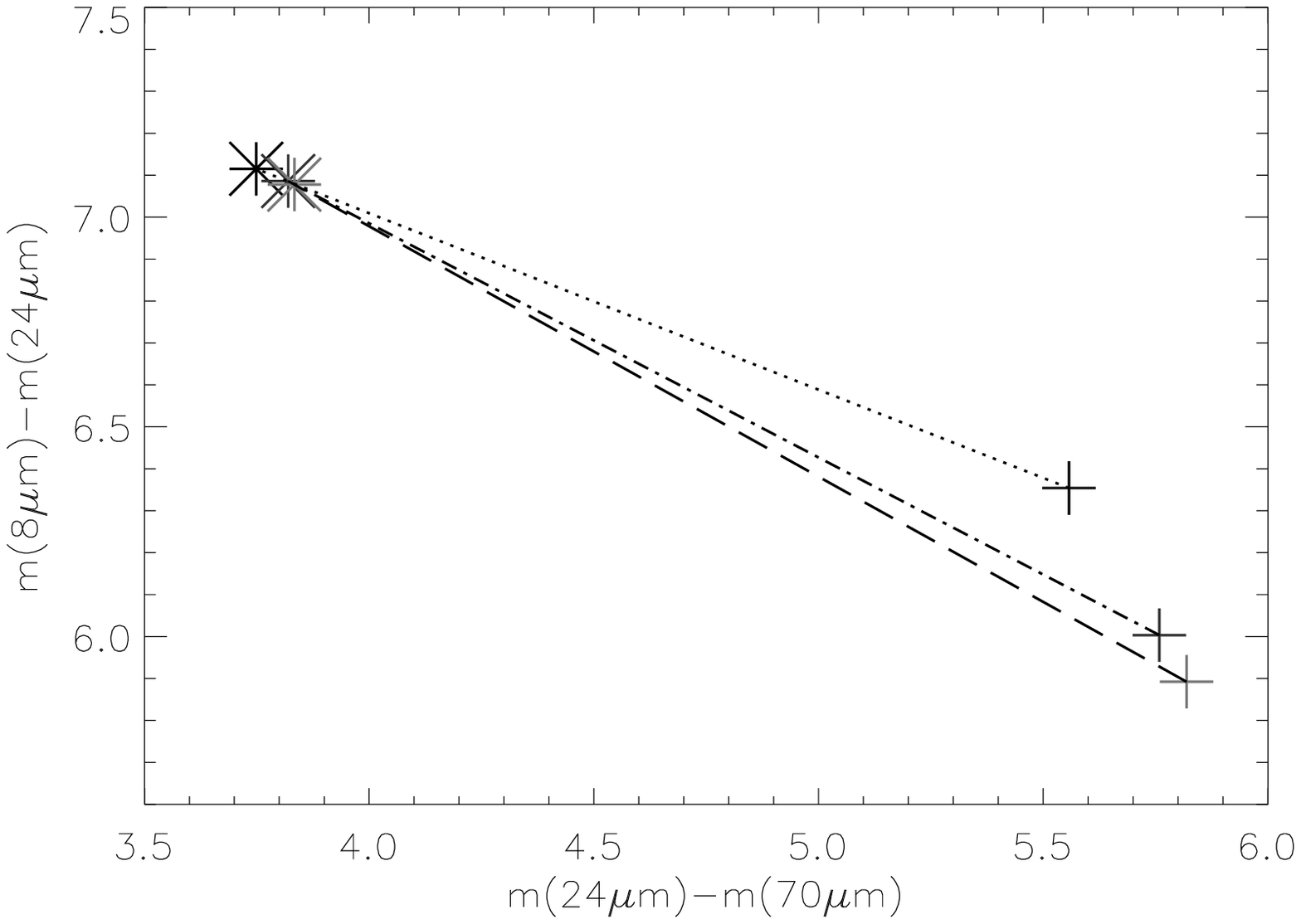}}
    \resizebox{0.49\hsize}{!}{\includegraphics{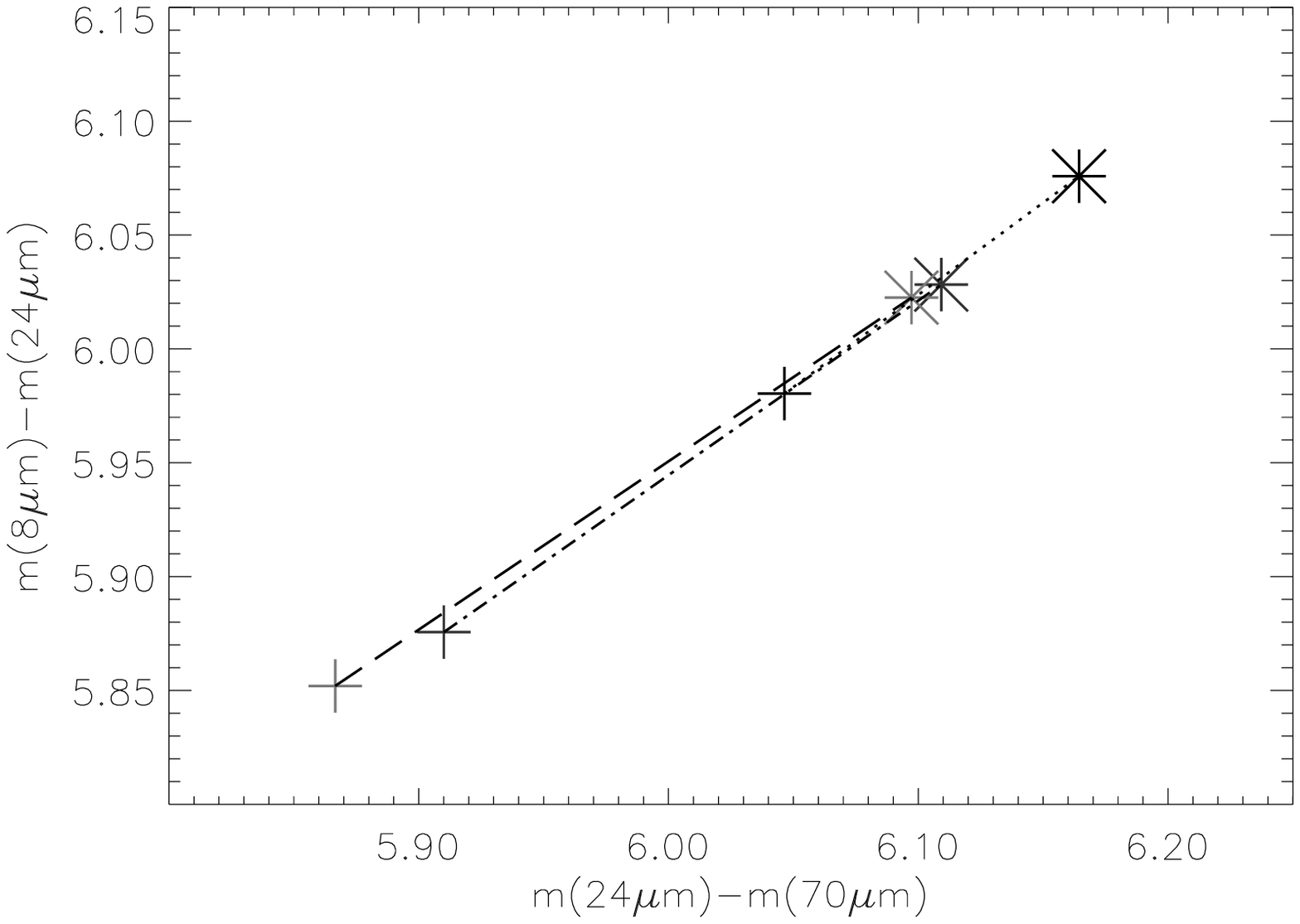}}
  \end{center}
  \caption{SIRTF color-color diagrams corresponding to the SEDs shown in Fig.~\ref{sidi}.
    {\em Left/Right:} Minimum grain radius amounts to 0.1$\mu$m / 10$\mu$m.
    Different line styles represent different maximum dust grain radii: 
     0.1mm (dotted),
     1mm   (dot-dashed), 
    10mm   (dashed).
    Different symbols represent different power-law exponents $p$ of the grain size distribution:
    $p=2.5$ ({\sl plus}), and
    $q=3.5$ ({\sl asterisk}).
  }
  \label{col2}
  \bigskip
\end{figure}


Based on selected SEDs presented in previous sections we derive the colors $m(8\mu$m$)-m(24\mu$m$)$
and $m(24\mu$m$)-m(70\mu$m$)$, applying the filter characteristics of the \SIRTF\ cameras
IRAC (8$\mu$m) and MIPS (24$\mu$m, 70$\mu$m).
The resulting color-color diagrams, shown in Fig.~\ref{col1}--\ref{col3}, are normalized
to a blackbody with a temperature of 9650\,K which sufficiently accurately represents the 
mid-infrared - to - submillimeter SED of the A0V standard star Vega (Castelli \& Kurucz~1994, Decin et al.~2003):
\begin{equation}
  m_1 - m_2 = 
  -2.5 
  \left(
  \log_{10}
  \frac{\int_{0.2\mu{\rm m}}^{200\mu{\rm m}}  T_1 S_{\lambda} d\lambda}{\int_{0.2\mu{\rm m}}^{200\mu{\rm m}} T_2 S_{\lambda} d\lambda}
  \right)
  + (m_{1,\,\rm Standard} - m_{2,\,\rm Standard}), 
\end{equation}
where
\begin{equation}
  m_{1,\,\rm Standard} - m_{2,\,\rm Standard} = -2.5 
  \log_{10}
  \frac{\int_{0.2\mu{\rm m}}^{200\mu{\rm m}} T_1 B_{\lambda}(9650K) d\lambda}{\int_{0.2\mu{\rm m}}^{200\mu{\rm m}} T_2 B_{\lambda}(9650K) d\lambda}.
\end{equation}
Here,
$S_{\lambda} = S_{\nu}\, d\nu/d\lambda$ is the dust scattering/reemission SED, 
$T_{1}$ and $T_{2}$ describe the wavelength-dependent filter 
transmission\footnote{\SIRTF\ instruments characteristics are available at {\tt http://sirtf.caltech.edu/SSC/obs/}.},
and $B_{\lambda}$ is the Planck function.
Note in these figures that only the dust emission / scattering spectrum is represented;
the stellar photosphere has been removed in the SEDs.

These diagrams reflect what has been outlined in the discussions about the possibility 
to distinguish between different disk models. 
For example, the color-color diagram for disks with different outer radii
(Fig.~\ref{col1}, upper right plot, based on the SEDs given in Fig.~\ref{rou})
shows a large dynamic range of colors and therefore clearly distinguishable models 
only in case of a small dust grains (here: 0.1$\mu$m).
In contrast to this, a grain size distribution with grain radii of 10$\mu$m and 1mm the almost constant
offset between the SEDs in the mid-infrared wavelength range results in a nearly constant value
of $m(8\mu$m$) - m(24\mu$m$)$.

\section{Summary and Conclusions}

In preparation for \SIRTF\ and {\em SOFIA} observations of circumstellar debris disks we have investigated
the influence of various disk and dust parameters on the resulting SED,
in particular the offset from the stellar photospheric flux due to dust reemission and scattering processes.
Based on an analytic disk model we considered parameter ranges as suggested from existing observations
of a small number of debris disk systems as well as constraints derived from theoretical approaches.
We restricted our investigations primarily to the wavelength range accessible
by \SIRTF. 

We draw the following main conclusions from our study of the parameter space:
\begin{enumerate}
\item The correct estimation/substraction of the stellar photospheric flux is 
  essential for the subsequent data analysis in two respects.
  First, it is important for the correct estimation of the dust continuum (not taking into account any
  particular dust emission features) which mainly constrains the disk mass,
  the upper grain size and grain size distribution exponent (\S~\ref{sizedis}),
  the outer disk radius (\S~\ref{outerradius}), and
  the overall radial disk density distribution exponent (\S~\ref{diskdendis}).
  Secondly, the strength of remaining stellar spectral features allows one to derive conclusions about the scattering
  to absorption efficiency (or albedo) of the dust grains and therefore
  - based on a simple silicate/carbon dust grain model -
  to derive constraints on the
  crystalline silicate / iron-poor amorphous silicate to
  iron-rich amorphous silicate / carbon ratio (\S~\ref{efficiencies}).

\item The characteristic amorphous silicate broad band solid state emission features are predicted
  to be of decisive importance for the characterization of the evolutionary state of debris disks. 
  The appearance of these features
  depends strongly on the existence of small grains (smaller than about 10$\mu$m), the abundance
  of which, however, is assumed to be small in a collisionless system
  (depending on the stellar SED; \S~\ref{efficiencies} and \S~\ref{sizedis}).
  The strength of these features will allow to decide if collision processes took place still recently
  in the disk of the particular star.
  However, one has to take into account that only ``snapshots'' of the current, in respect of the timescale
  of the collisional events arbitrarily selected states of debris disks can be obtained.
  The SED and thus the derived smallest size of highly abundant grains does therefore not necessarily
  describe the real, ``mean'' evolutionary state of a particular disk (if defined by the time-dependent
  collision rate).
  However, surveys based on a statistically large number of debris disk systems with large subsets of stars 
  with a similar age will allow to derive general characteristics of debris disks as a function
  of their (evolutionary) age.

\item The mid-infrared continuum will allow to conclude if an inner hole
  - due to the presence of at least one embedded planet - exists. Furthermore, the inner radius
  of that gap (= orbit of the planet) could be derived (\S~\ref{sec_gaps}).
\end{enumerate}

The presented study was performed to address the most fundamental questions one will want to answer with the analysis
of the SED of debris disk systems. Based on this ``first-level'' analysis, a more detailed investigation of particular
debris disk systems will be possible. For instance, 
the radial segregation of chemically different dust grains,
the spatial distribution of ice-mantled grains with spectral features in the range of 10-100$\mu$m range
(e.g., Warren~1984),
the possibility to distinguish qualitatively different planetary architectures embedded in debris disks 
(Moro-Mart\'{\i}n \& Malhotra~2002),
and the determination of characteristic wavelength ranges in which planetary radiation
-- in particular astrobiological tracers -- can be separated from the underlying dust continuum radiation 
(Beichmann~1996, Fischer \& Pfau~1997)
are tasks which are beyond the recent study and will be considered in a subsequent investigation.

\acknowledgments

S.~Wolf was supported through the NASA grant NAG5-11645 and through the SIRTF legacy science
program through an award issued by JPL/CIT under NASA contract 1407.
We are grateful for valuable discussions with Th. Henning, S. Weidenschilling, A. Moro-Mart\'{\i}n 
and the other members of the FEPS team led by M. Meyer.



\end{document}